\documentclass[12pt]{iopart}

\usepackage{graphicx}
\usepackage{dcolumn}
\usepackage{bm}
\usepackage{ccaption}
\usepackage{iopams}
\usepackage{setstack}
\usepackage[dvips]{color}

\newcommand{\vq}{\vec{q}}
\newcommand{\ve}{\vec{e}}

\newcommand{\vx}{\vec{x}}
\newcommand{\vy}{\vec{y}}
\newcommand{\vz}{\vec{z}}
\newcommand{\vr}{\vec{r}}

\newcommand{\eps}{\,\epsilon}

\newcommand{\sigl}{\sigma_l}
\newcommand{\sigd}{\sigma_\odot}

\newcommand{\ccr}{\,\hat{c}^{\dag}}
\newcommand{\can}{\,\hat{c}}

\newcommand{\hpsi}{\hat{\psi}}
\newcommand{\psid}{\hat{\psi}^{\dag}}
\newcommand{\hPsi}{\hat{\Psi}}
\newcommand{\Psid}{\hat{\Psi}^{\dag}}
\newcommand{\sgn}{{\rm{sgn}}}
\newcommand{\mod}{{\,\mathsf{mod}\,}}
\newcommand{\hrho}{\hat{\rho}}
\newcommand{\leads}{\mathsf{leads}}

\newcommand{\bias}{\mathsf{bias}}
\newcommand{\gate}{\mathsf{gate}}
\newcommand{\ket}{\rangle}
\newcommand{\bra}{\langle}
\newcommand{\kBT}{k_BT}
\newcommand{\dgr}{^{\circ}}

\newcommand{\pint}{\ \ \mathcal{P}\hspace{-1em}\int}

\newcommand{\simgr}{\ {>}_{{}_{\hspace{-0.7em}\sim}}\ }

\newfixedcaption{\imagecaption}{figure}
\newfixedcaption{\tabcaption}{table}

\begin{document}

\title{Spin transport across carbon nanotube quantum dots}

\author{Sonja Koller, Leonhard Mayrhofer and Milena Grifoni}
\address{Institut f\"ur Theoretische Physik, Universit\"at Regensburg, 93040 Regensburg, Germany}%
\ead{Sonja.Koller@physik.uni-regensburg.de}
\date{\today}

\begin{abstract}
We investigate linear and nonlinear transport in interacting single-wall carbon nanotubes (SWCNTs) that are weakly attached to ferromagnetic leads. For the reduced density matrix of a SWCNT quantum dot, equations of motion which account for an arbitrarily vectored magnetisation of the contacts are derived. We focus on the case of large diameter nanotubes where exchange effects emerging from short-ranged processes can be excluded and the four-electron periodicity at low bias can be observed. This yields in principle four distinct resonant tunnelling regimes, but due to symmetries in the involved groundstates, each two possess a mirror-symmetry. With a non-collinear configuration, we recover at the $4\mathbb{N}\leftrightarrow4\mathbb{N}\pm1$ resonances the analytical results known for the angular dependence of the conductance of a single level quantum dot or a metallic island. The two other cases are treated numerically and show on the first glance similar, yet not analytically describable dependences. In the nonlinear regime, negative differential conductance features occur for non-collinear lead magnetisations.
\end{abstract}

\maketitle

\section{Introduction}\label{intro}
Carbon nanotubes are promising candidates for constituent of tomorrow's electronic nanodevices \cite{GRA} and thus subject of today's theoretical \cite{EGGO,KBF,ORG,BE,THW,LEO2,LEO} and experimental investigations \cite{TANS,BOCK,TAA,LBP,KIM,SAH2,SAP,MIY,SAH,MAN}. A single-wall carbon
nanotube (SWCNT) is just formed by the planar honeycomb lattice of graphene, wrapped along one axis such that a closed, seamless cylindrical surface arises \cite{SAI}. To guarantee
this, the axis and the tube diameter cannot be chosen completely arbitrary, but in principle there are countless possible configurations. Only special ones, however,
inherit the metallic properties of graphene, which stem from the fact that in graphene conduction and valence band touch at the corner points of the first Brillouin zone. Two of
those points are independent and labelled Fermi points $\pm \vec{K}_0$. To be metallic, a SWCNT must contain the Fermi points in its reciprocal lattice of allowed momenta and
all armchair type tubes, as well as certain types of zig-zag configurations \cite{SAI} fulfil this condition (but curvature effects yield a small band gap for all SWCNT types except for the armchair ones). For low energy processes, just momenta close to the Fermi points can contribute, and as the SWCNT
diameter is normally a small fraction of the tube length, the size quantisation permits many longitudinal, but merely one radial mode. This fact makes the carbon
nanotube a one-dimensional conductor.  Moreover, the shape of the graphene bandstructure is very beneficial in the considered region: it develops linearly around the Fermi points,
which allows to describe the electronic properties of a nanotube by means of the Tomonaga-Luttinger model for interacting fermions \cite{EGGO,BE}, involving a bosonisation of
the electron operators.\smallskip\\A research topic in rapid expansion is spin-dependent transport. In particular, recent theoretical works have focussed on transport across interacting quantum dots \cite{KOE1,KOE,COT}, metallic islands \cite{WOU} and wires \cite{BE} with non-collinearly magnetised leads, offering the prospect of spin-sensitive single electron
transistors. It has been already realised that interactions \cite{KOE1,KOE,WOU} as well as reflection processes at the lead-system-interface \cite{COT2,COT} can strongly influence the spin-accumulation in the dot (island), and hence the properties of the spin-valve transistor. During the last years, there have been various experimental investigations on spin-dependent transport in SWCNTs \cite{TAA,KIM,SAH2,SAH,MAN}, and a possible measurement setup is sketched in Fig. \ref{expsetup}. Notice that we include an arbitrary relative magnetisation of the leads and do not restrict ourselves to a common limitation of nowadays experiments on SWCNTs, where merely the two discrete states of parallel (P) and antiparallel (AP) contact magnetisation can be realised (\footnote{To be able to obtain an AP configuration, usually source and drain contact are manufactured with different width so that the coercive fields emerging in the metals are not of equal strength. Performing a sweep of an external magnetic field, the wider contact switches later to the opposite magnetisation than the smaller one and for some range of the applied field the leads are thus polarised in antiparallel.}). There have already been experimental studies on non-collinear spin transport in magnetic multilayers \cite{URA}, so we can expect that in the near future it will be possible to attach arbitrarily polarised contacts to a SWCNTs.\\

\begin{center}
\includegraphics[width=0.72\textwidth]{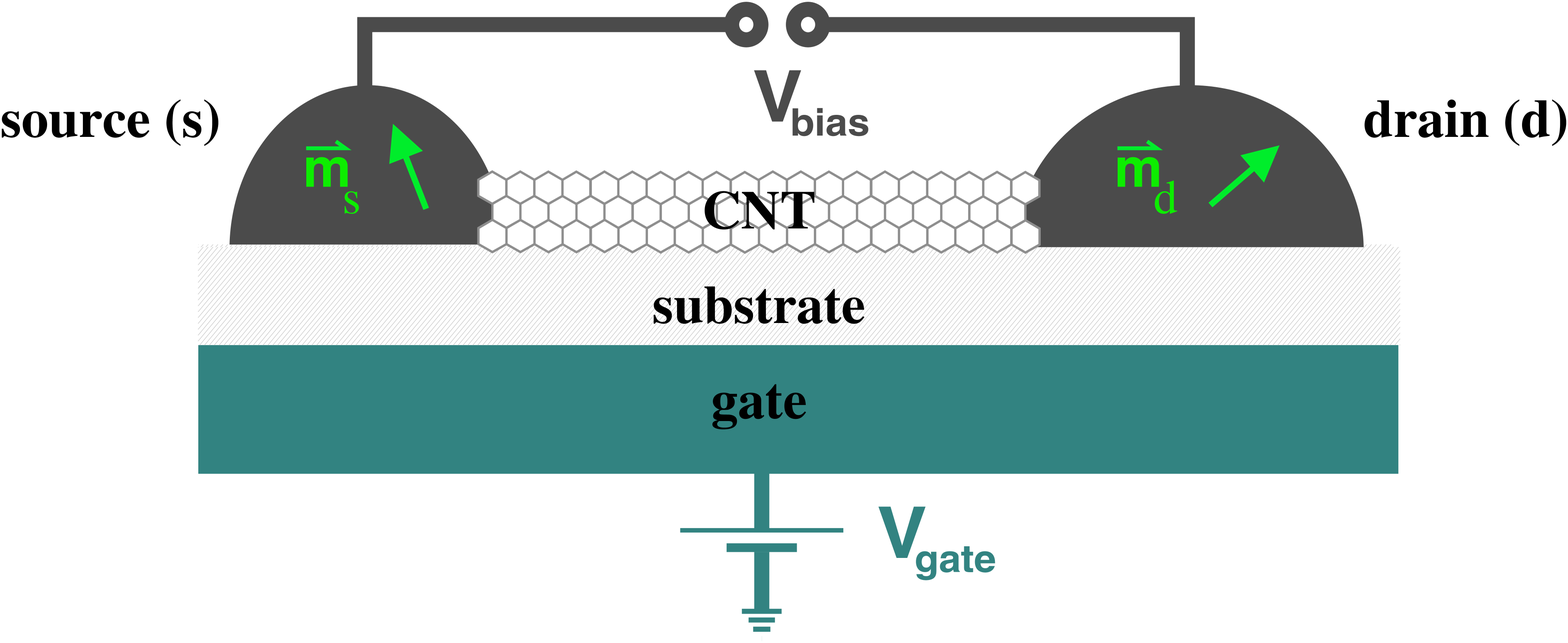}
\imagecaption{\small Ferromagnetic contacts are attached to the SWCNT. Besides a bias voltage $V_\bias$, a gate voltage $V_\gate$ can be applied via a back gate in order to shift the chemical potential inside the tube. The magnetisations $\vec{m}_s$ and $\vec{m}_d$ are controlled by varying an external magnetic field.}\label{expsetup}
\end{center}\vskip 0.4cm

A variable of special interest in spin-dependent transport experiments is the so-called \textit{tunnelling
magnetoresistance} (TMR), which we define for collinear contact magnetisations as (TMR)\,$=(G_{P}-G_{AP})/G_{P}$, where $G_{i}$ is meant to denote the conductance in the parallel ($i=P$) resp.
antiparallel ($i=AP$) configuration (\footnote{There are various other definitions of the TMR, e.g. $(R_{AP}-R_{P})/R_{P}$ \cite{KIM,SAH}, $(G_{P}-G_{AP})/(G_P+G_{AP})$ \cite{COT}, $2(R_{AP}-R_{P})/(R_P+R_{AP})$ \cite{SAH2}, where $R_i$ denotes the resistances in the parallel ($i=P$) resp. antiparallel ($i=AP$) configuration. The physical meaning, nevertheless, is always the same, as merely the normalisations differ.}). In Ref. \cite{SAH}, the observation of quite regular TMR oscillations with the gate voltage is reported, where large changes of almost $20\%$ in the conductance are reached. Moreover, the TMR acquires negative values, which means that the AP conductance can exceed the P one.\smallskip\\So far, spin-polarised transport in interacting SWCNT quantum dots has been considered only for very long tubes, which do not exhibit level quantisation and charging effects \cite{BE}. For this latter case, a non-trivial dependence of the current on the interaction strength was predicted. Recently, a single impurity Anderson model with four degenerate orbitals was proposed \cite{COT} as a minimal model to understand the aforesaid negative TMR effect reported in \cite{SAH}. Particularly, the authors could show that it can originate from multiple reflection processes at the SWCNT-contact interfaces.\smallskip\\In this work, a microscopic treatment of spin-dependent transport in SWCNT quantum dots is presented. Specifically, we focus on the \textit{low transparency} regime, where a weak coupling between tube and contacts is assumed. Therefore it is justified to treat all Hamiltonians associated with the tunnelling barrier in lowest nonvanishing order. The opposite case of high transparency, i.e. low ohmic contacts, has already been studied both experimentally \cite{MAN} and theoretically \cite{PEC,KRO}; it was found that phase shifts picked up during backscattering events at the tube ends yield a Fresnel/Fabry-Perot interference pattern in $V_\gate-V_\bias-$TMR plots .\smallskip\\Our model takes into account interface reflections, as well as virtual transition processes. Both are relevant exchange effects for spin-dependent transport \cite{WOU}, inducing a precession of the spin accumulating on the quantum dot.\smallskip\\For strict lowest order perturbation theory, nevertheless, the interface reflections cannot be source of any negative TMR, as it will become clear in the course of this article, which is structured as follows:\smallskip\\In section \ref{modham} we describe the model Hamiltonian we use for our system, section \ref{mastereq} explains how to derive the master equation for the SWCNT density matrix in the presence of arbitrary lead magnetisations. Section \ref{spinaxis} introduces the necessary coordinate transformations to proper spin quantisation axes and subsequently, section \ref{lintrans} contains the results we acquire for the current in the linear transport regime. Section \ref{nonlintrans} presents numerical data for the nonlinear case and finally section \ref{conclusion} gives a summary of our achievements. In the appendix, some explicit calculations skipped in section \ref{mastereq} can be found.

\section{Model Hamiltonian}\label{modham}
To build up a model for spin-dependent transport in the quantum dot regime of a SWCNT, we expand the theoretical work \cite{LEO} on correlated transport in carbon nanotube quantum dots to include spin polarisation of both contacts. The magnetisations $\vec{m}_s$ and $\vec{m}_d$ may enclose an arbitrary angle $\theta$, which is a possibility future experiments are likely to offer (see Fig. \ref{expsetup}).\\

The Hamilton operator we use for the setup in Fig. \ref{expsetup} is

\begin{equation}
 \hat{H}=\hat{H}_{\odot}+\sum_{l=s,d}\hat{H}_{l}+\hat{H}_{Tl}+\hat{H}_{Rl}\,,
\end{equation}where effects of external voltages have been absorbed into $\hat{H}_{\odot}$ and $\hat{H}_{l}$. The index $l$ labels the source ($l$=$s$) and the drain ($l$=$d$) contacts, which are metallic and thus characterised by a Hamiltonian

\begin{equation}\hat{H}_l =\sum_{\vq}\left(\eps_{\vq}-\sgn(\sigl)E_{\mathsf{Stoner}}-eV_l\right)\ccr_{l\sigl\vq}\can_{l\sigl\vq}\,.\label{l_ham}\end{equation}
In (\ref{l_ham}), $\can_{l\sigl\vq}$ is the fermionic annihilation operator for an electron of momentum $\vq$ in lead $l$ with spin $\sigl\in\{+_l,-_l\}$, where $\pm_l$ denotes the majority/minority spin species for a quantisation along $\vec{\sigma}_l$. For reasons of simplicity, we just apply the Stoner model for our ferromagnetic contacts, but any other description could be used as well. Both the bias voltage $V_l$ and the Stoner exchange splitting energy $E_{\mathsf{Stoner}}$ for the two spin species in the ferromagnetic leads add a shift to the kinetic energy $\eps_{\vq}$ of the particle. $e$ is the electron charge.\smallskip\\
The tunnelling processes at contact $l$ are modelled by

\begin{equation}\hat{H}_{Tl}=\sum_{\sigl} \int\!\rmd{}^3 r\ \left(T_l(\vr)\,\Psid_{\odot\sigl}(\vr)\hPsi_{l\sigl}(\vr)+h.c.\right)\,.\label{lt_ham}\end{equation}
The so-called \textit{transparency} $T_l(\vr)$ specifies the tunnelling properties of lead $l$, $\hPsi_{\odot\sigl}(\vr)$ and $\hPsi_{l\sigl}(\vr)$ are the electron annihilation operators in real space representation, for a particle of spin $\sigl$ in the tube resp. in lead $l$. For later purposes it is necessary to know the decomposition

\begin{equation}
\hPsi_{l\sigl}(\vr)=\int\!\rmd{}\epsilon\ D_{l\sigl}(E^{l\sigl}_{tot}|_{\epsilon})\sum_{\vq\left.\right|_{\epsilon}}\phi_{l\vq}(\vr)\can_{l\sigl\vq}\,,\label{hPsi_l}
\end{equation}where $D_{l\sigl}$ is the density of states for carriers of spin $\sigl$ in lead $l$, taking as an argument the total energy

\[E^{l\sigl}_{tot}|_{\epsilon}=\epsilon-eV_l-\sgn(\sigl)E_{\mathsf{Stoner}}\]of a particle. $\phi_{l\vq}(\vr)$ is just a wave function inside the lead.\smallskip\\Furthermore, we allow for boundary backscattering by introducing the `reflection' Hamiltonian

\begin{equation}\hat{H}_{Rl}=-\int\!\rmd{}^3r\ \Delta_{l}(\vr)\sum_{\sigl}\sgn(\sigl)\Psid_{\odot \sigl}(\vr)\hPsi_{\odot \sigl}(\vr)\,,\label{lr_ham}\end{equation}which is equivalent to the momentum space expression used e.g. in \cite{WOU} (where also the relation to the \textit{mixing conductance} introduced in \cite{BRA} is explained). An electron picks up some phase when it is scattered at the tube ends, which overall results in a certain energy shift, coming with a positive or a negative sign depending on the spin polarisation. In other words, being close to a contact, the electron feels a bit of the magnetic field inside it, causing a position dependent energy splitting $\Delta_{l}(\vr)$ for the two spin species.\\We assume the tunnelling barrier to be spin-independent and as there is only weak spin-orbit coupling in clean nanotubes, it is justified not to consider any spin-flip processes.\\At this stage, we still want to stay general and do not introduce any assumption on the position dependence of $T_{l}(\vr)$ or $\Delta_{l}(\vr)$. Later on, however, we will have to impose the restriction that both parameters are of relevant value only nearby the contacts.\smallskip\\
The most complex contribution to the system Hamiltonian is given by the terms which belong to the SWCNT itself, due to the presence of Coulomb interaction.\\The SWCNT Hamiltonian $\hat{H}_{\odot} = \hat{T}_\odot+\hat{V}_\odot$ reads 

\begin{eqnarray}
\fl\hat{H}_{\odot}=\hbar v_{F}\sum_{\tilde{r}\sigd}\sgn(\tilde{r})\sum_{n\in\mathbb{Z}}\,n\, \ccr_{\tilde{r}\sigd n}\can_{\tilde{r}\sigd n}+\nonumber\\+\frac{1}{2}\!\sum_{\sigd\sigd'}\int\!\!\int\rmd{}^{3}r\ \rmd{}^{3}r'\,\hPsi_{\odot\sigd}^{\dagger}(\vec{r})\hPsi_{\odot\sigd'}^{\dagger}(\vec{r}\,')V(\vec{r}-\vec{r}\,')\hPsi_{\odot\sigd'}(\vec{r}\,')\hPsi_{\odot\sigd}(\vec{r})\,.\label{hodot1}\end{eqnarray}
Notice that we index the electron operators by a spin $\sigd,\sigd'\in\{\uparrow,\downarrow\}$ which refers to an arbitrary, but fixed unique quantisation axis \textit{inside} the tube as e.g. sketched in Fig. \ref{kosys1}.\smallskip\\
The first term in (\ref{hodot1}), $\hat{T}_\odot$, collects the kinetic energy of particles in the nanotube. To recognise this, one needs some knowledge about the SWCNT bandstructure. We already pointed out in the introduction that from the graphene lattice, the reciprocal space of an armchair SWCNT inherits two Fermi points $\pm K_0$, where valence and conduction band touch.
Transport can take place in the vicinity of these points, around which the bands develop linearly and, due to our restriction to low energies, one-dimensional. Fig. \ref{linband} (left) shows the two branches we obtain at each Fermi point when periodic boundary conditions (PBCs) are employed \cite{SAI}: a left mover band $L$ of negative slope and a
right mover band $R$ of positive slope; both slopes have an absolute value $\hbar v_F$, where $v_F$ is the Fermi velocity. By the linear transformations

\begin{equation}
\eqalign{
\varphi^{OBC}_{\tilde{R}\kappa_n}(\vr):=\frac{1}{\sqrt{2}}\left(\varphi_{R\kappa_nK_0}(\vr)-\varphi_{L[-\kappa_n][-K_0]}(\vr)\right)\,,\\\varphi^{OBC}_{\tilde{L}\kappa_n}(\vr):=\frac{1}{\sqrt{2}}\left(\varphi_{L\kappa_nK_0}(\vr)-\varphi_{R[-\kappa_n][-K_0]}(\vr)\right)\,,}\label{phitrans}
\end{equation}wave functions $\varphi_{\tilde{r}\kappa_n}$ which fulfil open boundary conditions (OBCs, Fig. \ref{linband} (right)) --\,those are the appropriate ones for the finite-size system the tube represents\,-- are constructed from the usual PBC wave functions $\varphi_{r\kappa_nF}$ ($r\in\{L,R\}\,$). The latter can, as worked out in \cite{LEO}, be freed from the dependence on the momentum $\kappa_n$ by the approximation
\begin{equation}\label{phiappr}\varphi_{r\kappa_nF}(\vr)\approx \rme^{\rmi\kappa_n x}\varphi_{rF}(\vr)\,.\end{equation}The transformation (\ref{phitrans}) maps the four $rF$ branches onto two bands $\tilde L$ and $\tilde R$, whereas the number of admitted momenta $\kappa_n=\hbar v_f n\,,\ n\in \mathbb{Z}$ doubles \cite{LEO}.

\begin{center}
\vskip 0.6cm
\includegraphics[width=12cm]{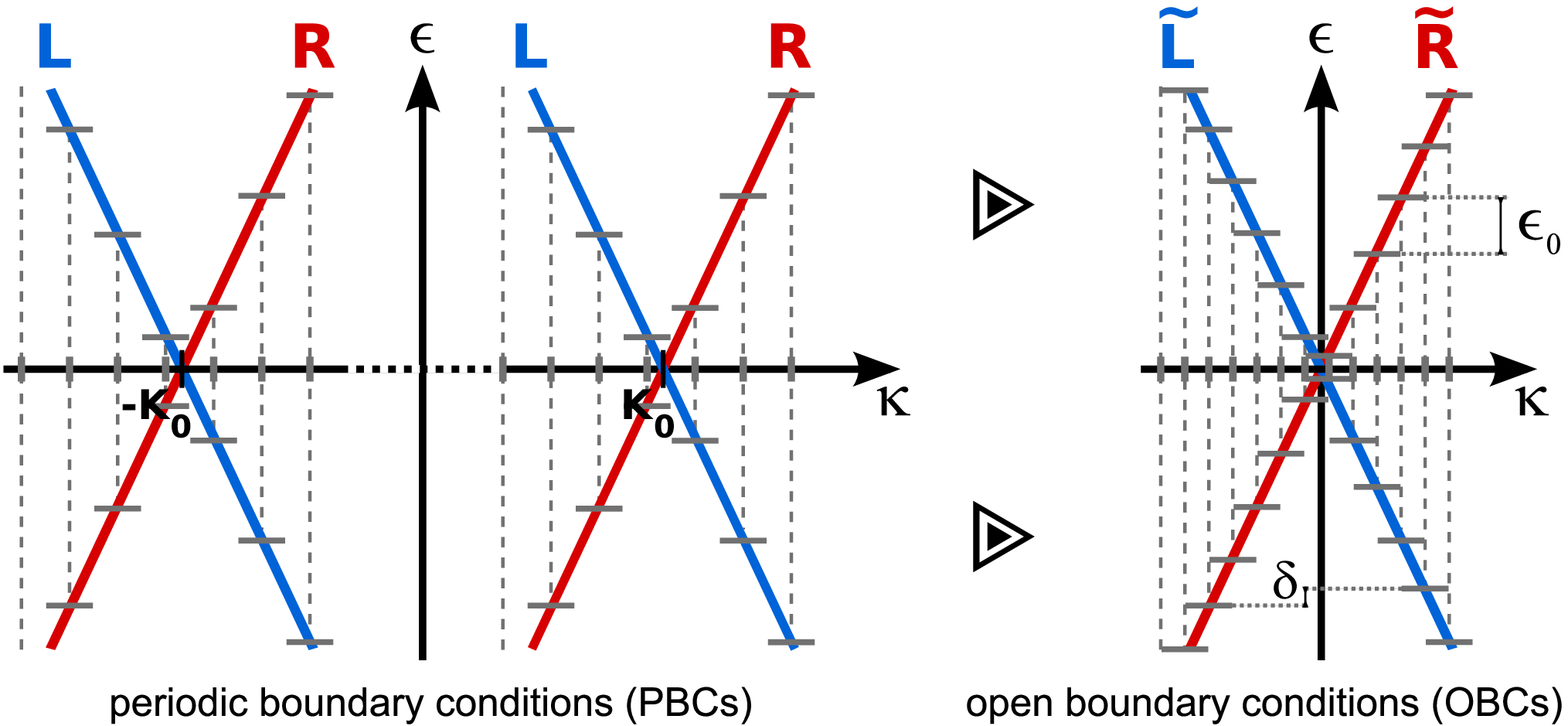}\vskip 0.4cm
\imagecaption{\small Band structure of a SWCNT with periodic (left, PBC) and open boundary conditions (right, OBC). For (non-interacting) wave functions which fulfil OBCs, the spectrum is characterised by two possible bands $\tilde{L}$ and $\tilde{R}$ and momenta $\kappa_n=\hbar v_f n\,,\ n\in \mathbb{Z}$. The energy levels of left- and right mover branches can be shifted with respect to each other, which results in a nonzero band offset $\delta$.}
\label{linband}
\vskip 0.4cm
\end{center}
The second part of (\ref{hodot1}), $\hat{V}_\odot$, arises from the strong electron-electron interactions inside the nanotube, where $V(\vr-\vr\,')$ is the Coulomb potential,
including screening. Under the assumption that forward scattering processes are the only relevant ones, this potential part can be simplified enormously. We are allowed to do so if interactions within one of the
two graphene sublattices, so-called \textit{intra-lattice} interactions, cannot be distinguished from \textit{inter-lattice} interactions. Such an approximation holds for
tubes with sufficiently large diameters ($\simgr 5$ lattice constants) and length ($\simgr 100\,$nm) \cite{LEO3}.\smallskip\\
By combining the OBCs with the expression (\ref{phiappr}) and the linear character of the bands around the Fermi points, the electron operator \[\hPsi_{\odot\sigd}(\vr)=\sum_{\tilde{r}\in\{\tilde{L},\tilde{R}\}}\sum_{n}\varphi^{OBC}_{\tilde{r}\kappa_n}\can_{\tilde{r}\sigd n}\] can be expressed in terms of the 1D operator $\hpsi_{\tilde{r}\sigd F}(x)=\frac{1}{\sqrt{2L_t}}\sum_n\rme^{\rmi\sgn(F)\kappa_n x}\can_{\tilde{r}\sigd n}$ as:
\begin{equation}
\hPsi_{\odot\sigd}(\vr)={\sqrt{L_t}}\,\sum_{r\tilde{r}F}\sgn(F)\,\delta_{r,\tilde{r}}\,\varphi_{[\sgn(F)r]F}(\vr)\,\hpsi_{\tilde{r}\sigd F}(x)\label{hPsi_r2}\,,
\end{equation}
with $r\in\{L,R\}$, $+r=r$, $-r=\bar{r}$ ($\bar{L}=R,\,\bar{R}=L$), $\delta_{r,\tilde{r}}\neq 0$ if $(r,\tilde{r})\in\{(L,\tilde{L}),(R,\tilde{R})\}$ and $L_t$ being the SWCNT length.\smallskip\\
Introducing $\hrho_{\tilde{r}\sigd F}(x):=\psid_{\tilde{r}\sigd F}(x)\hpsi_{\tilde{r}\sigd F}(x)$, $\hat{V}_\odot$ can be rewritten as:

\begin{equation}
\fl\hat{V}_\odot=\frac{1}{2}\sum_{\tilde{r}\tilde{r}'}\sum_{FF'}\sum_{\sigd\sigd'}\int_0^{L_t}\!\!\int_0^{L_t}\rmd{}x\,\rmd{}x' \hrho_{\tilde{r}\sigd F}(x)V(x,x')\hrho_{\tilde{r}'\sigd'F'}(x')\,.
\end{equation}
The one-dimensional density operator $\hrho_{\tilde{r}\sigd F}(x)$ has the convenient property that from its Fourier components, bosonic operators $\hat{b}_{\sigd n}\,,\ n\in\mathbb{Z}^{\pm}$ can be constructed, where the ones indexed with $n>0$ stem from $\hrho_{\tilde{R}\sigd F}$, the others with $n<0$ from $\hrho_{\tilde{L}\sigd F}$. Within the spin-charge-separation model, a unitary transformation to charge- and spin-like excitations is usually performed:\begin{equation*}\hat{b}_{cn}:=\frac{1}{\sqrt{2}}\left(\hat{b}_{\uparrow n}+\hat{b}_{\downarrow n}\right)\ ,\quad \hat{b}_{sn}:=\frac{1}{\sqrt{2}}\left(\hat{b}_{\uparrow n}-\hat{b}_{\downarrow n}\right)\,.\end{equation*}It is possible to reexpress $\hat{H}_{\odot}$ in terms of these new quantities and, by means of some more linear transformations ($\hat{b}_{jn}\to\cdots\to\hat{a}_{jn}\,,\ j\in\{c,s\}$) eventually diagonalise the SWCNT Hamiltonian.\\

The final form of $\hat{H}_{\odot}=\hat{T}_{\odot}+\hat{V}_{\odot}$ is:

\begin{equation}
\fl\hat{H}_{\odot} = \underbrace{\frac{1}{2}E_c\hat{\mathcal{N}}_{c}^{2}}_{\rm{charging}}+\,\epsilon_0\sum_{\tilde{r}\sigd}
\Biggl(\underbrace{\frac{\hat{\mathcal{N}}_{\tilde{r}\sigd}^{2}}{2}}_{\rm{Pauli}}+\underbrace{\delta\,\sgn(\tilde{r})
\hat{\mathcal{N}}_{\tilde{r}\sigd}}_{\rm{band\ offset}}\Biggr)+\underbrace{\sum_{n\neq0}\sum_{j=c,s}\epsilon_{jn}\hat{a}_{jn}^{\dagger}\hat{a}_{jn}}_{\rm{bosonic\ modes}}
  \,.\label{hodot2}\end{equation}
The first three contributions (\ref{hodot2}) contains are purely fermionic, as the operator $\hat{\mathcal{N}}_{\tilde{r}\sigd}$ is just defined
to count the particles in band $\tilde{r}\sigd\in\{\tilde{L}\!\uparrow,\tilde{L}\!\downarrow,\tilde{R}\!\uparrow,\tilde{R}\!\downarrow\}$ and
$\hat{\mathcal{N}}_c\equiv\sum_{\tilde{r}\sigd}\hat{\mathcal{N}}_{\tilde{r}\sigd}$.\\The three summands, as implied in (\ref{hodot2}), are the charging energy, a term accounting for the Pauli principle if more and more electrons are filled in the same band, and a correction for a potential band offset $\delta$, as the energy levels of $\tilde{L}$- and $\tilde{R}$-band might be shifted with respect to each other (illustrated in Fig. \ref{linband}). Here, $\epsilon_0= \pi\hbar v_f/L_t$ is the level spacing and $E_c=W_{00}$, where $W_{nn}\equiv\frac{1}{2L_t^2}\int_0^{L_t}\rmd{}x\int_0^{L_t}\rmd{}x'\,V(x,x')\left[\cos(n\{x+x'\})+\cos(n\{x-x'\})\right]\,$. For a typical SWCNT, $\epsilon_0$ and $E_c$ are both of the order of some meV.\smallskip\\The last term counts the energies of collective, bosonic excitations. The energies $\epsilon_{cn}$ are for $n>0$ dependent on the interaction, $\epsilon_{c+|n|}=|n|\epsilon_0\sqrt{1+8W_{nn}\epsilon_{0}^{-1}}$\,, and thus called \textit{charged} modes, while the other \textit{neutral} modes only scale with the level spacing: $\epsilon_{s\,\pm|n|}=\epsilon_{c\,-\!|n|}=|n|\epsilon_0$. None of the bosonic excitations influences the particle numbers in the single bands and that is why one can classify the eigenstates of the total SWCNT Hamiltonian by a vector $|\vec{N},\vec{m}\ket$ with $\vec{N}=(N_{\tilde{L}\downarrow},N_{\tilde{L}\uparrow},N_{\tilde{R}\downarrow},N_{\tilde{R}\uparrow})$ determining the fermionic configuration and $\vec{m}$ counting the bosonic modes. This uniquely fixes a state; without any bosonic excitations, no `holes' are allowed: all $N_{\tilde{r}\sigd}$ electrons in a certain band ${\tilde{r}\sigd}$ have to populate the $N_{\tilde{r}\sigd}$ lowermost states.\smallskip\\
With this background, we are now ready to start calculating the dynamics of our system.

\section{Equations of motion for the reduced density matrix}\label{mastereq}

We want to investigate the time evolution of our system consisting of SWCNT quantum dot and leads
by using the Liouville equation for its density matrix $\hrho^I(t)$ in the interaction picture.
This representation is well-suited, because we intend to treat both the tunnelling $\hat{H}_{T}\equiv \hat{H}_{Ts}+\hat{H}_{Td}$ and 
the reflection $\hat{H}_{R}\equiv \hat{H}_{Rs}+\hat{H}_{Rd}$ as a perturbation $\hat{H}_I=\hat{H}_{T}+\hat{H}_{R}$ to $\hat{H}_0:=\hat{H}_\odot+\sum_l\hat{H}_l$.\smallskip\\
Indeed it is a critical question whether or not $\hat{H}_{R}$ should be included in $\hat{H}_I$. Considerations in Ref. \cite{WOU} show that the values of the phase shifts picked up during boundary reflections are of the same order as the transmission coefficients, such that a weak conductive coupling will bring about a weak ferromagnetic coupling.\\In Ref. \cite{COT}, an Anderson model is used and both the reflection and the tunnelling are treated non-perturbatively.\\

The equation of motion reads:

\begin{equation}
\rmi\hbar\frac{\partial\hrho^{I}(t)}{\partial t}=\left[\hat{H}_{I}^{I}(t),\hrho^{I}(t)\right],\label{liouville-ia}\end{equation}
where $\hat{H}_I$ had to be transformed into the interaction picture according to $H_{I}^{I}(t)=\rme^{\frac{\rmi}{\hbar}\hat{H}_{0}(t-t_{0})}\hat{H}_{I}(t_0)\,\rme^{-\frac{\rmi}{\hbar}\hat{H}_{0}(t-t_{0})}\,$.\smallskip\\
Our final interest is dedicated to transport through the SWCNT quantum dot, thus we would not mind to lose all the information about the contacts contained in $\hrho^I(t)$: it is sufficient to calculate a \textit{reduced density matrix} (RDM) $\hrho_\odot^I(t)$, where the lead degrees of freedom have been traced out:
\begin{equation}
\fl\hrho_{\odot}^{I}(t):=\Tr_{\leads}\left(\hrho^{I}(t)\right)\,,\quad{\rm{yielding}}\quad\rmi\hbar\frac{\partial}{\partial t}\hrho^I_\odot(t)=\Tr_\leads\left[\hat{H}_{I}^{I}(t),\hrho^I(t)\right]\,.\label{eq2:lead_trace}\end{equation}\vskip 0.6cm

\subsection{Perturbation theory in the tunnelling and the reflection Hamiltonian}
In general, the contacts can be considered as large systems compared to the SWCNT.
Besides, in our case of low transparency, the influence of the nanotube on the leads is marginal.
That is why they can approximately be treated as reservoirs which stay in thermal
equilibrium all the time, and the following ansatz is valid:\vskip 0.4cm
\begin{equation}
\hrho^{I}(t)=\hrho_{\odot}^{I}(t)\hrho_{s}\hrho_{d}+\mathcal{O}(T_s,T_d)+\mathcal{O}(\Delta_{s},\Delta_{d})\,,\label{dm_factorized}
\end{equation}\vskip 0.2cm
where $\hrho_{l}\ (l=s,d)$ are the thermal equilibrium density matrices of the source and the drain contact. Putting (\ref{dm_factorized}) into (\ref{eq2:lead_trace}), we are allowed to neglect the corrections to the factorisation, because in the product with $\hat{H}_{I}^{I}$ they all produce a higher order remainder.\smallskip\\
Now $\hat{H}^I_T(t)$ contains operators belonging to the contacts: either $\hPsi_{l\sigl}(\vr)$ \textit{or} $\Psid_{l\sigl}(\vr)$ takes part in each term of $\hat{H}^I_T$ (see Eq. (\ref{lt_ham})), while $\hat{H}^I_R$ (Eq. (\ref{lr_ham})) involves no lead electron operator at all. Since $\hrho_s$ and $\hrho_d$ appear under the trace, they give us the thermally averaged expectation values of the source and drain operators. Here, due to Wick's theorem, all terms with an odd number of lead electron operators, such as $\Tr_{\leads}[\hat{H}^I_T(t),\hrho_{\odot}^{I}(t)\hrho_{s}\hrho_{d}]$ vanish.\smallskip\\$\Tr_{\leads}[\hat{H}^I_R(t),\hrho_{\odot}^{I}(t)\hrho_{s}\hrho_{d}]$ is nonzero and of first order in the reflection parameters. In order to go to the first nonvanishing order in the tunnelling, we reformulate (\ref{liouville-ia}) to an integral equation:\vskip 0.2cm
\begin{equation}
\hrho^{I}(t)=\hrho^{I}(t_{0})-\frac{\rmi}{\hbar}\int_{t_{0}}^{t}\!\rmd{}t_{1}\left[\hat{H}_{I}^{I}(t_{1}),\hrho(t_{1})\right]\,.\label{eq:liouville-int}\end{equation}\vskip 0.2cm
The time $t_0$ is chosen to be the time where the interaction is switched on, such that for $t'<t_0$ we have $\hat{H}^I_I(t')=\hat{H}^I_T(t')=\hat{H}^I_R(t')\equiv0$ and hence $\hrho^{I}(t')=\hrho_{\odot}^{I}(t')\hrho_{s}\hrho_{d}$ holds true accurately.\smallskip\\
Eq. (\ref{eq:liouville-int}) is exact and we are allowed to reinsert it into (\ref{liouville-ia}). As we are comfortable with $[\hat{H}_{R}^{I}(t),\hrho^{I}(t)]$, we only replace $\hrho^{I}(t)$ in the commutator term involving $\hat{H}^I_T(t)$ and find:

\begin{eqnarray}
\dot{\hrho}^I_\odot(t)&\!\approx&\!-\frac{\rmi}{\hbar}\Tr_\leads\left[\hat{H}_{R}^{I}(t),\hrho^I_\odot(t)\hrho_s\hrho_d\right]-\nonumber\\&&-\frac{1}{\hbar^{2}}\int_{t_{0}}^{t}\!\!\rmd{}t_1\,\Tr_\leads\left[\hat{H}_{T}^{I}(t),\left[\hat{H}_{T}^{I}(t_1),\hrho^{I}_\odot(t_1)\hrho_s\hrho_d\right]\right]\,.\label{eq_motion_rdm}\end{eqnarray}
A significant simplification of (\ref{eq_motion_rdm}) is achieved with the so-called Markoff approximation. We know that the contacts impose large reservoirs to the SWCNT and assume that they induce fast relaxations inside it. For this reason, the current state of the tube can only be influenced by its very latest history, on a timescale not exceeding a certain range $\tau$.
Stating that $\tau$ is sufficiently small, we may safely average $\hrho^I(t)$ over a time period of the order of $\tau$; this enables us to replace $\hrho^I_\odot(t_1)$ in the double commutator in (\ref{eq_motion_rdm}) by $\hrho^I_\odot(t)$. The approximation is also valid when respecting the fact that we examine a static DC circuit, where the detailed dynamics on short time intervals need not to be taken into account.\smallskip\\
We arrive at the equation of motion we want to work with when we send $t_{0}\rightarrow-\infty$ (because we are merely interested in the longterm behaviour of the system) and use the abbreviation $t_2=t-t_1$ :\vskip 0.2cm
\begin{equation*}
\fl\dot{\hrho}_{\odot}^{I}(t)=-\frac{\rmi}{\hbar}\Tr_{\leads}\left[\hat{H}_{R}^{I}(t),\hrho_{\odot}^{I}(t)\hrho_{s}\hrho_d\right]-\frac{1}{\hbar^{2}}\Tr_{\leads}\int_{0}^{\infty}\!\!\!\!\!\rmd{}t_2\left[\hat{H}_{T}^{I}(t),\left[\hat{H}_{T}^{I}(t\!-\!t_2),\hrho_{\odot}^{I}(t)\hrho_{s}\hrho_d\right]\right].\end{equation*}\vskip 0.2cm
\quad\\Removing the commutators, we obtain
\begin{eqnarray}
\fl\dot{\hrho}_{\odot}^{I}(t)=-\frac{\rmi}{\hbar}\Tr_{\leads}\left(\hat{H}_{R}^{I}(t)\hrho_{\odot}^{I}(t)\hrho_{s}\hrho_d+h.c.\right)-\label{eq:liouville1}\\\hspace{-1.9cm}-\frac{1}{\hbar^{2}}\Tr_{\leads}\!\!\int_{0}^{\infty}\!\!\!\!\!\!\rmd{}t_2\left(\left\{\hat{H}_{T}^{I}(t)\hat{H}_{T}^{I}(t-t_2)\hrho_{\odot}^{I}(t)\hrho_{s}\hrho_d-\hat{H}_{T}^{I}(t)\hrho_{\odot}^{I}(t)\hrho_{s}\hrho_d\hat{H}_{T}^{I}(t-t_2)\right\}+h.c.\right)\,.\nonumber\end{eqnarray}\quad\\We can now substitute the explicit form of $\hat{H}^I_T$ (given in Eq. (\ref{lt_ham})), and of $\hat{H}^I_R$ (from Eq. (\ref{lr_ham})) in Eq. (\ref{eq:liouville1}) and get rid of the traces. The terms containing $\hat{H}^I_R$ involve no lead operators, such that the trace just makes the product $\hrho_{s}\hrho_d$ vanish. For the contributions with $\hat{H}^I_T$, the lead operators must be brought next to the density matrix. This is done by using the cyclic property of the trace and commuting them past \textit{two} SWCNT operators in each term; this, remarkably, means that it does not play any role whether the operators are commuting or anticommuting.\smallskip\\Then one can introduce the correlation functions\vskip 0.5cm
\begin{equation}\left\langle \Psid_{l\sigma_l}\hPsi_{l\sigma_{l}}\right\rangle_{\mathsf{th}}:=\Tr_\leads\left(\Psid_{l\sigma_l}\hPsi_{l\sigma_{l}}\hrho_s\hrho_d\right)\,,\label{correl_func}\end{equation}\begin{eqnarray}
\fl\rm{and\ exploit}\quad&&\Tr_\leads\left(\hPsi_{l\sigma_l}\hPsi_{l'\sigma_{l'}}\hrho_s\hrho_d\right)=\Tr_\leads\left(\hPsi_{l\sigma_l}^{\dagger}\hPsi_{l'\sigma_{l'}}^{\dagger}\hrho_s\hrho_d\right)=0\,,\nonumber\\
\fl&&\Tr_\leads\left(\hPsi_{l\sigma_l}\Psid_{l'\sigma_{l'}}\hrho_s\hrho_d\right)=0\quad{\rm{for}}\ \ l\sigl\neq l'\sigma_{l'}\,.\label{correl_func2}\end{eqnarray}

\begin{eqnarray*}\fl\rm{With\ the\ abbreviation}\quad&&
\mathcal{E}_{l\sigl}(\vec{r},\vec{r}\,'\!\!,t_2):=T_{l}(\vec{r})T_{l}^{*}(\vec{r}\,')\left\langle \hPsi_{l\sigl}(\vec{r})\hPsi_{l\sigl}^{\dagger}(\vec{r}\,'\!\!,-t_2)\right\rangle _{\rm{th}}\!\!,\\
&&\mathcal{F}_{l\sigl}(\vec{r},\vec{r}\,'\!\!,t_2):=T_{l}^{*}(\vec{r})T_{l}(\vec{r}\,')\left\langle \hPsi_{l\sigl}^{\dagger}(\vec{r})\hPsi_{l\sigl}(\vec{r}\,'\!\!,-t_2)\right\rangle _{\rm{th}}\!\!,
\end{eqnarray*}
Eq. (\ref{eq:liouville1}) is changed to:

\begin{eqnarray}
\fl\dot{\hrho}_{\odot}^{I}(t)&=&\frac{\rmi}{\hbar}\sum_{l\sigl}\int\!\rmd^{3}r\ \Delta_l(\vr)\ \sgn(\sigl)\left(\hPsi_{\odot\sigl}^{\dagger}(\vec{r},t)\hPsi_{\odot\sigl}(\vec{r},t)\,\hrho_{\odot}^{I}(t)-h.c.\right)-\nonumber\\\fl
&&-\frac{1}{\hbar^{2}}\sum_{l\sigl}\int\!\rmd^{3}r\int\!\rmd^{3}r'\int_{0}^{\infty}\!\!\!\rmd{t}_2\times\nonumber\\\fl&&\hspace{1.8cm}\times\left(\ \left\{\mathcal{E}_{l \sigl}(\vec{r},\vec{r}\,'\!\!,t_2)\hPsi_{\odot\sigl}^{\dagger}(\vec{r},t)\hPsi_{\odot\sigl}(\vec{r}\,'\!\!,t-t_2)\,\hrho_{\odot}^{I}(t)\right.\right.\!\!+\nonumber\\\fl
&&\left.\hspace{2.8cm}+\mathcal{F}_{l\sigl}(\vec{r},\vec{r}\,'\!\!,t_2)\hPsi_{\odot\sigl}(\vec{r},t)\hPsi_{\odot\sigl}^{\dagger}(\vec{r}\,'\!\!,t-t_2)\,\hrho_{\odot}^{I}(t)\right\}+h.c.-\nonumber\\\fl
&&\hspace{2.3cm}-\left\{\mathcal{F}_{l\sigl}^{*}(\vec{r},\vec{r}\,'\!\!,t_2)\hPsi_{\odot\sigl}^{\dagger}(\vec{r},t)\,\hrho_{\odot}^{I}(t)\,\hPsi_{\odot\sigl}(\vec{r}\,'\!\!,t-t_2)+\right.\nonumber\\\fl
&&\hspace{3.0cm}+\left.\left.\mathcal{E}_{l\sigl}^{*}(\vec{r},\vec{r}\,'\!\!,t_2)\hPsi_{\odot\sigl}(\vec{r},t)\,\hrho_{\odot}^{I}(t)\,\hPsi_{\odot\sigl}^{\dagger}(\vec{r}\,'\!\!,t-t_2)\right\}-h.c.\right).\label{eq:hrho1}\end{eqnarray}\quad\\Moreover, we are able to write out the time dependences of the operators according to

\begin{equation}\fl\hPsi_f(t)=\rme^{+\frac{\rmi}{\hbar}\hat{H}_{0}(t-t')}\,\,\hPsi_f(t')\,\,\rme^{-\frac{\rmi}{\hbar}\hat{H}_0(t-t')}=\rme^{+\frac{\rmi}{\hbar}\hat{H}_{f}(t-t')}\,\,\hPsi_f(t')\,\,\rme^{-\frac{\rmi}{\hbar}\hat{H}_f(t-t')}\,,\label{Psi-time}
\end{equation}
independently of all indices but $f\in\{s,d,\odot\}$, because all parts of $\hat{H}_0$ besides $\hat{H}_f$ commute with $\hPsi_f$.

\subsection{Equations of motion in the SWCNT energy basis}\label{eom_SWCNT}
The starting point for this subsection is Eq. (\ref{eq:hrho1}), which is used to derive the equations of motion for a single element $\left({\hrho}_{\odot}^{I}(t)\right)_{nm}$ of the reduced density matrix in the SWCNT energy basis.\smallskip\\At first, we apply two more approximations that will ease our tasks considerably:\\
\begin{enumerate}
\begin{minipage}{0.95\textwidth}\item{We presume that a SWCNT generally is in a pure charge state, i.e. it is filled with a certain number of electrons $N$ and thus density matrix elements between states of distinct electron numbers are set to zero (we deal with bias voltages not exceeding the height of the Coulomb diamonds and permanently measure $N$ in our circuit via the gate voltage).}
\end{minipage}\vskip 0.18cm
\begin{tabular}{lr}
\hspace{-0.32cm}
\begin{minipage}[b]{0.4\textwidth}
\item{Being interested in the static behaviour of our system, we can neglect fast oscillating terms arising from the exponentials in (\ref{Psi-time}) for $f$=$\odot$. This \textit{secular approximation} completely decouples the time evolution of matrix elements between states degenerate in energy from all matrix elements between states non-degenerate in energy \mbox{(fig. \ref{blocks})}.}
\end{minipage}
&
\hspace{0.2cm}
\begin{minipage}[b]{0.45\textwidth}
\begin{center}
\includegraphics[width=0.5\textwidth]{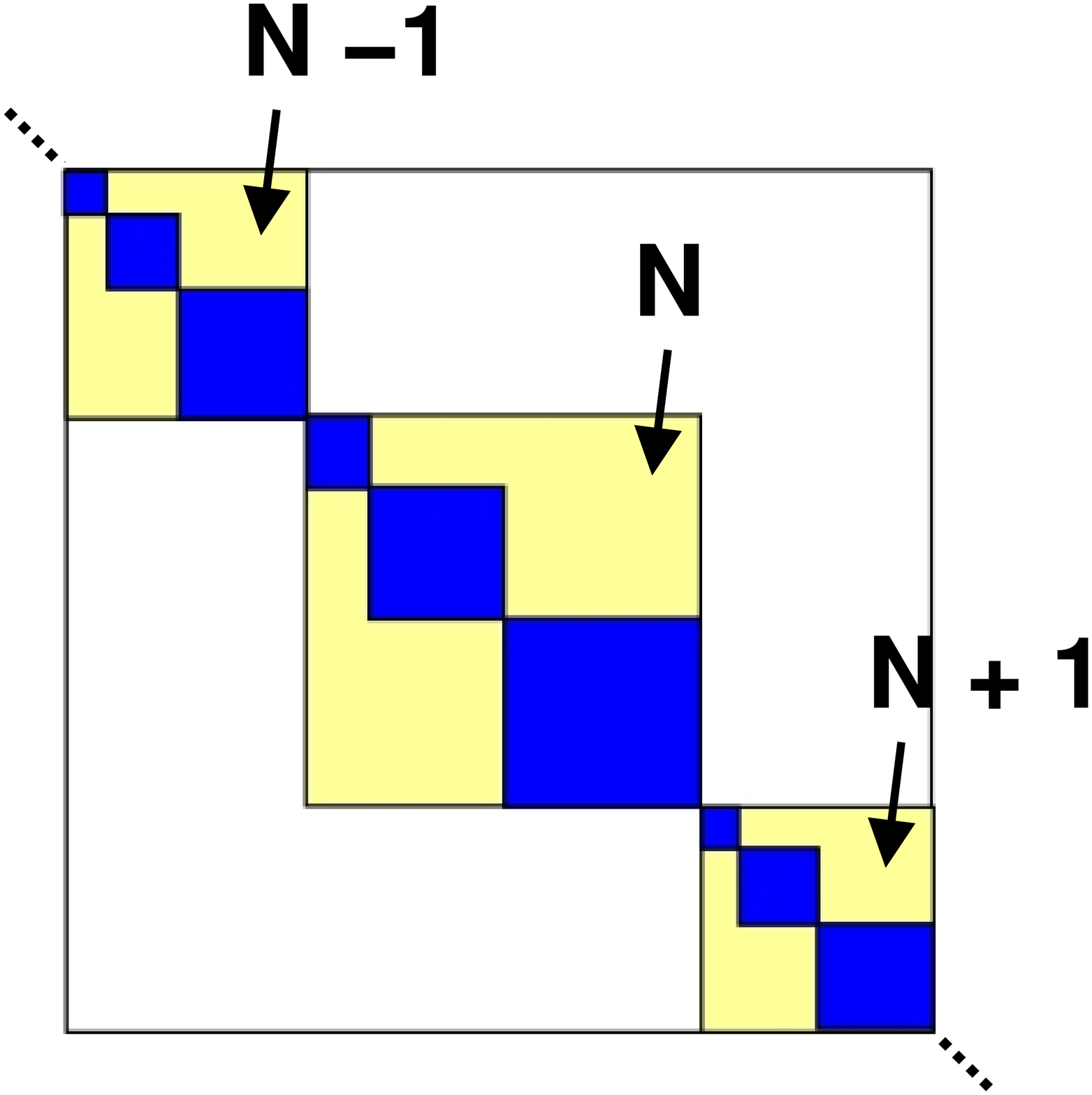}\vskip 0.2cm
\imagecaption{\small The RDM acquires a block diagonal form. Light squares: particle numbers fixed. Dark squares: also degenerate in energy}\label{blocks}
\end{center}
\end{minipage}
\end{tabular}
\end{enumerate}\vskip 0.2cm
To calculate the current, merely the diagonal elements of the density matrix, the \textit{populations}, are required, and as a consequence it is actually sufficient to deal with block matrices $\hrho^I_{\odot E_N N}$, which are restricted to the Hilbert space spanned by states

\[j\in{|E_NN\ket}:=\left\{|\vec{N}\,'\ket\,:\,|\vec{N}\,'|=N\,,\ \bra\vec{N}\,'|\hat{H}_\odot|\vec{N}\,'\ket=E_N\right\}\] with fixed energy $E_N$ and particle number $N$ (dark squares in Fig. \ref{blocks}).\smallskip\\
Eq. (\ref{eq:hrho1}) then reads, introducing \textit{Bloch-Redfield tensors} in order to simplify the notation,

\begin{eqnarray}
\fl\left(\dot{\hrho}_{\odot E_NN}^{I}(t)\right)_{nm}&=&-\sum_{jj'}R^{NN}_{nm,jj'}\left(\hrho_{\odot E_NN}^{I}(t)\right)_{jj'}+\sum_{kk'}R^{NN+1}_{nm,kk'}\left(\hrho_{\odot E_{k}N+1}^{I}(t)\right)_{kk'}+\nonumber\\\fl&&+\sum_{ii'}R^{NN-1}_{nm,ii'}\left(\hrho_{\odot E_{i}N-1}^{I}(t)\right)_{ii'}.\label{eq:hrho2}\end{eqnarray}
The sums in (\ref{eq:hrho2}) run over states with fixed particle numbers: $j,j'\{\in|E_NN\ket\}$, $k,k'\in\{|N$+$1\ket\}$, $i,i'\in\{|N$-$1\ket\}$, where it is the secular approximation which additionally fixes the energy of $j,j'$.\bigskip\\
The Bloch-Redfield tensors are defined as follows:
\begin{eqnarray}
\fl R^{NN}_{nm,jj'}&:=&\sum_l\!\!\sum_{\scriptsize \begin{array}{l}i\in\{|N\rm{-}1\ket\}\\k\in\{|N\rm{+}1\ket\}\end{array}}\!\!\left(\delta_{m,j'}\left[\Gamma^{(+)NN-1}_{l(niij)}+\Gamma^{(+)NN+1}_{l(nkkj)}\right]+\delta_{n,j}\left[\Gamma^{(-)NN-1}_{l(j'iim)}+\Gamma^{(-)NN+1}_{l(j'kkm)}\right]\right),\nonumber\\\fl\nonumber\\\fl R^{NN\pm1}_{nm,ss'}&:=&\sum_l\left(\Gamma^{(+)NN\mp1}_{l(s'mns)}+\Gamma^{(-)NN\mp1}_{l(s'mns)}\right)\,,\label{blochred}
\end{eqnarray}
with the rates
\begin{eqnarray*}
\fl\Gamma^{(\alpha)NN+1}_{l(nkk'j)}&:=&\frac{1}{\hbar^2}\sum_{\sigl}\!\int\!\!\rmd{}^3r\!\int\!\!\rmd{}^3r'\!\left(\!\hPsi_{\odot\sigl}(\vr)\!\right)_{nk}\!\left(\!\Psid_{\odot\sigl}(\vr\,')\!\right)_{k'j}\int_0^\infty\!\!\!\!\rmd{}t_2 \ \mathcal{F}_{l\sigl}(\vec{r},\vec{r}\,'\!\!,t_2) \ \rme^{\alpha\frac{\rmi}{\hbar}\left(E_j-E_k\right)t_2}\!,\end{eqnarray*}\begin{eqnarray}
\fl\Gamma^{(\alpha)NN-1}_{l(nii'j)}&:=&\frac{1}{\hbar^2}\sum_{\sigl}\!\int\!\!\rmd{}^3r\!\int\!\!\rmd{}^3r'\!\left(\!\Psid_{\odot\sigl}(\vr)\!\right)_{ni}\!\left(\!\hPsi_{\odot\sigl}(\vr\,')\!\right)_{i'j}\int_0^\infty\!\!\!\!\rmd{}t_2 \ \mathcal{E}_{l\sigl}(\vec{r},\vec{r}\,'\!\!,t_2) \ \rme^{\alpha\frac{\rmi}{\hbar}\left(E_j-E_i\right)t_2}-\nonumber\\\fl&&-\alpha\,\frac{\rmi}{\hbar}\sum_{\sigl}\!\int\!\rmd{}^3r\ \Delta_l(\vr)\ \sgn(\sigl)\left(\!\Psid_{\odot\sigl}(\vr)\!\right)_{ni}\!\left(\!\hPsi_{\odot\sigl}(\vr)\!\right)_{i'j}\,.\label{eq1:rates}\\\fl\nonumber\end{eqnarray}
We like to point out that up to now, the relations we deduced are very general ones, as we did not exploit any property specific for SWCNTs. Eq. (\ref{eq:hrho2}) already shows the features that are important for us: the time evolution of the diagonal matrix elements of $\hrho_{\odot E_NN}$ will couple to some elements of $\hrho_{\odot E_NN}$, $\hrho_{\odot E_i N-1}$ and $\hrho_{\odot E_k N+1}$. Which elements are actually involved can be mapped out by transforming the expression (\ref{eq1:rates}) further.\\First, the explicit form of the correlation functions (\ref{correl_func}) needs to be determined. This and some consecutive steps are carried out in the appendix, and as a result the rates are changed to (\ref{app_res}): 

\begin{eqnarray}\fl\Gamma^{(\alpha)NN+1}_{l(nkk'j)}&:=&\frac{\pi L_t}{\hbar}\sum_{\tilde{r}}\sum_{\sigl}\,\Phi_{l}\left(\hpsi_{\tilde{r}\sigl}\right)_{nk}\left(\psid_{\tilde{r}\sigl}\right)_{k'j}\times\nonumber\\\fl&&\hspace{3cm}\times\left[D_{l\sigl}(E_{kj})\,f_l(E_{kj})+\alpha\frac{\rmi}{\pi}\left(\!\!\pint_{-\infty}^{\infty}\!\!\!\rmd{}\epsilon\ \frac{D_{l\sigl}(\epsilon)\,f_l(\epsilon)}{\epsilon-E_{kj}}\right)\right]\,,\nonumber\\\fl&&\quad\nonumber\\\fl
\Gamma^{(\alpha)NN-1}_{l(nii'j)}&:=&\frac{\pi L_t}{\hbar}\sum_{\tilde{r}}\sum_{\sigl}\,\Phi_l\left(\psid_{\tilde{r}\sigl}\right)_{ni}\left(\hpsi_{\tilde{r}\sigl}\right)_{i'j}\times\nonumber\\\fl&&\hspace{0.2cm}\times\Biggl[D_{l\sigl}(E_{ji})\,\left(1-f_l(E_{ji})\right)-\alpha\frac{\rmi}{\pi}\left(\!\!\pint_{-\infty}^{\infty}\!\!\!\rmd{}\epsilon\ \frac{D_{l\sigl}(\epsilon)\,\left(1-f_l(\epsilon)\right)}{\epsilon-E_{ji}}+\frac{1}{\Phi_l}R_l\right)\Biggr]\nonumber\\\fl\nonumber\\\fl\label{eq3:rates}\\\fl\nonumber{\rm{Here,}}&& f_l(E^{l\sigl}_{tot}|_{\epsilon})=\Biggl(1+\exp{\frac{E^{l\sigl}_{tot}|_{\epsilon}+eV_l-\tilde{E}^{(zf)}_{F,l}}{\kBT}}\Biggr)^{-1}\end{eqnarray}
is the Fermi function in lead $l$ and has arisen from the correlation functions. $\tilde{E}^{(zf)}_{F,l}$ is the common Fermi level for the two spin species $\sigl=+_l$ and $\sigl=-_l$ in contact $l$ without any bias voltage applied (for calculations we will assume that source and drain are made of the same metal, which allows to shift the energy scale such that $\tilde{E}^{(zf)}_{F,s}=\tilde{E}^{(zf)}_{F,d}=0$).\smallskip\\Further, we have abbreviated $E_{ab}\equiv E_a-E_b$ and $\Phi_l\,$, $R_l$ (\ref{eq:transparencyreflection}) are real values into which the integrations over space, the tunnelling and reflection parameters as well as the wave functions from the decompositions (\ref{hPsi_l}),(\ref{hPsi_r2}) of the electron operators have been absorbed. To do so, the assumption that both $T_l(\vr)$ and $\Delta_l(\vr)$ are small away from the tunnelling contacts was needed. The new parameters $\Phi_l$ and $R_l$ scale with the strength of the tunnelling, respectively the reflection.\\Finally, {\small{$\mathcal{P}$}}\hspace{-0.8em}{\Large{$\int$}} denotes a principal part integration.\medskip\\
Now let us have a closer look to the rates (\ref{eq3:rates}). As indicated by the notation, $\Gamma^{NN+1}$ is related to transitions $N\to N$+$1$; that is why its real part involves the product of the density of states $D_{l\sigl}$ and the Fermi function $f_l$, which refers to the number of electrons that can potentially leave the contact. $\Gamma^{NN-1}$ correspondingly contains $1-f_l$, which accounts for vacancies in the lead.\\The imaginary parts also include a product of Fermi function and density of states, but the energy arguments of the functions are integrated over: there is no restriction to energetically permitted transitions. That is why we call these \textit{virtual}; the real parts, however, represent processes where the energy is conserved.\smallskip\\
Still, the electron operators in Eqns. (\ref{eq3:rates}) are given for spins along the quantisation axes of the two distinct lead coordinate systems. To go on, we must reexpress them in a common basis.

\section{Spin quantisation axes}\label{spinaxis}

\subsection{Non-collinear magnetisations}
A clever choice for the SWCNT spin quantisation axis $z_\odot$ is the direction
perpendicular to the plane spanned by $\vec{m}_s$ and $\vec{m}_d$  \cite{KOE}:
using the coordinate systems drawn in Fig. \ref{kosys1}, we will get
particularly nice matrices $U_{l,\odot}$, which transfer the electron operators $\psid_{\tilde{r}\sigl}$ to their representation $\psid_{\tilde{r}\sigd}$, where the spin is quantised along $z_{\odot}$.
\begin{center}
\includegraphics[width=0.67\textwidth]{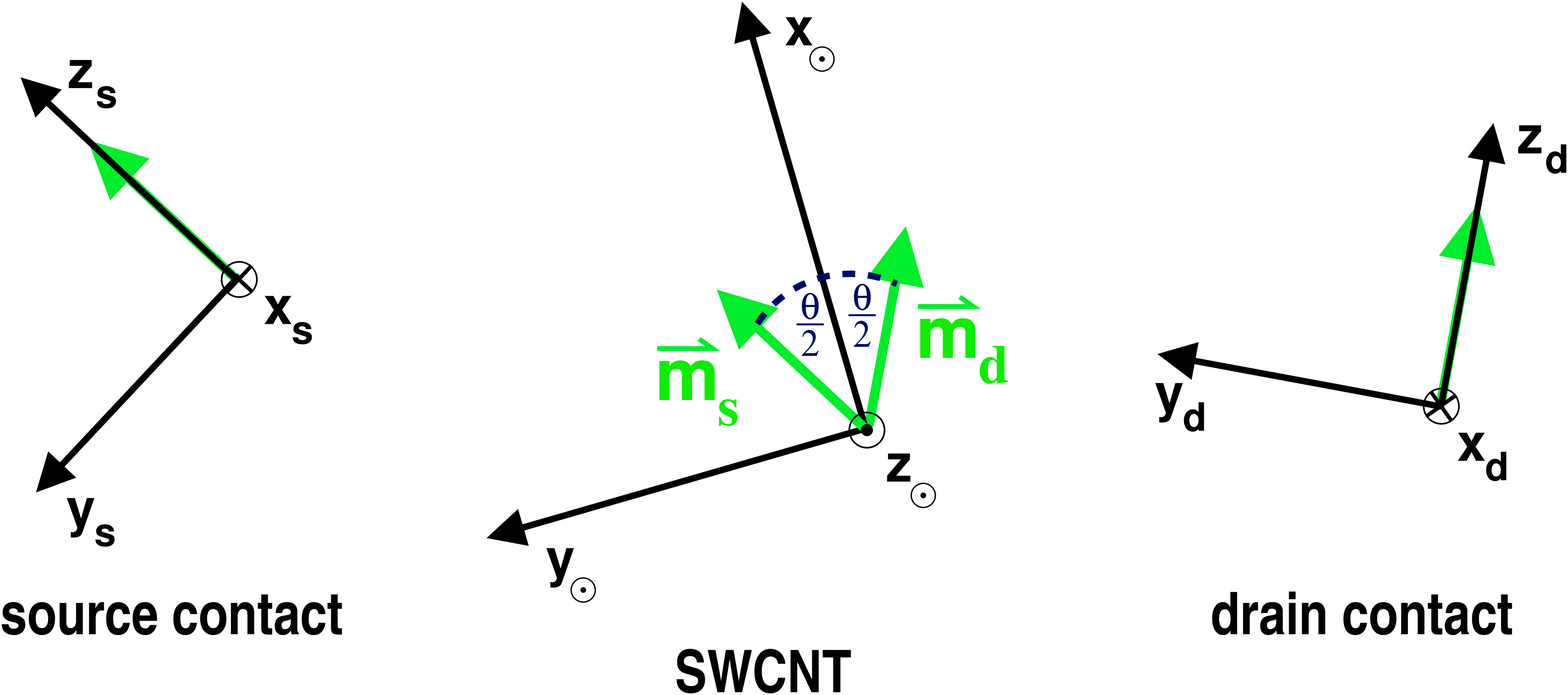}\vskip 0.3cm
\imagecaption{\small The $x_\odot$-axis is chosen such that
it bisects the angle $\theta$ between $\vec{m}_s$ and $\vec{m}_d$. In each case, $z$ is the quantisation axis in the corresponding part of the system.}\label{kosys1}
\vskip 0.4cm
\end{center}
To calculate $U_{l,\odot}$, we rewrite the  three-dimensional coordinate system basis
$\{\vx_l,\vy_l,\vz_l\}$ of contact $l$ in the nanotube basis:
\begin{equation*}
(\ve_l)_{\odot}=U_{l,\odot}\,(\ve_l)_l=U_{l,\odot}\,(\ve_{\odot})_{\odot}\,,
\end{equation*}
where $\ve\in\{\vx,\vy,\vz\}\, ,\ U_{l,\odot}\in SO(3)\,$ and $(\ve)_{l/\odot}$ indicates that $\ve$ is given in
the lead/SWCNT coordinate representation.\\
The second equality makes clear that the coordinate transformation matrix $U_{l,\odot}$ is just the 3D rotation matrix that rotates the basis $\{\vx_{\odot},\vy_\odot,\vz_\odot\}$ of the nanotube coordinate system onto the one of lead $l$,\,\ $\{\vx_l,\vy_l,\vz_l\}\,$.\\
$U_{l,\odot}$ itself is simply a product of two rotation matrices of different angles
around different axes. Carefully considering the directions and the resulting signs for the angles of the necessary
rotations, we find:
\begin{equation*}
U_{l,\odot}=U_{\vx'_{\odot}}(\theta_l)U_{\vy_\odot}({\pi}/{2})\quad{\rm{with}}\quad\theta_l:=\left\lbrace\begin{array}{cc}-\theta/2&l=s\,,\\+\theta/2&l=d\,.\end{array}\right.
\end{equation*}
Here $\vx'_\odot$ denotes the $\vx_\odot$ axis, having undergone the first rotation. With some basic quantum mechanics we obtain the corresponding transformation matrices for the electron operator $\psid_{\tilde{r}\sigl}$:
\vspace{0.15cm}
\begin{eqnarray}
 U_{l,\odot}&=&\Bigl(\mathbf{1}\cos(-\theta_l/2)+\rmi\sigma_x\sin(-\theta_l/2)\Bigr)\Bigl(\mathbf{1}\cos(45°)+\rmi\sigma_y\sin(45°)\Bigr)=\nonumber\quad\quad\quad\\
 &=&\frac{1}{\sqrt{2}}\left(\begin{array}{rr}+\rme^{+\rmi\theta_l/2}&+\rme^{-\rmi\theta_l/2}\\-\rme^{+\rmi\theta_l/2}&+\rme^{-\rmi\theta_l/2}\end{array}\right)\,,\label{S_l}
\end{eqnarray}
such that
\begin{equation}
\left(\begin{array}{c}\psid_{\tilde{r}\,+_l}\\\psid_{\tilde{r}\,-_l}\end{array}\right)=\underbrace{\frac{1}{\sqrt{2}}\left(\begin{array}{cc}+\rme^{-\rmi\theta_l/2}&+\rme^{+\rmi\theta_l/2}\\-\rme^{-\rmi\theta_l/2}&+\rme^{+\rmi\theta_l/2}\end{array}\right)}_{=U^{-1}_{l,\odot}}\left(\begin{array}{c}\psid_{\tilde{r}[\sigd=\uparrow]}\\\psid_{\tilde{r}[\sigd=\downarrow]}\end{array}\right)\,.\label{transf}\end{equation}\quad\\
Using (\ref{transf}) it is straightforward to evaluate

\begin{eqnarray}\sum\limits_{\sigl}D_{l\sigl}\hpsi^{\dag}_{\sigl}\hpsi_{\sigl}&=&\frac{1}{2}\,\Bigl(D_{l+_l}+D_{l-_l}\Bigr)\,\left[\hpsi^{\dag}_{\uparrow}\hpsi_{\uparrow}+\hpsi^{\dag}_{\downarrow}\hpsi_{\downarrow}\right]+\nonumber\\&&+\frac{1}{2}\,\Bigl(D_{l+_l}-D_{l-_l}\Bigr)\,\left[\rme^{-\rmi\theta_l}\hpsi^{\dag}_{\uparrow}\hpsi_{\downarrow}+\rme^{\rmi\theta_l}\hpsi^{\dag}_{\downarrow}\hpsi_{\uparrow}\right]\,,\label{sumtrans1}\end{eqnarray}where all uninvolved indices and arguments have been dropped for clarity. Defining

\begin{equation}\fl\label{Phi_sigd}
\tilde{\Phi}_{l\sigd\sigd'}:=\left\lbrace\begin{array}{l@{\quad}c}
1 & \sigd=\sigd'\\
\rme^{\rmi\theta_l} & \sigd=\uparrow,\sigd'=\downarrow\\
\rme^{-\rmi\theta_l} & \sigd=\downarrow,\sigd'=\uparrow
\end{array}\right.\quad{\rm{and}}\quad D_{l\sigd\sigd'}:=\left\lbrace\begin{array}{l@{\quad}c}
D_{l+_l}+D_{l-_l} & \sigd=\sigd'\\
D_{l+_l}-D_{l-_l} & \sigd\neq\sigd'
\end{array}\right.
\end{equation}the rates in Eq. (\ref{eq3:rates}) are reformulated for one last time:

\begin{eqnarray}\fl\Gamma^{(\alpha)NN+1}_{l(nkk'j)}&:=&\frac{\pi L_t}{\hbar}\sum_{\tilde{r}}\sum_{\sigd\sigd'}\,\tilde{\Phi}_{l\sigd\sigd'}\left(\hpsi_{\tilde{r}\sigd}\right)_{nk}\!\left(\psid_{\tilde{r}\sigd'}\right)_{k'j}\Biggl[D_{l\sigd\sigd'}(E_{kj})\,f_l(E_{kj})+\Biggr.\nonumber\\\fl&&\hspace{6.6cm}\Biggl.+\alpha\frac{\rmi}{\pi}\left(\pint_{-\infty}^{\infty}\!\!\rmd{}\epsilon\ \frac{D_{l\sigd\sigd'}(\epsilon)\,f_l(\epsilon)}{\epsilon-E_{kj}}\right)\Biggr]\,,\nonumber\\\fl\Gamma^{(\alpha)NN-1}_{l(nii'j)}&:=&\frac{\pi L_t}{\hbar}\sum_{\tilde{r}}\sum_{\sigl}\,\tilde{\Phi}^*_{l\sigd\sigd'}\left(\psid_{\tilde{r}\sigd}\right)_{ni}\!\left(\hpsi_{\tilde{r}\sigd'}\right)_{i'j}\Biggl[D_{l\sigd\sigd'}(E_{ji})\,\left(1-f_l(E_{ji})\right)-\Biggr.\nonumber\\\fl&&-\Biggl.\alpha\frac{\rmi}{\pi}\left(\pint_{-\infty}^{\infty}\!\!\rmd{}\epsilon\ \frac{D_{l\sigd\sigd'}(\epsilon)\,\left(1-f_l(\epsilon)\right)}{\epsilon-E_{ji}}+\frac{1}{\Phi_l}R_{l}\left(\delta_{\sigd\uparrow}\delta_{\sigd'\downarrow}+\delta_{\sigd\downarrow}\delta_{\sigd'\uparrow}\right)\right)\Biggr].\nonumber\\\fl\nonumber\\\fl\label{eq4:rates}\end{eqnarray}

\clearpage

\subsection{Collinear magnetisations}
Before we explain how to solve the final master equation, we should have a look at the special case of collinear contact configurations, where it is clever to use the common quantisation axis of the leads also for the coordinate system inside the SWCNT (fig. \ref{kosys2}).

\begin{center}\vskip 0.6cm
\includegraphics[width=0.67\textwidth]{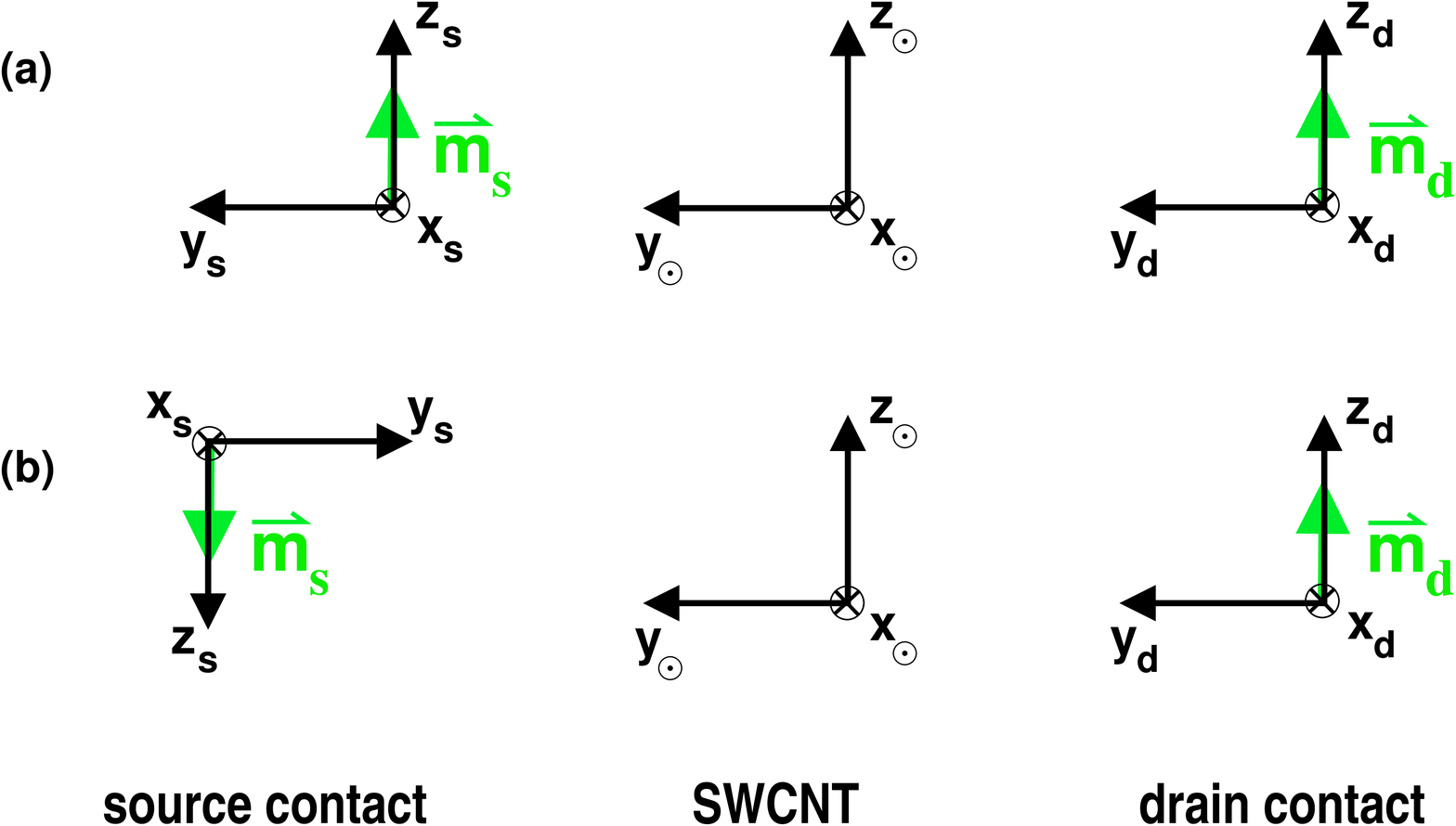}\vskip 0.3cm
\imagecaption{\small Considering merely collinear contact magnetisations, it holds advantages just to choose one of the lead coordinate systems as SWCNT coordinate system, too.\newline(a) depicts the parallel, (b) the antiparallel case.}\label{kosys2}
\end{center}\quad\\
The matrices we need for replacing Eq. (\ref{S_l}) when we work with the coordinate systems sketched in Fig. \ref{kosys2} are
\begin{equation}\fl U_{l,\odot}=\left\lbrace\begin{array}{c}\mathbf{1}\cos(0°)+\rmi\sigma_x\sin(0°)\\\mathbf{1}\cos(-\pi/2)+\rmi\sigma_x\sin(-\pi/2)\end{array}\right\rbrace=\left\lbrace\begin{array}{cl}\left(\begin{array}{rr}1\,&\ \,0\ \\\,\ 0\ &1\,\end{array}\right)&\quad l=d\,\ \mathrm{or}\,\ l=s\,\wedge\,\theta=0\,,\\\left(\begin{array}{rr}0&-\rmi\\-\rmi&0\end{array}\right)&\quad l=s\, \wedge\, \theta=\pi\,.\end{array}\right.\label{S_2}\end{equation}\quad\\
The electron operators transform accordingly,\\
\begin{equation}
\left(\begin{array}{c}\psid_{\tilde{r}\,+_s}\\\psid_{\tilde{r}\,-_s}\end{array}\right)=\left\lbrace\begin{array}{rl}\left(\begin{array}{c}\psid_{\tilde{r}[\sigd=\uparrow]}\\\psid_{\tilde{r}[\sigd=\downarrow]}\end{array}\right)&\quad l=d\,\ \mathrm{or}\,\ l=s\,\wedge\,\theta=0\,,\\-\rmi\left(\begin{array}{c}\psid_{\tilde{r}[\sigd=\downarrow]}\\\psid_{\tilde{r}[\sigd=\uparrow]}\end{array}\right)&\quad l=s\, \wedge\, \theta=\pi\,,\end{array}\right.\end{equation}
and instead of (\ref{sumtrans1}) we obtain

\begin{equation}\label{sumtrans2}
\sum\limits_{\sigma_l}D_{l\sigma_l}\hpsi^{\dag}_{\sigma_l}\hpsi_{\sigma_l}=\left\lbrace\begin{array}{cl}D_{l+_l}\hpsi^{\dag}_{\uparrow}\hpsi_{\uparrow}+D_{l-_l}\hpsi^{\dag}_{\downarrow}\hpsi_{\downarrow}&\quad l=d\,\ \mathrm{or}\,\ l=s\,\wedge\,\theta=0\,,\\D_{l-_l}\hpsi^{\dag}_{\uparrow}\hpsi_{\uparrow}+D_{l+_l}\hpsi^{\dag}_{\downarrow}\hpsi_{\downarrow}&\quad l=s\, \wedge\, \theta=\pi\,.\end{array}\right.\end{equation}
The corresponding rates are not much different from (\ref{eq3:rates}):

\begin{eqnarray}\fl\Gamma^{(\alpha)NN+1}_{l(nkk'j)}&:=&\frac{\pi L_t}{\hbar}\sum_{\tilde{r}}\sum_{\sigd}\,\Phi_{l}\left(\hpsi_{\tilde{r}\sigd}\right)_{nk}\left(\psid_{\tilde{r}\sigd}\right)_{k'j}\Biggl[D_{l\sigd}(E_{kj})\,f_l(E_{kj})+\Biggr.\nonumber\\\fl&&\hspace{6.6cm}\Biggl.+\alpha\frac{\rmi}{\pi}\left(\pint_{-\infty}^{\infty}\!\!\rmd{}\epsilon\ \frac{D_{l\sigd}(\epsilon)\,f_l(\epsilon)}{\epsilon-E_{ji}}\right)\Biggr]\,,\nonumber\\\fl\Gamma^{(\alpha)NN-1}_{l(nii'j)}&:=&\frac{\pi L_t}{\hbar}\sum_{\tilde{r}}\sum_{\sigd}\,\Phi_l\left(\psid_{\tilde{r}\sigd}\right)_{ni}\left(\hpsi_{\tilde{r}\sigd}\right)_{i'j}\Biggl[D_{l\sigd}(E_{ji})\,\left(1-f_l(E_{ji})\right)-\Biggr.\nonumber\\\fl&&\hspace{3.0cm}\Biggl.-\alpha\frac{\rmi}{\pi}\left(\pint_{-\infty}^{\infty}\!\!\rmd{}\epsilon\ \frac{D_{l\sigd}(\epsilon)\,\left(1-f_l(\epsilon)\right)}{\epsilon-E_{ji}}+\frac{(-1)^{\gamma_l(\theta)}}{\Phi_l}R_l\right)\Biggr],\nonumber\\\fl\nonumber\\\fl\label{eq5:rates}\end{eqnarray}
with $\gamma_l(\theta):=\delta_{l,s}\delta_{\theta,\pi}$ and\\
\begin{equation}
D_{l\sigd}=\left\lbrace\begin{array}{rc}\left.\begin{array}{rc}D_{l+_l}&\ \sigd=\uparrow\\D_{l-_l}&\ \sigd=\downarrow\end{array}\right\}&\quad l=d\,\ \mathrm{or}\,\ l=s\,\wedge\,\theta=0\,,\\\left.\begin{array}{rc}D_{l-_l}&\ \sigd=\uparrow\\D_{l+_l}&\ \sigd=\downarrow\end{array}\right\}&\quad l=s\, \wedge\, \theta=\pi\,.\end{array}\right.
\end{equation}\vskip 0.4cm
The important point is that in Eq. (\ref{sumtrans2}), no terms with mixed spins occur at all, which will make setting up the necessary equations easier than in the general case for an arbitrary $\theta$. In particular, all coherences drop out and for this reason, also the imaginary parts the rates (\ref{eq5:rates}) still contain will vanish: in our regime of weak coupling, neither the virtual transitions nor the boundary reflections have any influence for collinear configurations.\smallskip\\Now we can start an analytical examination of the low bias regime, i.e. the linear transport.

\section{Linear transport}\label{lintrans}

What we are going to do analytically is to consider linear transport, where $eV_l\ll k_BT\,$, at thermal energies $k_BT\ll\epsilon_0\leq\epsilon_{jn}\ \forall jn$. In this case, both \textit{fermionic} (\footnote{In the presence of fermionic excitations the four $\tilde{r}\sigd$ bands are no longer filled as equally as possible: there are at least two band whose electron fillings differ by more than one. Still, however, it is the lowermost states of each band which must be populated. "Holes" can solely be produced by bosonic excitations}) and bosonic excitations can only be created virtually. That is why $|\vec{N},\vec{m}\ket=|\vec{N},\vec{0}\ket$ is certainly true for all states contributing to transport in our limit, and caring only about different fermionic configurations we symbolise the associated eigenstates by $|\vec{N}\ket$. Due to the absence of any excitations, the four bands of the SWCNT have to be populated as equally as possible, i.e. their electron fillings $N_{\tilde{L}\downarrow}\,,\ N_{\tilde{L}\uparrow}\,,\ N_{\tilde{R}\downarrow}\,,\ N_{\tilde{R}\uparrow}$ may at most differ by one.\smallskip\\Involved states $|\vec{N}\ket$ are thus for fixed $|\vec{N}|$ characterised by the energy

\[E_{\vec{N}}=E^{(0)}_N:=\min\{E_{\vec{N}}: |\vec{N}|=N\}\,.\]
They are the groundstates, i.e. the states with lowest possible energy for a fixed tube filling, and therefore we call them \textit{lowest energy (ground)states} (LEGs). For a SWCNT without a band offset $\delta$, Fig. \ref{combpic} shows all these LEGs which can, under the given conditions, a priori be involved in transport. We will soon explain that also for the virtual processes, only transitions between the LEGS contribute in the end. \begin{center}
\vskip -0.1cm
\includegraphics[width=0.70\textwidth]{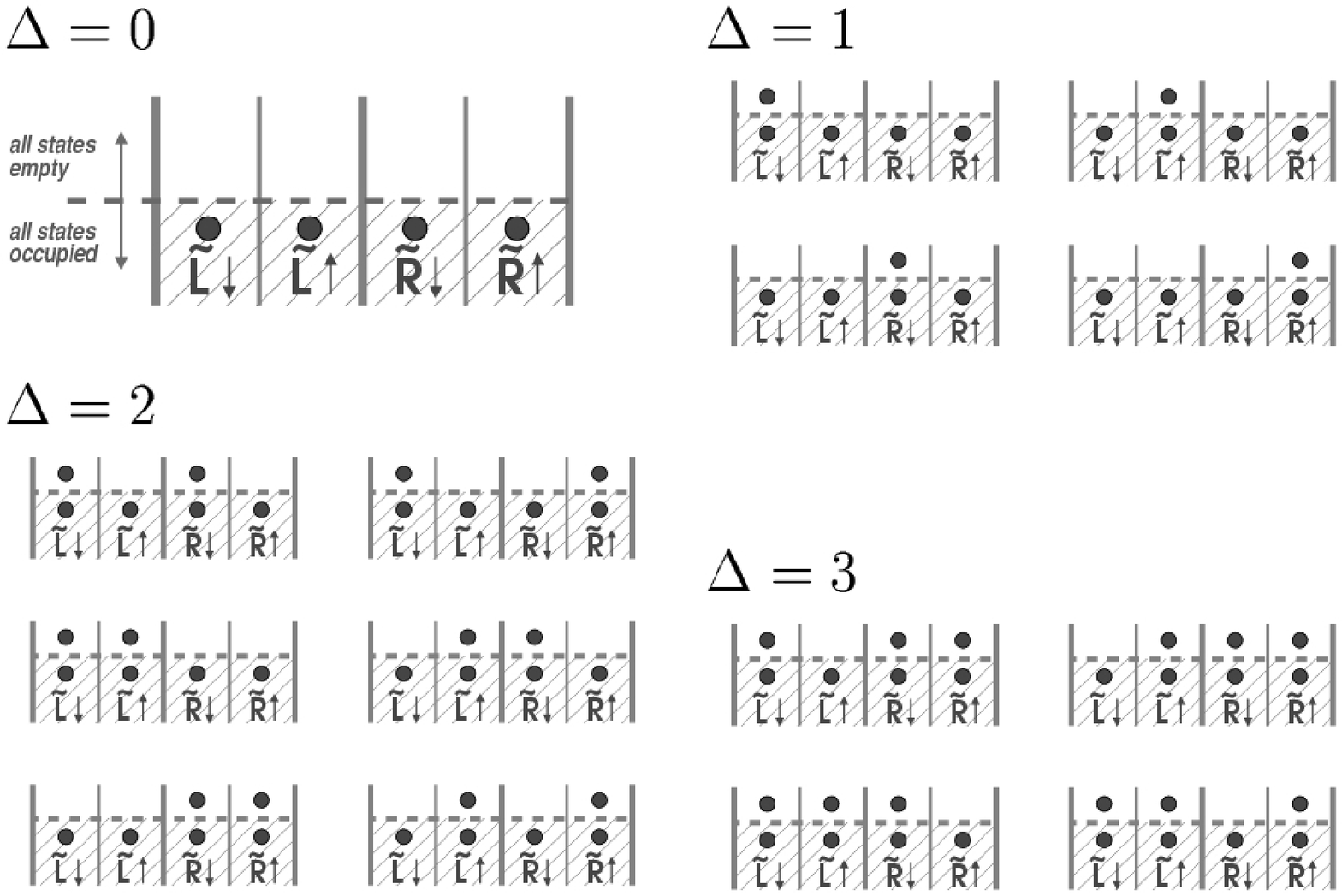}\vskip 0.3cm\imagecaption{\small All possible \textit{lowest energy (ground)states} for a SWCNT without band offset, filled with $4\mathbb{N}+\Delta$ electrons.}\label{combpic}\vskip 0.4cm
\end{center}
At low bias voltages, the current flow across the SWCNT is Coulomb blocked, unless the applied gate voltage $V_\gate$ aligns the states $|E^{(0)}_NN\ket$ and $|E^{(0)}_{N+1}N$+$1\ket$: resonant tunnelling then permits to add or remove the $N+1$st inside the SWCNT without any energy cost.\smallskip\\In this article we will just consider the situation $\delta=0$ as shown in Fig. \ref{combpic}, because this is sufficient for describing a SWCNT also in the case of a finite band offset $\delta>k_BT$ completely. Namely, if the thermal energy cannot overcome the energy gap between the $\tilde{R}$- and the $\tilde{L}$-band, resonant tunnelling can solely be allowed for one of those two bands: at each value of the gate voltage, we just deal with a specific of two independent single level quantum dots, while the other one is not of importance. Any regime of such a system can be mapped on either the resonance $\Delta=0\leftrightarrow\Delta=1$ or $\Delta=3\leftrightarrow\Delta=0$ for a zero band mismatch.

To calculate the current, we need the \textit{populations} of the LEGs,

\begin{equation}
P_{N}=\sum_{n\in\{|E^{(0)}_NN\ket\}} \left(\hrho_{\odot E^{(0)}_{N}N}^I(t)\right)_{nn}\,,
\label{population}\end{equation}
which represent the probabilities to find the SWCNT filled with $N$ electrons. As we want to examine the DC long term behaviour of the system, we claim that the populations do not change with time. If the value of $V_\gate$ is tuned to admit resonant tunnelling, this stationarity demands:
\begin{equation}\eqalign{
0=\dot{P}_{N}=-\sum_{l=s,d}A_{l}^{N\rightarrow N+1}P_{N}+\sum_{l=s,d}A_{l}^{N+1\rightarrow N}P_{N+1}\,,\\
0=\dot{P}_{N+1}=-\sum_{l=s,d}A_{l}^{N+1\rightarrow N}P_{N+1}+\sum_{l=s,d}A_{l}^{N\rightarrow N+1}P_{N}\,.}\label{pops}\end{equation}
Here, $A^{N\to N+1}_l\,/\,A^{N+1\to N}_l$ denote the rates by which electrons
tunnel into ($N\to N+1$)\,/\,out of ($N+1\to N$) the nanotube at contact $l$.\smallskip\\
The question is now to which elements of $\hrho_{\odot N}^{I}(t)$, $\hrho_{\odot N+1}^{I}(t)$, $\hrho_{\odot N-1}^{I}(t)$ the time evolutions of the populations, $\dot{P}_{N}$ and $\dot{P}_{N+1}$, couple. For an answer, the rates --\,either Eq. (\ref{eq4:rates}) or (\ref{eq5:rates})\,-- must be consulted: we find that their real parts always include Fermi functions with an energy difference as an argument. While $V_\gate$ is chosen such that \[\mu_{N+1}:=E^{(0)}_{N+1}-E^{(0)}_{N} < k_BT\,,\] all other energy differences are at least of the order of the charging energy $E_c$ and due to the condition $k_BT \ll \epsilon_0 \approx E_c$, the real parts of the corresponding rates are suppressed enormously. Not the rates, but the Bloch-Redfield coefficients (\ref{blochred}) enter the equation of motion for the density matrix. By its definition, $R^{NN\pm1}$ is real and thus the only nonvanishing contribution is the one of $R^{NN+1}_{nm,kk'}$ with $k,k'$ being LEGs: merely elements of the density matrix $\hrho_{\odot E^{(0)}_{N+1} N+1}^{I}(t)$ are involved. For $\hrho_{\odot N}^{I}(t)$, the Kronecker deltas in $R^{NN}_{nm,jj'}$ demand either $n=j$ or $m=j'$. Hence, $E_n=E_m=E^{(0)}_N$ and consequently $E_j=E_j'=E^{(0)}_N$: no other element than the ones of $\hrho_{\odot E^{(0)}_NN}^{I}(t)$ itself are participating.\\
With the help of Fig. \ref{combpic}, finally the concerned Bloch-Redfield coefficients can be figured out. Before, however, the role of the electron operators must be revealed. From \cite{LEO}, it can be extracted that:
\begin{equation}
\left(\hpsi_{\tilde{r}\sigd }\right)_{ab}\left(\psid_{\tilde{r}'\sigd'}\right)_{bc}=\left\lbrace\begin{array}{cr}\pm1/\sqrt{2L_t}  & {\rm{if}}\quad \vec{N}_c-\vec{N}_a=\ve_{\tilde{r}\sigd}-\ve_{\tilde{r}'\sigd'}\,,\\0& {\rm{else}}\,.\end{array}\right.\label{exp_elop}
\end{equation}
If nonzero, the sign of (\ref{exp_elop}) depends on whether an even (+) or an odd (-) number of electrons with the same energy sits in the bands prior to $\tilde{r}\sigd$. To define prior, one has to choose an arbitrary but fixed ordering of the bands, e.g. $\tilde{L}$ be prior to $\tilde{R}$ and $\tilde{r}\downarrow$ be prior to $\tilde{r}\uparrow$.\smallskip\\
For our case, $a,c\in\{|E^{(0)}_{N}N\ket\}$ in Eq. (\ref{exp_elop}). What about $b\in\{|N$+$1\ket\}$? Still there is a sum over all $\Gamma_{niim}$ and $\Gamma_{nkkm}$ in $R^{NN}_{nm,jj'}\,$, with $i\in\{|N$-$1\ket$ and $\Gamma_{nkkm}$ with $k\in|N$+$1\ket$. For the real parts of the rates, the energy argument of the Fermi functions restricts these $i$ and $k$ to LEGs, but for the imaginary parts, this energy argument is integrated over and so there is no forbiddance for fermionic or bosonic excitations a priori. After some closer examination, however, it turns out that really all non-LEG contributions of the four rates $R^{NN}_{nm,jj'}$ contains drop out exactly. Especially, one has to comprehend that if for example in

\[\left(\hpsi_{\tilde{r}\sigd}\right)_{jk}\left(\psid_{\tilde{r}'\sigd'}\right)_{kj'}\,,\] $k\in\{N$+$1\ket\}$ was a non-LEG while $j$ and $j'$ are LEGs, the expression can be nonzero only for $\tilde{r}\sigd=\tilde{r}'\sigd'$: if we create an electron to reach an excited state and wish to return to a LEG, there is actually no other possibility than to remove again the electron we added. Of course, $j=j'$ then, such that the concerned rates $\Gamma^{(+)NN+1}_{l(nkkj)}$ and $\Gamma^{(-)NN+1}_{l(j'kkm)}$ in Eq. (\ref{blochred}) become $\Gamma^{(+)NN+1}_{l(nkkn)}$ and $\Gamma^{(-)NN+1}_{l(mkkm)}$; the imaginary parts of the latter, however, cancel each other.\\

After all, we can set up Eqns. (\ref{pops}) for the four different tunnelling regimes $\Delta\leftrightarrow \Delta+1$, where $\Delta:=N\mod4$. Once we have done so, it is easy to extract the current by just omitting the sum over one of the leads:

\begin{equation}
\fl\left|I_{\Delta\Delta+1}\right|=e\left|A_{s}^{\Delta\rightarrow \Delta+1}P_{\Delta}-A_{s}^{\Delta+1\rightarrow \Delta}P_{\Delta+1}\right|=e\left|A_{d}^{\Delta\rightarrow \Delta+1}P_{\Delta}-A_{d}^{\Delta+1\rightarrow \Delta}P_{\Delta+1}\right|\,.\label{curr}
\label{I_NN+1}\end{equation}
Notice that from here on, we replace the indices $N$ by $\Delta$, because for all the properties we will study just $N\mod4=\Delta$ and not the actual value of $N$ matters.\smallskip\\For the calculations we state

\[D_{l+_l}(\mu_\Delta)+D_{l-_l}(\mu_\Delta)=:D_l(\mu_\Delta)\equiv D_l\,;\] this is justified since the densities have to be taken at the Fermi edge $E_{F,l}$ of the lead metal and $E_{F,l}\gg \mu_\Delta$, so that we can assume them to be approximately constant within the range we consider.\\Moreover we define the \textit{contact polarisation}:

\[\mathcal{P}_l=\frac{D_{l+_l}-D_{l-_l}}{D_{l}}\ .\]

\subsection{Resonant tunnelling regime $\Delta=0\leftrightarrow\Delta=1$}
If $\Delta=0$, there is only one LEG, with all bands equally occupied (see Fig. \ref{combpic}). We have

\begin{equation*}
\hat{\rho}^I_{\Delta=0}(t)=\left(\hat{\rho}^I_{\Delta=0}(t)\right)_{11}=:P_0\,,
\end{equation*}
which is the probability to find the SWCNT filled with $N=4\tilde{N}\,,\ \tilde{N}\in\mathbb{N}$ electrons. Note that for the new variables we drop the argument $(t)$ to save some space in the following.\medskip\\
For $\Delta=1$, things are a bit more complicated, because the LEG is fourfold degenerate: the excess electron can be placed in each of the four bands $\tilde{L}\uparrow,\ \tilde{L}\downarrow,\ \tilde{R}\uparrow,\ \tilde{R}\downarrow\,$; consequently, the density matrix consists of sixteen elements. Eight of them, however, can immediately be set to
zero, as we know that our system is unpolarised with respect to $\tilde{L}$- and $\tilde{R}$-band. This forbids any transitions and thus coherences between those two
bands:

\begin{equation*}
\hat{\rho}^I_{\Delta=1}(t)=
\left(
\begin{array}{cc}
\hat{\rho}^I_{\Delta=1,\tilde{L}}(t) & 0\\
0 & \hat{\rho}^I_{\Delta=1,\tilde{R}}(t)
\end{array}
\right)\,.
\end{equation*}
Here, due to the indistinguishability of $\tilde{L}$- and $\tilde{R}$-band, we are allowed to set

\begin{equation*}
\hat{\rho}^I_{\Delta=1,\tilde{L}}(t)=\hat{\rho}^I_{\Delta=1,\tilde{R}}(t)=\left(
\begin{array}{cc}
P^{(1)}_{\downarrow} & p^{(1)}\rme^{\rmi\alpha^{(1)}}\\
p^{(1)}\rme^{-\rmi\alpha^{(1)}} & P^{(1)}_{\uparrow}
\end{array}
\right)\,,
\end{equation*}
where $P^{(1)}_{\downarrow}\,,\ P^{(1)}_{\uparrow}$ are the probabilities to find the SWCNT in a single-electron spin-down ($\downarrow$) respectively spin-up ($\uparrow$) state. Correspondingly,

\begin{equation*}P_1:=2P^{(1)}_{\downarrow}+2P^{(1)}_{\uparrow}\end{equation*} is the total occupation probability for one electron.\smallskip\\
Furthermore, the density matrix is hermitian and that is why we could define

\[\left(\hat{\rho}^I_{\Delta=1,\tilde{r}}(t)\right)_{12}=
\left(\hat{\rho}^I_{\Delta=1,\tilde{r}}(t)\right)^{*}_{21}=:p^{(1)}\rme^{\rmi\alpha^{(1)}}
\]
for the off-diagonal elements.\\ The meaning of these quantities is revealed when we extract the information about the average spin $\vec{S}^{(1)}$ on the quantum dot from $\hat{\rho}^I_{\Delta=1}$; a single spin-$1/2$ particle that can either be in the spin-up or the spin-down state would be described by a $2\times2$ density matrix and the average spin would be given by a trace with the Pauli matrices. Now we have additionally the two $\tilde{r}$-bands, but those decouple completely, such that we can obtain the components of the average spin by

\begin{equation*}S^{(1)}_j=\frac{1}{2}\,\Tr\left(\left(\begin{array}{cc}\sigma_j&0\\0&\sigma_j\end{array}\right)\hrho^I_{\Delta=1}\right)\ ,\quad j\in\left\{x,y,z\right\}\,,\end{equation*}
where $\sigma_j$ are the Pauli matrices, and therefore
\begin{eqnarray*}
S^{(1)}_x&=&\frac{1}{2}\left(2p^{(1)}\rme^{\rmi\alpha^{(1)}}+2p^{(1)}\rme^{-\rmi\alpha^{(1)}}\right)=2p^{(1)}\cos(\alpha^{(1)})\,,\\
S^{(1)}_y&=&\frac{\rmi}{2}\left(2p^{(1)}\rme^{\rmi\alpha^{(1)}}-2p^{(1)}\rme^{-\rmi\alpha^{(1)}}\right)=-2p^{(1)}\sin(\alpha^{(1)})\,,\\
S^{(1)}_z&=&\frac{1}{2}\left(2P^{(1)}_\uparrow-2P^{(1)}_\downarrow\right)=P^{(1)}_\uparrow-P^{(1)}_\downarrow\,.
\end{eqnarray*}
If we chose our spin quantisation axis inside the SWCNT such that $\vec{z}_\odot\,||\vec{S}^{(1)}$, then $S^{(1)}_x$ and $S^{(1)}_y$ and with them all off-diagonal elements of $\hrho_{\Delta=1}$ would vanish: we had a coordinate system where the hermitian density matrix is diagonalised. For arbitrary $\theta$\,, we are not able to give the direction of $\vec{S}$ a priori and thus cannot make any use of this insight. For $\theta=0°$ and $\theta=\pi$, however, we have already found the right axis and could obtain diagonal equations when using the rates (\ref{eq5:rates}) instead of (\ref{eq4:rates}). But actually the $\Delta=0\leftrightarrow\Delta=1$ regime is analytically accessible for any $\theta$ and as it is instructive, we want to set up the general equations, in terms of the physical quantities $P_0$, $P_1$, $S_x$, $S_y$ and $S_z$, which are five independent variables determining the density matrices $\hrho_{\Delta=0}$ and $\hrho_{\Delta=1}$\,.\smallskip\\The equations for $\dot{P}_0$ and $\dot{P}_1$ are not independent ($\dot{P}_0=-\dot{P}_1$), but together with the normalisation condition \begin{equation}P_0+P_1=1\label{eqa:0}\end{equation}
we have the necessary set of five equations, where the physical meaning of the single terms becomes obvious:

\begin{eqnarray}
\fl\frac{d}{dt} P_1 =\frac{\pi}{\hbar}\sum\limits_{l=s,d}\Phi_lD_l\Biggl[\underbrace{4f_l(\mu_{\Delta=1})P_0}_{\scriptsize \begin{array}{c}\rm{electrons}\\\rm{tunnelling\,\,in}\end{array}}\underbrace{-\left(1-f_l(\mu_{\Delta=1})\right)P_1}_{\scriptsize \begin{array}{c}\rm{electrons}\\\rm{tunnelling\,\,out}\end{array}}\underbrace{-\ 2\mathcal{P}_l\left(1-f_l(\mu_{\Delta=1})\right)\left(\vec{S}^{(1)}\cdot\vec{m}_l\right)}_{\scriptsize \begin{array}{c}\rm{difference\,in\,chemical\,potential}\\\rm{for\,spin\!\!-\!\!up\,and\,spin\!\!-\!\!down}\end{array}}\Biggr]\,,\nonumber\\\fl
\frac{d}{dt}\vec{S}^{(1)}=\frac{\pi}{\hbar}\sum\limits_{l=s,d}\Phi_lD_l\Biggl[\underbrace{\mathcal{P}_l\left(2f_l(\mu_{\Delta=1})P_0-\frac{1}{2}\left(1-f_l(\mu_{\Delta=1})\right)P_1\right)\vec{m}_l}_{\rm{spin\,\,accumulation}}-\Biggr.\nonumber\\\fl\hspace{4.4cm}\Biggl.\underbrace{-\left(1-f_l(\mu_{\Delta=1})\right)\vec{S}^{(1)}}_{\rm{spin\,\,relaxation}}-\underbrace{\frac{\mathcal{P}_l}{\pi}\,\mathfrak{P}_l(\mu_{\Delta=1},\mu_{\Delta=2})\left(\vec{m}_l\times\vec{S}^{(1)}\right)}_{\rm{spin\,\,precession}}\,\Biggr]\,.\nonumber\\\fl\label{01eqns}
\end{eqnarray}The probability $P_1$ of the SWCNT to be occupied with $4\tilde{N}+1$ electrons grows with the rate of electrons tunnelling into the tube already containing $4\tilde{N}$ electrons and decays with electrons leaving the quantum dot occupied with $4\tilde{N}+1$. Additionally, the average spin of the electrons inside the SWCNT interacts with the magnetic fields in the contacts, which yields a special term accounting for the difference in the chemical potential for spin-up and spin-down electrons.\smallskip\\That is why we have also to solve the second equation for $\vec{S}^{(1)}$. The time evolution of the average spin is affected by three contributions: the net number of particles accumulating on the tube brings a spin polarised along the lead magnetisations with it; electrons tunnelling out take some spin with them such that the spin inside the tube relaxes, and finally the electrons feel the magnetic field of the contacts, what makes them precess a little bit.\\Besides of a factor $2$ (due to the fact that we have the distinct $\tilde{r}$ bands) and the precise form of $\mathfrak{P}_l$, these are the very same equations obtained in \cite{KOE} for a single level quantum dot. The function

\begin{equation*}\mathfrak{P}_l(\mu_{\Delta=1},\mu_{\Delta=2})\equiv\pint\rmd{}\epsilon\left(\frac{1-f_l(\epsilon)}{\epsilon-\mu_{\Delta=1}}+\frac{f_l(\epsilon)}{\epsilon-\mu_{\Delta=2}}\right)+\frac{\Phi^{-1}_lR_l}{\mathcal{P}_l D_l}\end{equation*}
summarises all the imaginary contributions the rates contain; this means that it merges our two exchange effects: the interface backscattering processes --\,which have not been included in \cite{KOE}\,-- and the virtual transitions. The exchange is responsible for the precession of the total spin on the quantum dot.\\The principal part integration appearing in $\mathfrak{P}_l$ can be evaluated by a trick:\\for a Lorentzian-like shaped function $L(\epsilon)=L\,\frac{E^2_W}{\epsilon^2+E_W^2}$\,,

\[\pint d\epsilon \frac{L(\epsilon)f(\pm\epsilon)}{\epsilon-E}\approx \mp L \ln\frac{E_W}{\max\left(\kBT,|E|\right)}\,.\]For $E_W\to\infty$, $L(\epsilon)$ is constant and approximately we may set our function $D_{l\uparrow\downarrow}(\epsilon)=D_{l+_l}(\epsilon)-D_{l-_l}(\epsilon)\equiv \mathcal{P}_lD_l$ constant under the principal part integral. This yields for $\mathfrak{P}_l$:
\begin{equation}\label{digammaweg}\mathfrak{P}_l(\mu_{\Delta=1},\mu_{\Delta=2})=\underbrace{\ln\frac{\max\left(\kBT,|\mu_{\Delta=2}|\right)}{\max\left(\kBT,|\mu_{\Delta=1}|\right)}}_{=:\mathfrak{P}_0}+\frac{R_{l}}{\Phi_l\mathcal{P}_lD_l}\,.\end{equation}
Why do we employ this estimation for the digamma function \cite{KOE1}? In our analytical results, we will find $\mathfrak{P}_l$ accompanied by a Fermi function, which at low temperatures dominates the gate voltage evolution of the product. Therefore, we will, though (\ref{digammaweg}) cuts the peaks of the digamma function, without any exception obtain smooth curves (which indicates that we make no noticeable error).\\

For the simplest case of identical source and drain tunnelling contacts,

\[D_s=D_d=:D\,,\ \mathcal{P}_s=\mathcal{P}_d=:\mathcal{P}\,,\  \Phi_s=\Phi_d=:\Phi\,,\  R_s=R_d=:R\]--\,and thus $\mathfrak{P}_s=\mathfrak{P}_d=:\mathfrak{P}$\,-- we can give a nice analytical expression for the low bias current in dependence on both the magnetisation angle $\theta$ and the bias voltage $V_\gate$. The zero point of the latter we fix for each resonance $\Delta\leftrightarrow\Delta+1$ such that $V_\gate=0\Leftrightarrow E^{(0)}_{\Delta+1}-E^{(0)}_\Delta=0\,$, which means $V_\gate=-\mu_{\Delta+1}$:

\numparts
\begin{eqnarray}
I_{01}(\theta,V_\gate,V_\bias)=\tilde{I}_{01}\left(1-\frac{\mathcal{P}^2\sin^2\left({\frac{\theta}{2}}\right)}{1+\mathcal{P}^2\frac{\mathfrak{P}^2(\mu_{\Delta=1},\mu_{\Delta=2})}{\pi^2f^2(-\mu_{\Delta=1})}\,\cos^2\left({\frac{\theta}{2}}\right){}}\right)V_\bias\,,\label{I01}
\end{eqnarray}where $V_d=-V_s=\frac{V_\bias}{2}$ and\begin{equation}\tilde{I}_{01}=\frac{2\pi^2e^2}{hk_BT}\,\Phi D\,\frac{f(\mu_{\Delta=1})f(-\mu_{\Delta=1})}{1+3f(\mu_{\Delta=1})}\,,\label{I01b}
\end{equation}
\endnumparts
just as it was derived in \cite{LEO} for a SWCNT attached to unpolarised leads. Fig. \ref{graph1}a shows $I_{01}$ in dependence on $V_\gate$ around its resonance for several $\theta$ and the corresponding TMR curves

\begin{equation}{\rm{TMR}}\,(\theta)=\frac{I_{\Delta\Delta+1}(0,V_\gate)-I_{\Delta\Delta+1}(\theta,V_\gate)}{I_{\Delta\Delta+1}(0,V_\gate)}\label{TMR}\end{equation} at $\Delta=0$. Notice that here we have defined a TMR for \textit{arbitrary} angles. Eq. (\ref{TMR}) expands the definition we gave in the introduction for the special case of $\theta=\pi$.\\In Fig. \ref{graph2}a, $V_\gate$ is fixed at three different values and the normalised current $I_{01}(\theta)/I_{01}(0)$ is plotted against $\theta$. The parameters we employ are given beneath the plots.\\
Figs. \ref{graph1}b and \ref{graph2}b belong to the $\Delta=3\leftrightarrow\Delta=0$ resonance and are mirror-symmetric to \ref{graph1}a and \ref{graph2}a with respect to $V_\gate=0$. This fact results from the symmetry in the LEGs: for $\Delta=1$, there is one excess electron that can be put in any of the four bands, and for $\Delta=3$ it is just the same situation with one hole instead of an electron. The resulting formula for the current is actually:

\numparts
\begin{eqnarray}
I_{30}(\theta,V_\gate,V_\bias)=\tilde{I}_{30}\left(1-\frac{\mathcal{P}^2\sin^2\left({\frac{\theta}{2}}\right)}{1+\mathcal{P}^2\frac{\mathfrak{P}^2(\mu_{\Delta=0},\mu_{\Delta=1})}{\pi^2f^{2}(\mu_{\Delta=0})}\,\cos^2\left({\frac{\theta}{2}}\right){}}\right)V_\bias\,,\label{I30}
\end{eqnarray}and  \begin{equation}\tilde{I}_{30}=\frac{2\pi^2e^2}{hk_BT}\,\Phi D\,\frac{f(-\mu_{\Delta=0})f(\mu_{\Delta=0})}{1+3f(-\mu_{\Delta=0})}\label{I30b}\,.\end{equation}\endnumparts

\vskip 0.5cm
Unless specified differently, we will employ for these and all following plots the parameters listed below in Tab. \ref{parm}.
\begin{center}\vskip 0.3cm
\fbox{
\begin{minipage}{10cm}
\hspace{-0.1cm}\begin{tabular}{rl}
Polarisation & $\mathcal{P}=0.6$\\
Reflection & $R/(\Phi\mathcal{P}D)=0.1\mathfrak{P}_0$\\
Tube length & $L_t=580\,$nm\\
Temperature & $T=20\,$mK\\
Thermal energy & $\kBT\approx 1.73\,\mu$eV\\
SWCNT charging energy & $E_c=9.49\,$meV\\
SWCNT level spacing & $\epsilon_0=2.89\,$meV
\end{tabular}
\end{minipage}}
\vskip 0.6cm
\tabcaption{Parameters used for the plots (if not specified differently)}\label{parm}
\end{center}

\begin{center}
\hspace{-5.6cm}\begin{minipage}{0.5\textwidth}
\begin{footnotesize}
\textsf{
\begin{picture}(0,0)%
\includegraphics{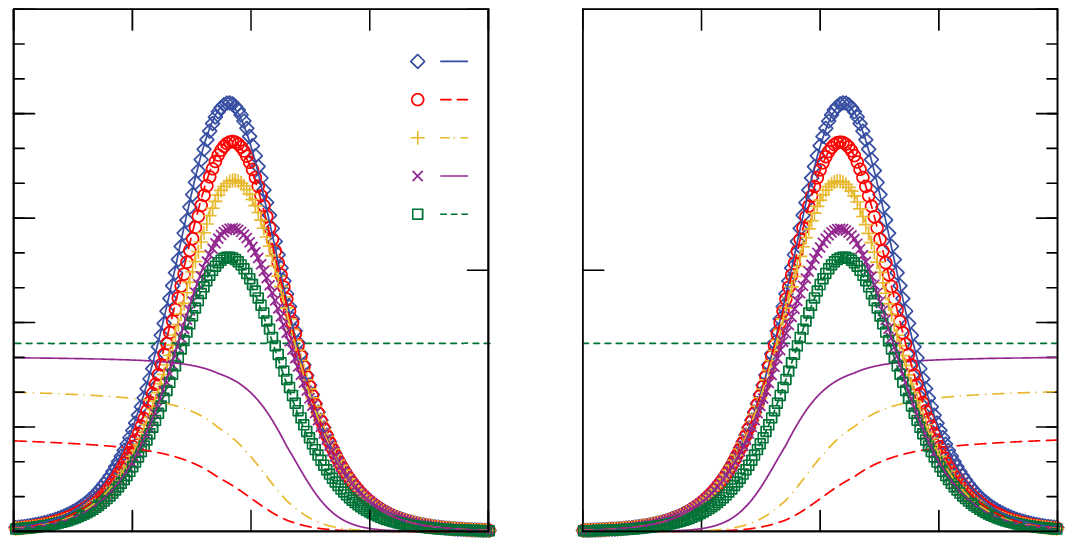}%
\end{picture}%
\begingroup
\setlength{\unitlength}{0.0200bp}%
\begin{picture}(18000,10800)(0,0)%
\put(1396,5410){\rotatebox{90}{\makebox(0,0){\strut{}\quad}}}%
\put(5364,275){\makebox(0,0){\strut{}\quad}}%
\put(7277,8418){\makebox(0,0)[r]{\strut{}\quad}}%
\put(7277,7868){\makebox(0,0)[r]{\strut{}\quad}}%
\put(7277,7318){\makebox(0,0)[r]{\strut{}\quad}}%
\put(7277,6768){\makebox(0,0)[r]{\strut{}\quad}}%
\put(7277,6218){\makebox(0,0)[r]{\strut{}\quad}}%
\put(1396,5410){\rotatebox{90}{\makebox(0,0){\strut{}\quad}}}%
\put(5364,275){\makebox(0,0){\strut{}\quad}}%
\put(1650,1650){\makebox(0,0)[r]{\strut{} 0}}%
\put(1650,3154){\makebox(0,0)[r]{\strut{} 0.3}}%
\put(1650,4658){\makebox(0,0)[r]{\strut{} 0.6}}%
\put(1650,6162){\makebox(0,0)[r]{\strut{} 0.9}}%
\put(1650,7666){\makebox(0,0)[r]{\strut{} 1.2}}%
\put(1650,9170){\makebox(0,0)[r]{\strut{} 1.5}}%
\put(8761,1100){\makebox(0,0){\strut{}12}}%
\put(7052,1100){\makebox(0,0){\strut{}6}}%
\put(5343,1100){\makebox(0,0){\strut{}0}}%
\put(3634,1100){\makebox(0,0){\strut{}-6}}%
\put(1925,1100){\makebox(0,0){\strut{}-12}}%
\put(9036,9170){\makebox(0,0)[l]{\strut{}\,\ 1}}%
\put(9036,5410){\makebox(0,0)[l]{\strut{}0.5}}%
\put(9036,1650){\makebox(0,0)[l]{\strut{}\,\ 0}}%
\put(550,5410){\rotatebox{90}{\makebox(0,0){\strut{}Current $I_{01}$ [nA]}}}%
\put(9366,4310){\rotatebox{90}{\makebox(0,0){\strut{}TMR}}}%
\put(9468,275){\makebox(0,0){\strut{}Gate voltage $e\alpha V_\gate$ [$\mu$eV]}}%
\put(2882,8418){\makebox(0,0){\textbf{\textsf{(a)}}}}%
\put(5616,8418){\makebox(0,0)[l]{\strut{}$\theta=$}}%
\put(7802,8418){\makebox(0,0)[r]{\strut{}\scriptsize{0}$\dgr\quad$}}%
\put(7802,7868){\makebox(0,0)[r]{\strut{}\scriptsize{115}$\dgr\quad$}}%
\put(7802,7318){\makebox(0,0)[r]{\strut{}\scriptsize{140}$\dgr\quad$}}%
\put(7802,6768){\makebox(0,0)[r]{\strut{}\scriptsize{160}$\dgr\quad$}}%
\put(7802,6218){\makebox(0,0)[r]{\strut{}\scriptsize{180}$\dgr\quad$}}%
\put(9845,5410){\makebox(0,0)[r]{\strut{}\quad}}%
\put(16956,1100){\makebox(0,0){\strut{}12}}%
\put(15247,1100){\makebox(0,0){\strut{}6}}%
\put(13538,1100){\makebox(0,0){\strut{}0}}%
\put(11829,1100){\makebox(0,0){\strut{}-6}}%
\put(10120,1100){\makebox(0,0){\strut{}-12}}%
\put(17231,1650){\makebox(0,0)[l]{\strut{} 0}}%
\put(17231,3154){\makebox(0,0)[l]{\strut{} 0.3}}%
\put(17231,4658){\makebox(0,0)[l]{\strut{} 0.6}}%
\put(17231,6162){\makebox(0,0)[l]{\strut{} 0.9}}%
\put(17231,7666){\makebox(0,0)[l]{\strut{} 1.2}}%
\put(17231,9170){\makebox(0,0)[l]{\strut{} 1.5}}%
\put(18330,5410){\rotatebox{90}{\makebox(0,0){\strut{}Current $I_{30}$ [nA]}}}%
\put(16288,275){\makebox(0,0){\strut{}\quad}}%
\put(11077,8418){\makebox(0,0){\textbf{\textsf{(b)}}}}%
\end{picture}%
\endgroup
}
\end{footnotesize}
\end{minipage}
\vskip 0.2cm
\imagecaption{\small Currents $I_{01}(V_\gate)$ and $I_{30}(V_\gate)$ for different values of $\theta$, together with the TMR (\ref{TMR}). The symbols represent numerical data, which is perfectly fit by the analytical lines. The corresponding angle-dependent TMR curves are plotted as lines as well.}\label{graph1}
\end{center}

\pagebreak
\quad
\vspace{-2.8cm}
\quad
\begin{center}
\hspace{-5.5cm}\begin{minipage}{0.5\textwidth}
\begin{footnotesize}
\textsf{
\begin{picture}(0,0)%
\includegraphics{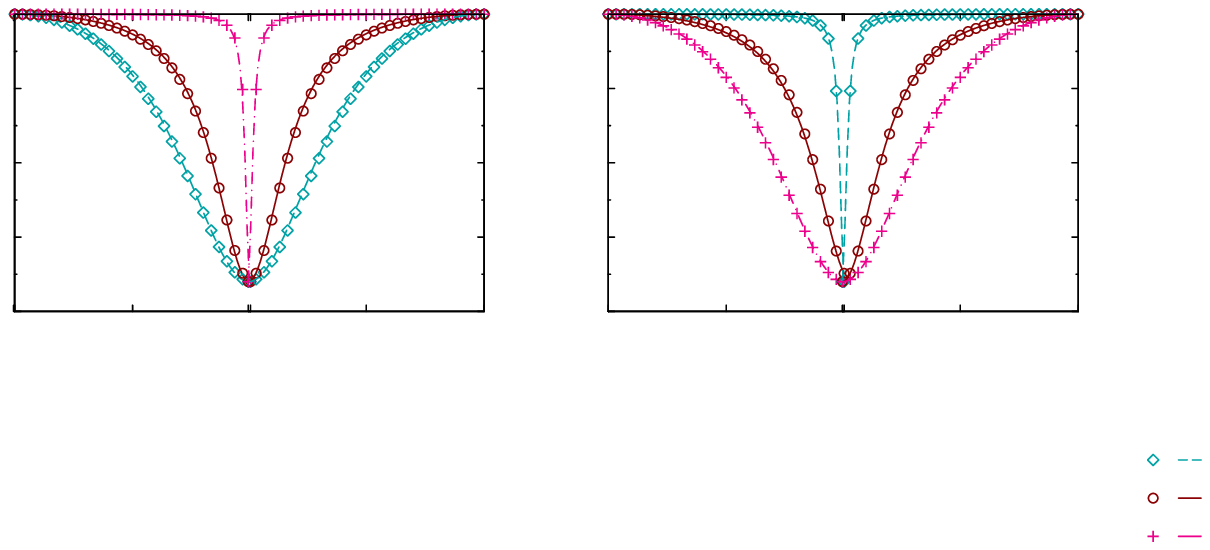}%
\end{picture}%
\begingroup
\setlength{\unitlength}{0.0200bp}%
\begin{picture}(18000,10800)(0,0)%
\put(1910,1100){\makebox(0,0){\strut{}}}%
\put(1360,3790){\rotatebox{90}{\makebox(0,0){\strut{}\quad}}}%
\put(5855,3790){\rotatebox{90}{\makebox(0,0){\strut{}\quad}}}%
\put(3608,275){\makebox(0,0){\strut{}\quad}}%
\put(18411,-489){\makebox(0,0)[r]{\strut{}\scriptsize{$e\alpha V_\gate=-5.2\,\mu$eV\quad }}}%
\put(18411,-1039){\makebox(0,0)[r]{\strut{}\scriptsize{$e\alpha V_\gate=\,0\mu$eV\quad }}}%
\put(18411,-1589){\makebox(0,0)[r]{\strut{}\scriptsize{$e\alpha V_\gate=+5.2\,\mu$eV\quad }}}%
\put(1650,1650){\makebox(0,0)[r]{\strut{} 0.6}}%
\put(1650,2720){\makebox(0,0)[r]{\strut{} 0.7}}%
\put(1650,3790){\makebox(0,0)[r]{\strut{} 0.8}}%
\put(1650,4860){\makebox(0,0)[r]{\strut{} 0.9}}%
\put(1650,5930){\makebox(0,0)[r]{\strut{} 1}}%
\put(5321,1100){\makebox(0,0){\strut{}$\mathsf{\pi}$}}%
\put(3623,1100){\makebox(0,0){\strut{}$\mathsf{\frac{\pi}{2}}$}}%
\put(1925,1100){\makebox(0,0){\strut{}0}}%
\put(550,3240){\rotatebox{90}{\makebox(0,0){\strut{}Normalised current $I_{01}(\theta)/I_{01}(0)$}}}%
\put(5870,3790){\rotatebox{90}{\makebox(0,0){\strut{}\quad}}}%
\put(9618,275){\makebox(0,0){\strut{}Magnetisation angle $\theta$ [rad]}}%
\put(2672,2249){\makebox(0,0){\textbf{\textsf{(a)}}}}%
\put(17883,-489){\makebox(0,0)[r]{\strut{}\quad}}%
\put(17883,-1039){\makebox(0,0)[r]{\strut{}\quad}}%
\put(17883,-1589){\makebox(0,0)[r]{\strut{}\quad}}%
\put(8685,1100){\makebox(0,0){\strut{}$2\pi$}}%
\put(6987,1100){\makebox(0,0){\strut{}$\frac{3\pi}{2}$}}%
\put(5289,1100){\makebox(0,0){\strut{}}}%
\put(8960,1650){\makebox(0,0)[l]{\strut{}}}%
\put(8960,2720){\makebox(0,0)[l]{\strut{}}}%
\put(8960,3790){\makebox(0,0)[l]{\strut{}}}%
\put(8960,4860){\makebox(0,0)[l]{\strut{}}}%
\put(8960,5930){\makebox(0,0)[l]{\strut{}}}%
\put(9509,3790){\rotatebox{90}{\makebox(0,0){\strut{}\quad}}}%
\put(12982,275){\makebox(0,0){\strut{}\quad}}%
\put(10198,1650){\makebox(0,0)[r]{\strut{}}}%
\put(10198,2720){\makebox(0,0)[r]{\strut{}}}%
\put(10198,3790){\makebox(0,0)[r]{\strut{}}}%
\put(10198,4860){\makebox(0,0)[r]{\strut{}}}%
\put(10198,5930){\makebox(0,0)[r]{\strut{}}}%
\put(13869,1100){\makebox(0,0){\strut{}$\pi$}}%
\put(12171,1100){\makebox(0,0){\strut{}$\frac{\pi}{2}$}}%
\put(10473,1100){\makebox(0,0){\strut{}$0$}}%
\put(14418,3790){\rotatebox{90}{\makebox(0,0){\strut{}\quad}}}%
\put(18166,275){\makebox(0,0){\strut{}\quad}}%
\put(11220,2249){\makebox(0,0){\textbf{\textsf{(b)}}}}%
\put(17240,1100){\makebox(0,0){\strut{}$2\pi$}}%
\put(15542,1100){\makebox(0,0){\strut{}$\mathsf{\frac{3\pi}{2}}$}}%
\put(13844,1100){\makebox(0,0){\strut{}}}%
\put(17515,1650){\makebox(0,0)[l]{\strut{}0.6}}%
\put(17515,2720){\makebox(0,0)[l]{\strut{}0.7}}%
\put(17515,3790){\makebox(0,0)[l]{\strut{}0.8}}%
\put(17515,4860){\makebox(0,0)[l]{\strut{}0.9}}%
\put(17515,5930){\makebox(0,0)[l]{\strut{}1}}%
\put(13294,3790){\rotatebox{90}{\makebox(0,0){\strut{}\quad}}}%
\put(18614,3240){\rotatebox{90}{\makebox(0,0){\strut{}Normalised current $I_{30}(\theta)/I_{30}(0)$}}}%
\put(21537,275){\makebox(0,0){\strut{}\quad}}%
\end{picture}%
\endgroup
}
\end{footnotesize}
\end{minipage}
\vskip 1.4cm\imagecaption{\small Normalised currents $I_{01}(\theta)/I_{01}(0)$ and $I_{30}(\theta)/I_{30}(0)$ for different values of $V_\gate$.}\label{graph2}
\end{center}\vskip 0.8cm

Besides for the mirror-symmetry with respect to $V_\gate=0$, there is only one more difference between (\ref{I01}) and (\ref{I30}): the argument of $\mathfrak{P}$ is $\mu_{\Delta=2}=E_c-\mu_{\Delta=1}$ for the $\Delta=0\leftrightarrow\Delta=1$ regime, but at $\Delta=3\leftrightarrow\Delta=0$, additionally the level spacing comes in: $\mu_{\Delta=1}=E_c+\epsilon_0-\mu_{\Delta=1}$. The resulting deviations, however, are washed out by the squares of the trigonometric functions and the polarisation in our formula and thus are not noticeable in figs. \ref{graph1} and \ref{graph2}. As already mentioned, the dominating gate voltage dependence of the Fermi function by which $\mathfrak{P}$ is divided regulates many features of the plotted curves. For instance, it determines the roaming of the peak maxima towards $V_\gate=0$ for $\theta\not\in\{0,\pi\}$. In the case of collinear magnetisations ($\theta\in\{0,\pi\})\,$, the maxima positions, which are identical, can be calculated easily by differentiating (\ref{I01b}). They lie a bit off-resonance, at
\begin{equation*}e\alpha V_\gate=\left\lbrace\begin{array}{cl}\!\!- \kBT\ln{2}\approx -1.2 \,\mu{\rm{eV}}&{\rm{for}}\ \Delta=0\leftrightarrow\Delta=1\,,\\\!\!+ \kBT\ln{2}\approx+1.2\,\mu{\rm{eV}}&{\rm{for}}\ \Delta=3\leftrightarrow\Delta=0\,.\end{array}\right.\label{eq:peakmax}\end{equation*} Plugging the currents into Eq. (\ref{TMR}), the evolution of the TMR can be explained. For $\theta=\pi\,$, it is constant at the value $1-\mathcal{P}^2$. For non-collinear contact magnetisations, again the Fermi function in question is the decisive factor: the larger the inverse of its square becomes, the closer gets $I_{\Delta\Delta+1}(\theta)$ to $I_{\Delta\Delta+1}(0)$ and the TMR vanishes. In a physical sense, we can imagine the following picture: below the $\Delta=0\leftrightarrow\Delta=1$ resonance, a spin-polarised electron which is transferred across the quantum dot will not stay inside the SWCNT too long and thus not have time to equilibrate its spin, which hinders the transport, yielding a nonzero TMR. Above the resonance, the tube is mostly populated with $4\mathbb{N}+1$ electrons, such that for all but exactly antiparallel configurations, the spin of the excess electron will have equilibrated before it tunnels out, and consequently the non-collinear TMR decreases around the resonance with growing gate voltage. In analogue, the mirror-symmetric behaviour of the $\Delta=3\leftrightarrow\Delta=0$ regime can be understood, when thinking in terms of holes rather than electrons.\smallskip\\This also explains the dependence of the normalised currents on $\theta$, fig \ref{graph2}. Transport is blocked by an antiparallel polarisation of the contacts, and that is why the plots exhibit a dip at $\theta=\pi$. The more likely a spin-equilibration, the closer is $I_{\Delta\Delta+1}(\theta)$ to $I_{\Delta\Delta+1}(0)$, i.e. the narrower get the curves. So for $\Delta=0$, the width of the dip must shrink when raising the gate voltage, while for $\Delta=3$ it is just the other way round.\quad\\\vskip 0.3cm

\begin{tabular}{ll}
\hspace{-0.9cm}
\begin{minipage}{0.41\textwidth}
\vspace{-1.1cm}
\begin{footnotesize}
\textsf{
\hspace{-0.5cm}
\begin{picture}(0,0)%
\includegraphics{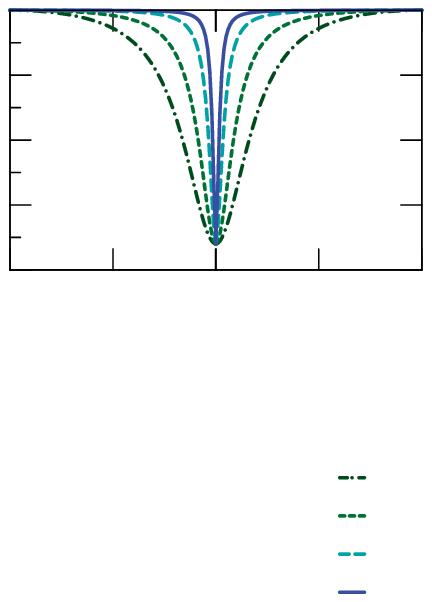}%
\end{picture}%
\begingroup
\setlength{\unitlength}{0.0200bp}%
\begin{picture}(18000,10800)(0,0)%
\put(2129,5257){\makebox(0,0)[r]{\strut{} 0.6}}%
\put(2129,6192){\makebox(0,0)[r]{\strut{} 0.7}}%
\put(2129,7127){\makebox(0,0)[r]{\strut{} 0.8}}%
\put(2129,8062){\makebox(0,0)[r]{\strut{} 0.9}}%
\put(2129,8997){\makebox(0,0)[r]{\strut{} 1}}%
\put(5372,4707){\makebox(0,0){\strut{}$\pi$}}%
\put(3888,4707){\makebox(0,0){\strut{}$\frac{\pi}{2}$}}%
\put(2404,4707){\makebox(0,0){\strut{}$0$}}%
\put(479,7127){\rotatebox{90}{\makebox(0,0){\strut{}Normalised current $I_{01}(\theta)/I_{01}(0)$}}}%
\put(5263,3882){\makebox(0,0){\strut{}Magnetisation angle $\theta$ [rad]}}%
\put(1899,3013){\makebox(0,0)[l]{\strut{}\scriptsize{$e\alpha V_\gate=0\,\mu$eV}}}%
\put(5669,3013){\makebox(0,0)[l]{\strut{}\scriptsize{$R/(\Phi\mathcal{P}D)=$}}}%
\put(6878,2355){\makebox(0,0)[r]{\strut{}\scriptsize{$0$}}}%
\put(6878,1805){\makebox(0,0)[r]{\strut{}\scriptsize{$\mathfrak{P}_0$}}}%
\put(6878,1265){\makebox(0,0)[r]{\strut{}\scriptsize{$4\,\mathfrak{P}_0$}}}%
\put(6878,705){\makebox(0,0)[r]{\strut{}\scriptsize{$10\,\mathfrak{P}_0$}}}%
\put(8334,4707){\makebox(0,0){\strut{}$2\pi$}}%
\put(6850,4707){\makebox(0,0){\strut{}$\frac{3\pi}{2}$}}%
\put(5366,4707){\makebox(0,0){\strut{}}}%
\put(8609,8997){\makebox(0,0)[l]{\strut{}}}%
\put(8609,8062){\makebox(0,0)[l]{\strut{}}}%
\put(8609,7127){\makebox(0,0)[l]{\strut{}}}%
\put(8609,6192){\makebox(0,0)[l]{\strut{}}}%
\put(8609,5257){\makebox(0,0)[l]{\strut{}}}%
\put(4816,7127){\rotatebox{90}{\makebox(0,0){\strut{}\quad}}}%
\put(9158,7127){\rotatebox{90}{\makebox(0,0){\strut{}\quad}}}%
\put(8225,3882){\makebox(0,0){\strut{}\quad}}%
\end{picture}%
\endgroup
}
\vskip 0.1cm
\imagecaption{In the case of symmetric tunnelling contacts, the dip width of the normalised current $I_{01}(\theta)/I_{01}(0)$ shrinks with growing $R$.}\label{varyR}
\end{footnotesize}
\end{minipage}
&
\hspace{0.56cm}
\begin{minipage}{0.5\textwidth}
We see from Eq. (\ref{digammaweg}) that $\mathfrak{P}$ grows (and hence the dips widths would shrink) with $R$: the reflection processes fortify the equilibration of spins. As a result, the width of the normalised current curves $I_{01}(\theta)/I_{01}(0)$ must shrink with increasing $R$, which can clearly be seen in Fig. \ref{varyR}.\\One should mention at this point that we chose the value of $R$ (Tab. \ref{parm}) small enough to see the differences for distinct angles in figs. \ref{graph1} and \ref{graph2}. For real systems, the actual value of $R$ might very well depend on how the different domains of the ferromagnetic contacts couple to the graphene sublattices and probably $R$ differs from sample to sample. At the current state of the experimental art, where adjusting a well-defined angle between the lead magnetisations is still a big challenge, we find it important to visualise the qualitative impact of the different quantities.
\end{minipage}
\end{tabular}\bigskip\\\\

\begin{tabular}{lc}
\hspace{-1.20cm}
\begin{minipage}{7cm}
We can furthermore give the analytical expressions for the probabilities $P_0$ and $P_1$ to find the SWCNT populated with $N=4\tilde{N}$ respectively $N=4\tilde{N}+1$ electrons (fig. \ref{graph3}):\\
\begin{equation}\hspace{-0.6cm}\eqalign{
P_0=\frac{f(-\mu_{\Delta=1})}{1+3f(\mu_{\Delta=1})}\,,\\P_1=\frac{4f(\mu_{\Delta=1})}{1+3f(\mu_{\Delta=1})}\,.}\label{eq:pops}\quad\quad\end{equation}
\end{minipage}
&
\hspace{0.6cm}
\begin{minipage}{7.4cm}
\begin{footnotesize}
\textsf{
\begin{picture}(0,0)%
\includegraphics{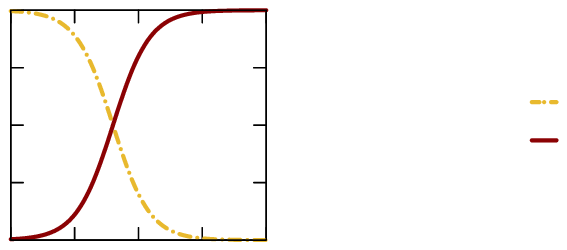}%
\end{picture}%
\begingroup
\setlength{\unitlength}{0.0200bp}%
\begin{picture}(9000,5400)(0,0)%
\put(1925,1539){\makebox(0,0)[r]{\strut{} 0}}%
\put(1925,2367){\makebox(0,0)[r]{\strut{} 0.25}}%
\put(1925,3195){\makebox(0,0)[r]{\strut{} 0.5}}%
\put(1925,4022){\makebox(0,0)[r]{\strut{} 0.75}}%
\put(1925,4850){\makebox(0,0)[r]{\strut{} 1}}%
\put(5878,989){\makebox(0,0){\strut{}12}}%
\put(4958,989){\makebox(0,0){\strut{}6}}%
\put(4039,989){\makebox(0,0){\strut{}0}}%
\put(3119,989){\makebox(0,0){\strut{}-6}}%
\put(2200,989){\makebox(0,0){\strut{}-12}}%
\put(550,3194){\rotatebox{90}{\makebox(0,0){\strut{}Occupation probabilty}}}%
\put(4039,275){\makebox(0,0){\strut{}Gate voltage $e\alpha V_\gate$ [$\mu$eV]}}%
\put(9428,3526){\makebox(0,0)[r]{\strut{}$N=4\tilde{N}$}}%
\put(9428,2976){\makebox(0,0)[r]{\strut{}$N=4\tilde{N}+1$}}%
\end{picture}%
\endgroup
}
\end{footnotesize}
\vskip 0.4cm
\imagecaption{\small Populations $P_0$ and $P_1$ around the resonance $\Delta=0\leftrightarrow\Delta=1$}\label{graph3}
\vskip 0.2cm
\end{minipage}
\\
\quad&\quad
\end{tabular}\quad\\
Of course, the occupation probability for an additional electron grows with the
gate voltage and it is worth to be stressed that the Eqns. (\ref{eq:pops}) do not depend on anything else than $V_\gate$.
The populations become $P_0=P_1=\frac{1}{2}$ at $e\alpha V_\gate=-\kBT\frac{\ln4}{2}$, where also the maximum of the resonance is located.\\
The components of the average spin in the $\Delta=1$ state follow simple dependences as well:
\begin{eqnarray}
\left\langle S^{(1)}_x \right\rangle&=&0\nonumber\\
\left\langle S^{(1)}_y \right\rangle&=&\frac{e}{4k_BT} V_\bias\, P_1\, \mathcal{P}\sin\frac{\theta}{2}\left(1+\mathcal{P}^2\frac{\mathfrak{P}^2(\mu_{\Delta=1},\mu_{\Delta=2})\cos^2\frac{\theta}{2}}{\pi^2f^{2}(-\mu_{\Delta=1})}\right)^{-1},\nonumber\\
\left\langle S^{(1)}_z \right\rangle&=&  \mathcal{P}\,\frac{\mathfrak{P}(\mu_{\Delta=1},\mu_{\Delta=2})\cos\frac{\theta}{2}}{-\pi f(-\mu_{\Delta=1})}\left\langle S^{(1)}_y \right\rangle\,.\label{eq:spin}\\\nonumber
\end{eqnarray}
In Fig. \ref{graph4} we show $\bra S^{(1)}_y\ket$ and  $\bra S^{(1)}_z\ket$ for $V_\gate=0$ and $V_\gate=\pm5.2\,\mu$eV.\\
\begin{center}
\vskip 0.1cm
\begin{minipage}{0.5\textwidth}
\begin{footnotesize}
\textsf{
\hspace{-4cm}
\begin{picture}(0,0)%
\includegraphics{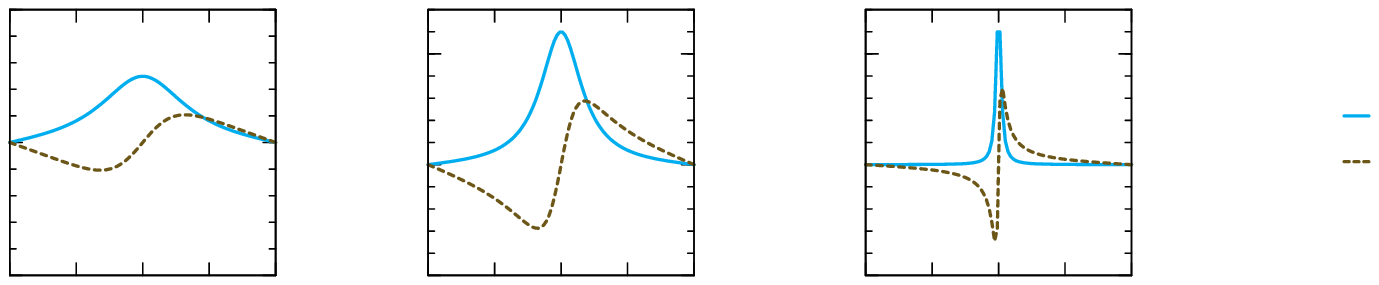}%
\end{picture}%
\begingroup
\setlength{\unitlength}{0.0200bp}%
\begin{picture}(11880,7128)(0,0)%
\put(1925,1650){\makebox(0,0)[r]{\strut{}-0.05}}%
\put(1925,3564){\makebox(0,0)[r]{\strut{} 0}}%
\put(1925,5478){\makebox(0,0)[r]{\strut{} 0.05}}%
\put(6028,1100){\makebox(0,0){\strut{}$\mathsf{2\pi}$}}%
\put(5071,1100){\makebox(0,0){\strut{}$\mathsf{\frac{3\pi}{2}}$}}%
\put(4114,1100){\makebox(0,0){\strut{}$\mathsf{\pi}$}}%
\put(3157,1100){\makebox(0,0){\strut{}$\mathsf{\frac{\pi}{2}}$}}%
\put(2200,1100){\makebox(0,0){\strut{} 0}}%
\put(0,3564){\rotatebox{90}{\makebox(0,0){\strut{}Average spin components}}}%
\put(600,3564){\rotatebox{90}{\makebox(0,0){\strut{}[$eV_\bias/(k_BT)$]}}}%
\put(4114,275){\makebox(0,0){\strut{}\quad}}%
\put(4114,6303){\makebox(0,0){\strut{}$e\alpha V_\gate=-5.2\,\mu$eV}}%
\put(7950,1650){\makebox(0,0)[r]{\strut{}-0.1}}%
\put(7950,3245){\makebox(0,0)[r]{\strut{} 0}}%
\put(7950,4840){\makebox(0,0)[r]{\strut{} 0.1}}%
\put(12053,1100){\makebox(0,0){\strut{}$\mathsf{2\pi}$}}%
\put(11096,1100){\makebox(0,0){\strut{}$\mathsf{\frac{3\pi}{2}}$}}%
\put(10139,1100){\makebox(0,0){\strut{}$\mathsf{\pi}$}}%
\put(9182,1100){\makebox(0,0){\strut{}$\mathsf{\frac{\pi}{2}}$}}%
\put(8225,1100){\makebox(0,0){\strut{} 0}}%
\put(10139,275){\makebox(0,0){\strut{}Magnetisation angle $\theta$ [rad]}}%
\put(10139,6303){\makebox(0,0){\strut{}$e\alpha V_\gate=0\,\mu$eV}}%
\put(14250,1650){\makebox(0,0)[r]{\strut{}-0.1}}%
\put(14250,3245){\makebox(0,0)[r]{\strut{} 0}}%
\put(14250,4840){\makebox(0,0)[r]{\strut{} 0.1}}%
\put(18353,1100){\makebox(0,0){\strut{}$\mathsf{2\pi}$}}%
\put(17396,1100){\makebox(0,0){\strut{}$\mathsf{\frac{3\pi}{2}}$}}%
\put(16439,1100){\makebox(0,0){\strut{}$\mathsf{\pi}$}}%
\put(15482,1100){\makebox(0,0){\strut{}$\mathsf{\frac{\pi}{2}}$}}%
\put(14525,1100){\makebox(0,0){\strut{} 0}}%
\put(16439,275){\makebox(0,0){\strut{}\quad}}%
\put(16439,6303){\makebox(0,0){\strut{}$e\alpha V_\gate=+5.2\,\mu$eV}}%
\put(21140,3947){\makebox(0,0)[r]{\strut{}$\langle S^{(1)}_{y}\rangle$}}%
\put(21140,3288){\makebox(0,0)[r]{\strut{}$\langle S^{(1)}_{z}\rangle$}}%
\end{picture}%
\endgroup
}
\end{footnotesize}
\end{minipage}
\vskip 0.4cm
\imagecaption{\small The peaks of $\langle S^{(1)}_y \rangle$ and $\langle S^{(1)}_z \rangle$ get more and more pronounced when raising the gate voltage.}\label{graph4}
\vskip 0.6cm
\end{center}
The total spin grows with the polarisation $\mathcal{P}$ and linearly with both the bias voltage $V_\bias$ and the occupation probability $P_1$.
Due to the choice for our coordinate system (remember Fig. \ref{kosys1}), the $x_\odot$-component of the average spin is zero. The fact that $\langle S^{(1)}_y \rangle$ peaks at $\theta=\pi$ is clear, because the $y_\odot$-components of the lead magnetisations are opposite and scaling with $\sin\frac{\theta_l}{2}$. It is the spin precession, which tilts the accumulated spin out of plane and therewith gives rise to its nonzero $z_\odot$-component; consequently $\langle S^{(1)}_z \rangle$ is proportional to $\mathfrak{P}$; its sign changes with the one of the cross product in (\ref{01eqns}), at $\theta=\pi$ .\\
\clearpage
What happens if the transparencies of the two contacts differ? Let us set
\[D_s=D_d=D\,,\ \mathcal{P}_s=\mathcal{P}_d=\mathcal{P}\,,\ \Phi_s=\varepsilon\Phi_d=\varepsilon\Phi\,.\]
Only for the parallel and the antiparallel current, a nice analytical expression can be given:
\vskip -2cm
\begin{equation}\eqalign{
I_{01}(\theta=0,V_\gate,\varepsilon)=\frac{2\varepsilon}{1+\varepsilon}\,I_{01}(\theta=0,V_\gate)\,,\\
I_{01}(\theta=\pi,V_\gate,\varepsilon)=\frac{2\varepsilon(1+\varepsilon)\,I_{01}(\theta=\pi,V_\gate)}{\left((1-\mathcal{P})+\varepsilon(1+\mathcal{P})\right)\left((1+\mathcal{P})+\varepsilon(1-\mathcal{P})\right)}\,.}\label{paap_asym}\end{equation}
Obviously, the ratio $\frac{I_{01}(\theta=\pi,\varepsilon)}{I_{01}(\theta=0,\varepsilon)}$ is again constant with $V_\gate$ and the positions of the peak maxima are still the same as in the symmetric case.\smallskip\\
Fig. \ref{5_01}a presents the numerical results for the normalised current at two gate voltages ($\pm5.2\mu$eV), for two strongly differing values of $R$ (for simplicity we assumed that the reflection scales with the transparency, i.e. $R_s/\Phi_s=R_d/\Phi_d=:R/\Phi$).\\We see that the four curves can hardly be distinguished from each other, and especially $I_{01}(0\dgr,\varepsilon)\geq I_{01}(\theta,\varepsilon)\ \ \forall\theta\in[0\dgr,360\dgr]\,$: there cannot appear any negative TMR. Fig. \ref{5_01}b shows the analytical peaks for $\theta=0\dgr$ and $\theta=180\dgr$, along with the numerical data, which additionally gives the peak for $\theta=140\dgr$. $I_{01}(0\dgr,\varepsilon)$ and $I_{01}(180\dgr,\varepsilon)$ will, as Eq. (\ref{paap_asym}) tells us, not differ for distinct $R$. Also for the non-collinear magnetisation, altering $R$ over four orders of magnitude obviously has no notable effect. No negative TMR can be found throughout the full range of the gate voltage, and in particular, the dependence of the non-collinear TMR on the gate voltage is even suppressed in comparison with the symmetric coupling.
\begin{center}
\vskip 0.4cm
\begin{minipage}{0.5\textwidth}
\begin{footnotesize}
\textsf{
\hspace{-3cm}
\begin{picture}(0,0)%
\includegraphics{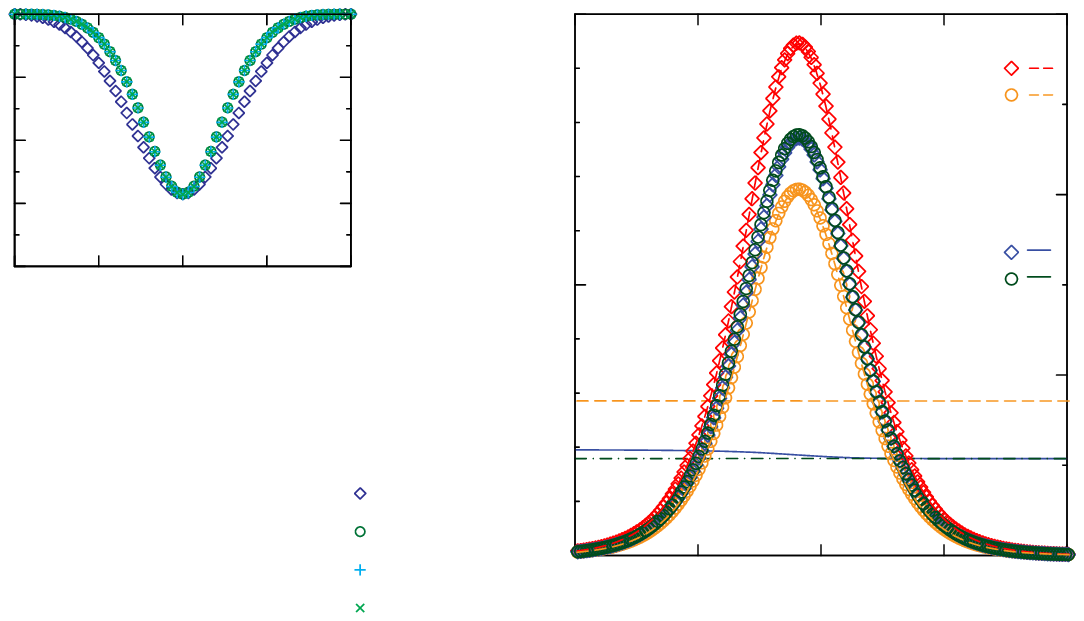}%
\end{picture}%
\begingroup
\setlength{\unitlength}{0.0200bp}%
\begin{picture}(16199,9719)(0,0)%
\put(9719,728){\makebox(0,0)[r]{\strut{} 0}}%
\put(9719,4625){\makebox(0,0)[r]{\strut{} 0.5}}%
\put(9719,8522){\makebox(0,0)[r]{\strut{} 1}}%
\put(17079,178){\makebox(0,0){\strut{}12}}%
\put(15308,178){\makebox(0,0){\strut{}6}}%
\put(13536,178){\makebox(0,0){\strut{}0}}%
\put(11765,178){\makebox(0,0){\strut{}-6}}%
\put(9994,178){\makebox(0,0){\strut{}-12}}%
\put(17354,728){\makebox(0,0)[l]{\strut{} 0}}%
\put(17354,3326){\makebox(0,0)[l]{\strut{} 0.2}}%
\put(17354,5924){\makebox(0,0)[l]{\strut{} 0.4}}%
\put(17354,8522){\makebox(0,0)[l]{\strut{} 0.6}}%
\put(8290,5725){\rotatebox{90}{\makebox(0,0){\strut{}TMR}}}%
\put(18453,4625){\rotatebox{90}{\makebox(0,0){\strut{}Current $I_{01}(V_\gate)$ [nA]}}}%
\put(13536,-372){\makebox(0,0){\strut{}\quad}}%
\put(16237,5124){\makebox(0,0)[r]{\strut{}\scriptsize{$10^{-2}\,\mathfrak{P}_0$}\ }}%
\put(16237,4740){\makebox(0,0)[r]{\strut{}\scriptsize{$10^{2}\,\mathfrak{P}_0$}\ }}%
\put(9694,5725){\rotatebox{90}{\makebox(0,0){\strut{}\quad}}}%
\put(17107,4625){\rotatebox{90}{\makebox(0,0){\strut{}\quad}}}%
\put(13565,-372){\makebox(0,0){\strut{}\quad}}%
\put(9694,5725){\rotatebox{90}{\makebox(0,0){\strut{}\quad}}}%
\put(17107,4625){\rotatebox{90}{\makebox(0,0){\strut{}\quad}}}%
\put(13565,-372){\makebox(0,0){\strut{}\quad}}%
\put(15841,7743){\makebox(0,0)[r]{\strut{}\quad}}%
\put(15841,7359){\makebox(0,0)[r]{\strut{}\quad}}%
\put(9694,5725){\rotatebox{90}{\makebox(0,0){\strut{}\quad}}}%
\put(17107,4625){\rotatebox{90}{\makebox(0,0){\strut{}\quad}}}%
\put(13565,-372){\makebox(0,0){\strut{}\quad}}%
\put(14515,6184){\makebox(0,0)[l]{\strut{}\scriptsize{$\mathsf{\theta=140\dgr\,,}$}}}%
\put(14557,5716){\makebox(0,0)[l]{\strut{}\scriptsize{$R/(\Phi\mathcal{P}D)=$}}}%
\put(15841,5093){\makebox(0,0)[r]{\strut{}\quad}}%
\put(15841,4709){\makebox(0,0)[r]{\strut{}\quad}}%
\put(9694,5725){\rotatebox{90}{\makebox(0,0){\strut{}\quad}}}%
\put(17107,4625){\rotatebox{90}{\makebox(0,0){\strut{}\quad}}}%
\put(13565,-372){\makebox(0,0){\strut{}Gate voltage $e \alpha V_\gate$ [$\mu$eV]}}%
\put(10873,7898){\makebox(0,0){\textbf{\textsf{(b)}}}}%
\put(16266,7743){\makebox(0,0)[r]{\strut{}\scriptsize{$\theta=$\quad\,0}$\dgr$\ }}%
\put(16266,7359){\makebox(0,0)[r]{\strut{}\scriptsize{180}$\dgr$\ }}%
\put(1650,4890){\makebox(0,0)[r]{\strut{}0.6}}%
\put(1650,5798){\makebox(0,0)[r]{\strut{}0.7}}%
\put(1650,6706){\makebox(0,0)[r]{\strut{}0.8}}%
\put(1650,7614){\makebox(0,0)[r]{\strut{}0.9}}%
\put(1650,8522){\makebox(0,0)[r]{\strut{}1}}%
\put(6767,4340){\makebox(0,0){\strut{}$2\pi$}}%
\put(5557,4340){\makebox(0,0){\strut{}$\frac{3\pi}{2}$}}%
\put(4346,4340){\makebox(0,0){\strut{}$\pi$}}%
\put(3136,4340){\makebox(0,0){\strut{}$\frac{\pi}{2}$}}%
\put(1925,4340){\makebox(0,0){\strut{}$0$}}%
\put(550,5881){\rotatebox{90}{\makebox(0,0){\strut{}Normalised current $I_{01}(\theta)/I_{01}(0)$}}}%
\put(4318,3515){\makebox(0,0){\strut{}Magnetisation angle $\theta$ [rad]}}%
\put(2603,5398){\makebox(0,0){\textbf{\textsf{(a)}}}}%
\put(1441,2348){\makebox(0,0)[l]{\strut{}\scriptsize{$e\alpha V_\gate=$}}}%
\put(4346,2348){\makebox(0,0)[l]{\strut{}\scriptsize{$R/(\Phi\mathcal{P}D)=$}}}%
\put(1731,1367){\makebox(0,0)[l]{\strut{}\scriptsize{$=-5.2\,\mu$eV}}}%
\put(1731,205){\makebox(0,0)[l]{\strut{}\scriptsize{$=+5.2\,\mu$eV}}}%
\put(6492,1641){\makebox(0,0)[r]{\strut{}\scriptsize{$10^{-2}\,\mathfrak{P}_0$}}}%
\put(6492,1091){\makebox(0,0)[r]{\strut{}\scriptsize{$10^2\,\mathfrak{P}_0$}}}%
\put(6492,541){\makebox(0,0)[r]{\strut{}\scriptsize{$10^{-2}\,\mathfrak{P}_0$}}}%
\put(6492,-9){\makebox(0,0)[r]{\strut{}\scriptsize{$10^{2}\,\mathfrak{P}_0$}}}%
\put(16440,4800){\colorbox{white}{\makebox(200.00,260.00)[cc]{ }}} 
\end{picture}%
\endgroup
}
\end{footnotesize}
\end{minipage}
\vskip 0.6cm
\imagecaption{\small (a) $I_{01}(\theta)/I_{01}(0)$ and (b) $I_{01}(V_\gate)$ together with the related TMR for $\Phi_s=0.3\,\Phi_d$. The normalised current is plotted at different gate voltages $V_\gate$ and hugely varying reflection parameters $R$ to show that neither $V_\gate$ nor $R$ influence its angular evolution much.}\label{5_01}\vskip 0.4cm
\end{center}

\subsection{Resonant tunnelling regime $\Delta=1\leftrightarrow\Delta=2$}
\begin{center}
\vskip -1.2cm
\begin{footnotesize}
\textsf{
\hspace{-0.5cm}
\begin{picture}(0,0)%
\includegraphics{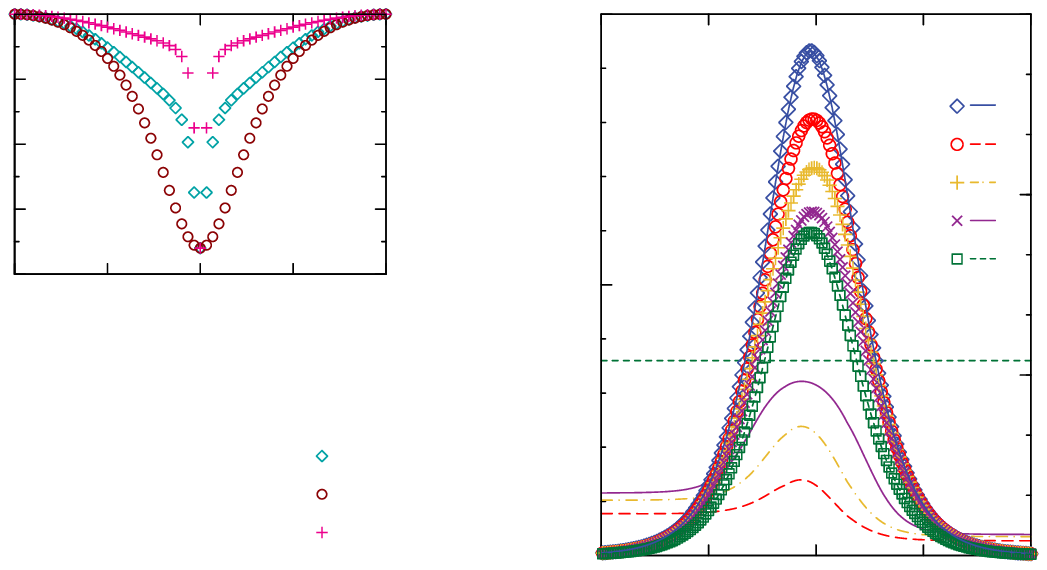}%
\end{picture}%
\begingroup
\setlength{\unitlength}{0.0200bp}%
\begin{picture}(18000,10800)(0,0)%
\put(10097,-351){\makebox(0,0)[r]{\strut{} 0}}%
\put(10097,3545){\makebox(0,0)[r]{\strut{} 0.5}}%
\put(10097,7442){\makebox(0,0)[r]{\strut{} 1}}%
\put(16557,-902){\makebox(0,0){\strut{}12}}%
\put(15011,-902){\makebox(0,0){\strut{}6}}%
\put(13465,-902){\makebox(0,0){\strut{}0}}%
\put(11918,-902){\makebox(0,0){\strut{}-6}}%
\put(10372,-902){\makebox(0,0){\strut{}-12}}%
\put(16832,-351){\makebox(0,0)[l]{\strut{} 0}}%
\put(16832,2246){\makebox(0,0)[l]{\strut{} 0.6}}%
\put(16832,4844){\makebox(0,0)[l]{\strut{} 1.2}}%
\put(16832,7442){\makebox(0,0)[l]{\strut{} 1.8}}%
\put(8668,4645){\rotatebox{90}{\makebox(0,0){\strut{}TMR}}}%
\put(17931,3545){\rotatebox{90}{\makebox(0,0){\strut{}Current $I_{12}(V_\gate)$ [nA]}}}%
\put(13464,-1452){\makebox(0,0){\strut{}Gate voltage $e\alpha V_\gate$ [$\mu$eV]}}%
\put(11114,6818){\makebox(0,0){\textbf{\textsf{(b)}}}}%
\put(15416,5571){\makebox(0,0)[r]{\strut{}\scriptsize{115}\ }}%
\put(15416,5021){\makebox(0,0)[r]{\strut{}\scriptsize{140}\ }}%
\put(15416,4471){\makebox(0,0)[r]{\strut{}\scriptsize{160}\ }}%
\put(15416,3921){\makebox(0,0)[r]{\strut{}\scriptsize{180}\ }}%
\put(10054,4645){\rotatebox{90}{\makebox(0,0){\strut{}\quad}}}%
\put(16567,3545){\rotatebox{90}{\makebox(0,0){\strut{}\quad}}}%
\put(13475,-1452){\makebox(0,0){\strut{}\quad}}%
\put(14341,6663){\makebox(0,0)[l]{\strut{}\scriptsize{$\theta=$}}}%
\put(15044,6117){\makebox(0,0)[r]{\strut{}\quad}}%
\put(15044,5567){\makebox(0,0)[r]{\strut{}\quad}}%
\put(15044,5017){\makebox(0,0)[r]{\strut{}\quad}}%
\put(15044,4467){\makebox(0,0)[r]{\strut{}\quad}}%
\put(15044,3917){\makebox(0,0)[r]{\strut{}\quad}}%
\put(10054,4645){\rotatebox{90}{\makebox(0,0){\strut{}\quad}}}%
\put(16567,3545){\rotatebox{90}{\makebox(0,0){\strut{}\quad}}}%
\put(13475,-1452){\makebox(0,0){\strut{}\quad}}%
\put(15415,6133){\makebox(0,0)[r]{\strut{}\scriptsize{0}\ }}%
\put(10054,4645){\rotatebox{90}{\makebox(0,0){\strut{}\quad}}}%
\put(16567,3545){\rotatebox{90}{\makebox(0,0){\strut{}\quad}}}%
\put(13475,-1452){\makebox(0,0){\strut{}\quad}}%
\put(1650,3699){\makebox(0,0)[r]{\strut{} 0.6}}%
\put(1650,4635){\makebox(0,0)[r]{\strut{} 0.7}}%
\put(1650,5570){\makebox(0,0)[r]{\strut{} 0.8}}%
\put(1650,6506){\makebox(0,0)[r]{\strut{} 0.9}}%
\put(1650,7442){\makebox(0,0)[r]{\strut{} 1}}%
\put(7272,3149){\makebox(0,0){\strut{}$2\pi$}}%
\put(5935,3149){\makebox(0,0){\strut{}$\frac{3\pi}{2}$}}%
\put(4598,3149){\makebox(0,0){\strut{}$\pi$}}%
\put(3262,3149){\makebox(0,0){\strut{}$\frac{\pi}{2}$}}%
\put(1925,3149){\makebox(0,0){\strut{}$0$}}%
\put(550,4745){\rotatebox{90}{\makebox(0,0){\strut{}Normalised current $I_{12}(\theta)/I_{12}(0)$}}}%
\put(4598,2435){\makebox(0,0){\strut{}Magnetisation angle $\theta$ [rad]}}%
\put(1925,1116){\makebox(0,0)[l]{\strut{}\scriptsize{$e\alpha V_\gate=$}}}%
\put(2567,4148){\makebox(0,0){\textbf{\textsf{(a)}}}}%
\put(5928,1079){\makebox(0,0)[r]{\strut{}\scriptsize{$-5.2\,\mu$eV}}}%
\put(5928,529){\makebox(0,0)[r]{\strut{}\scriptsize{$0\,\mu$eV}}}%
\put(5928,-21){\makebox(0,0)[r]{\strut{}\scriptsize{$+5.2\,\mu$eV}}}%
\end{picture}%
\endgroup
}
\end{footnotesize}
\vskip 2cm
\imagecaption{\small (a) $I_{12}(\theta)/I_{12}(0)$ and (b) $I_{12}(V_\gate)$ together with the related TMR around the resonance $\Delta=1\leftrightarrow\Delta=2$. Notice that in contrast to the $\Delta=0\leftrightarrow\Delta=1$ transition, the TMR exhibits a maximum in the vicinity of the conductance maximum.}\label{8_12}
\vskip 0.6cm
\end{center}
Because of the large degeneracy of the involved states, the $\Delta=1\leftrightarrow\Delta=2$ is not even for the symmetric case accessible analytically if $\theta$ is arbitrary. We therefore show numerical data for $\theta\not\in\{0,\pi\}$. The regime $\Delta=2\leftrightarrow\Delta=3$ has again a mirror-symmetry to the one we treat here and therefore omitted. For the P and the AP current, an analytical solution can be provided:

\begin{equation}\eqalign{
I_{12}(\theta=0,V_\gate)=\frac{3\pi^2e^2}{hk_BT}\,\Phi D\,\frac{f(\mu_{\Delta=2})f(-\mu_{\Delta=2})}{2+f(\mu_{\Delta=2})}\,,\\
I_{12}(\theta=\pi,V_\gate)=\left(1-\mathcal{P}^2\right)I_{12}(\theta=0,V_\gate)\,,}\label{I12}
\end{equation}
in the symmetrical case.\smallskip\\The TMR at $\theta=\pi$ is consequently constant, at the same value as in the $\Delta=0\leftrightarrow\Delta=1$\,/\,$\Delta=3\leftrightarrow\Delta=0$ regimes. The curves for $\theta\neq\pi$, however, qualitatively differ, due to degeneracy in the $\tilde{r}$ bands, which allows a nonzero average spin in the $\Delta=2$ state. Below the resonance, there is mainly one single excess electron inside the SWCNT, with equilibrated spin. During charge transfer, an additional, spin-polarised electron is entering, but it can be the first electron (the one with relaxed spin) which leaves the tube. That is why polarised non-collinear transport is easier, and hence the TMR much smaller, than below the $\Delta=0\leftrightarrow\Delta=1$ peak. Approaching the resonance, however, the time intervals between the described process and the next charge transfer shortens so that the $\Delta=1$ state does not persist long enough to allow a sufficient equilibration of the excess electron's spin. Therefore the TMR increases. Beyond the resonance, the $\Delta=2$ is stable long enough let the total spin of the two electrons equilibrate. Both excess electrons can now be involved in polarised transport and so the TMR drops to a value lower than below the resonance. Still it stays nonzero, as an outtunnelling particle always leaves a single unequilibrated spin inside the SWCNT.\smallskip\\
Again, the current peaks are located off-resonance, at 
\begin{equation}e\alpha V_\gate=-\kBT\frac{\ln1.5}{2}\approx-0.35\,\mu{\rm{eV}}\label{eq:peakmax12}\end{equation} 
for $\theta\in\{0,\pi\}$, as the derivative of (\ref{I12}) reveals. For angles in between, the peak maximum again moves slightly towards $e\alpha V_\gate=0$.\smallskip\\
The normalised current $I_{12}(\theta)$ in Fig. \ref{8_12}(a) looks rather familiar at $e\alpha V_\gate=0$, but nevertheless does not follow the analytical law we found for $I_{01}(\theta)$. Off-resonance, $I_{12}(\theta)$ obviously no longer exhibits any handsome dependence and as explained, the polarised non-collinear transport is hindered at the resonance so that the width of the dip is for $e\alpha V_\gate=0\,\mu$eV larger than for both $e\alpha V_\gate=\pm5.2\,\mu$eV.\\For an asymmetric setup, one finds for varying $R$ as less deviations as in the $\Delta=0\leftrightarrow\Delta=1$ regime. Again, the current in the nearly antiparallel setup with $\theta=170\dgr$ hardly differs from the one for an exact AP configuration with $\theta=180\dgr$ and the gate voltage dependence of the TMR is suppressed. Still, $I_{12}(0\dgr,\varepsilon)>I_{12}(170\dgr,\varepsilon)$ is guaranteed; the equations for the P and AP configurations can again be solved analytically and we recover our previous finding:
\begin{equation}
\frac{I_{\Delta\Delta+1}(\theta=\pi)}{I_{\Delta\Delta+1}(\theta=0)}=\frac{(1-\mathcal{P}^2)(1+\varepsilon)^2}{\left((1-\mathcal{P})+\varepsilon(1+\mathcal{P})\right)\left((1+\mathcal{P})+\varepsilon(1-\mathcal{P})\right)}\,.\label{PAAP_curr}\end{equation}
So the ratio of P to AP current is independent of the gate voltage and constant even for a possible band offset.

\section{Nonlinear transport}\label{nonlintrans}
Being interested in the nonlinear transport behaviour of the ferromagnetically contacted SWCNT quantum dot, one has to take into account the various non-LEG states of the system, together with all the arising coherences. This makes an analytical calculation impossible, but the numerics delivers the relevant data respecting the influence of both real and virtual transitions with the energy $3\epsilon_0$ of up to three (bosonic or fermionic) neutral excitations.\\An inclusion of higher excited states would multiply the computational cost, but anyway, their contribution for bias voltages below the second excitation is rather small: $k_BT\ll\epsilon_0$, so that thermal excitation is inhibited and the probability of bias voltage driven multiple subsequent excitations naturally decreases with the amount of required subsequent excitations \cite{LEO}.\\Again, we can confine ourselves to distinguishing between the two regimes $\Delta=0\leftrightarrow\Delta=1$ and $\Delta=1\leftrightarrow\Delta=2$, because the symmetry arguments still hold for all states.

\subsection{Resonant tunnelling regime $\Delta=0\leftrightarrow\Delta=1$}\label{VIA}

\begin{center}
\vskip -0.8cm
\begin{minipage}{0.5\textwidth}
\begin{footnotesize}
\textsf{
\hspace{-1.2cm}
\begin{picture}(0,0)%
\includegraphics{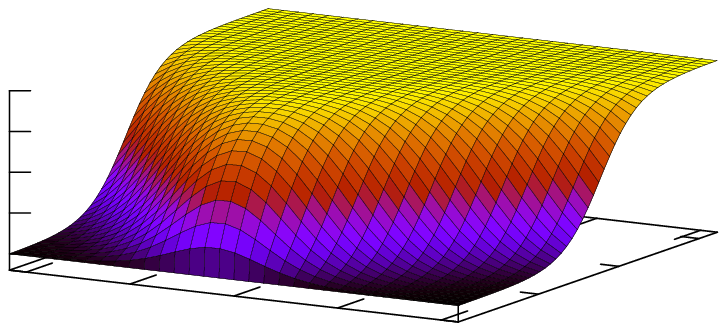}%
\end{picture}%
\begingroup
\setlength{\unitlength}{0.0200bp}%
\begin{picture}(14759,8855)(0,0)%
\put(8159,2181){\makebox(0,0)[r]{\strut{}12}}%
\put(6665,2354){\makebox(0,0)[r]{\strut{}6}}%
\put(5170,2526){\makebox(0,0)[r]{\strut{}0}}%
\put(3675,2699){\makebox(0,0)[r]{\strut{}-6}}%
\put(2180,2871){\makebox(0,0)[r]{\strut{}-12}}%
\put(12448,3537){\makebox(0,0){\strut{}45}}%
\put(11293,3137){\makebox(0,0){\strut{}30}}%
\put(10138,2736){\makebox(0,0){\strut{}15}}%
\put(8982,2336){\makebox(0,0){\strut{}0}}%
\put(1683,3372){\makebox(0,0)[r]{\strut{} 0}}%
\put(1683,3960){\makebox(0,0)[r]{\strut{} 1}}%
\put(1683,4547){\makebox(0,0)[r]{\strut{} 2}}%
\put(1683,5133){\makebox(0,0)[r]{\strut{} 3}}%
\put(1683,5721){\makebox(0,0)[r]{\strut{} 4}}%
\put(4306,1891){\makebox(0,0){\strut{}Gate voltage $eV_\gate$ [$\mu$eV]}}%
\put(12500,2081){\makebox(0,0){\strut{}Bias voltage $eV_\bias$ [$\mu$eV]}}%
\put(2282,6367){\makebox(0,0){\strut{}Current $I_{01}$ [mA]}}%
\end{picture}%
\endgroup
}
\end{footnotesize}
\end{minipage}
\vskip -0.8cm
\imagecaption{\small Nonlinear current $I_{01}(\theta=0)$ below the first excitation. For the computation of this plot, transitions with a maximum energy $\epsilon_0$ were taken into account.}\label{nonlin1_01}
\vskip 0.4cm
\end{center}
Fig. \ref{nonlin1_01} shows the current for bias voltages $eV_l\ll\epsilon_0\,$, i.e. far below the first excitation. We can nicely see how the Coulomb blockade diamonds for $\Delta=0$ and $\Delta=1$ emerge; the higher the bias, the wider gets the range of the gate voltage within which transport is allowed. This brings about the typical rhomb shapes.
\begin{center}
\begin{minipage}{0.5\textwidth}
\begin{footnotesize}
\textsf{
\hspace{-2.2cm}
\begin{picture}(0,0)%
\includegraphics{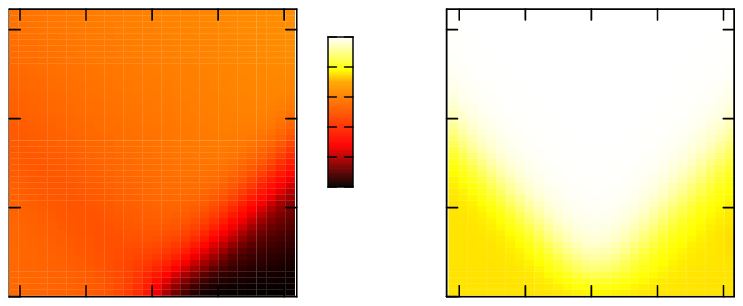}%
\end{picture}%
\begingroup
\setlength{\unitlength}{0.0200bp}%
\begin{picture}(12960,7776)(0,0)%
\put(9635,3672){\makebox(0,0)[l]{\strut{} 0}}%
\put(9635,4104){\makebox(0,0)[l]{\strut{} 0.1}}%
\put(9635,4536){\makebox(0,0)[l]{\strut{} 0.2}}%
\put(9635,4968){\makebox(0,0)[l]{\strut{} 0.3}}%
\put(9635,5400){\makebox(0,0)[l]{\strut{} 0.4}}%
\put(9635,5832){\makebox(0,0)[l]{\strut{} 0.5}}%
\put(6480,553){\makebox(0,0){\strut{}Gate voltage $eV_\gate$ [$\mu$eV]}}%
\put(2816,4163){\rotatebox{90}{\makebox(0,0){\strut{}Bias voltage $eV_\bias$ [$\mu$eV]}}}%
\put(4575,6895){\makebox(0,0)[l]{\strut{}\textbf{(a)}\ \,TMR\,$(\theta=140\dgr)$}}%
\put(8369,1378){\makebox(0,0){\strut{}12}}%
\put(7417,1378){\makebox(0,0){\strut{}6}}%
\put(6467,1378){\makebox(0,0){\strut{}0}}%
\put(5515,1378){\makebox(0,0){\strut{}-6}}%
\put(4564,1378){\makebox(0,0){\strut{}-12}}%
\put(4053,5938){\makebox(0,0)[r]{\strut{}45}}%
\put(4053,4656){\makebox(0,0)[r]{\strut{}30}}%
\put(4053,3375){\makebox(0,0)[r]{\strut{}15}}%
\put(4053,2093){\makebox(0,0)[r]{\strut{}0}}%
\put(12779,553){\makebox(0,0){\strut{}Gate voltage $eV_\gate$ [$\mu$eV]}}%
\put(10874,6895){\makebox(0,0)[l]{\strut{}\textbf{(b)}\ \,TMR\,$(\theta=180\dgr)$}}%
\put(14695,1378){\makebox(0,0){\strut{}12}}%
\put(13744,1378){\makebox(0,0){\strut{}6}}%
\put(12792,1378){\makebox(0,0){\strut{}0}}%
\put(11842,1378){\makebox(0,0){\strut{}-6}}%
\put(10890,1378){\makebox(0,0){\strut{}-12}}%
\put(10352,5938){\makebox(0,0)[r]{\strut{}}}%
\put(10352,4656){\makebox(0,0)[r]{\strut{}}}%
\put(10352,3375){\makebox(0,0)[r]{\strut{}}}%
\put(10352,2093){\makebox(0,0)[r]{\strut{}}}%
\end{picture}%
\endgroup
}
\end{footnotesize}
\end{minipage}
\vskip 0.3cm
\imagecaption{\small TMR vs. gate and bias voltage for $\theta=140\dgr$ and $\theta=180\dgr$ around the $\Delta=0\leftrightarrow\Delta=1$ resonance. As we have found it in the linear bias regime, Fig. \ref{graph1}, the non-collinear TMR ($\theta=140\dgr$) drops from a constant value to zero after the resonance, following the current. For $\theta=180\dgr$, the TMR is practically symmetric with respect to the resonant tunnelling regime, where it acquires an increased value.}\label{nonlin1_01_TMR}
\vskip 0.6cm
\end{center}
Figs. \ref{nonlin1_01_TMR}a and \ref{nonlin1_01_TMR}b are colourmap TMR plots for the polarisation angles $\theta=140\dgr$ and $\theta=180\dgr$ in the same regime Fig. \ref{nonlin1_01} covers. In Fig. \ref{nonlin1_01_TMR}a we regain what we already knew from Fig. \ref{graph1} for the non-collinear magnetisations, namely that the TMR drops from a constant value to zero after the resonant tunnelling regime. Fig. \ref{nonlin1_01_TMR}b, however, shows that in the antiparallel configuration, for higher bias voltages the TMR is not everywhere constant with the gate voltage but increases outside the Coulomb diamonds. Nevertheless, just as for the low bias, TMR\,$(\theta=140\dgr)<\,$TMR\,$(\theta=180\dgr)$.
Moreover, both cases lack any negative TMR and actually the TMR gets widely constant for high bias voltages, as the nonzero currents in between the Coulomb diamonds acquire a constant value. That is why for $eV_l>\epsilon_0$, we only need to plot the current at one special gate voltage, e.g. $V_\gate=0$ (fig. \ref{nonlin2_01}).
\begin{center}
\begin{minipage}{0.5\textwidth}
\begin{footnotesize}
\textsf{
\hspace{-0.6cm}
\begin{picture}(0,0)%
\includegraphics{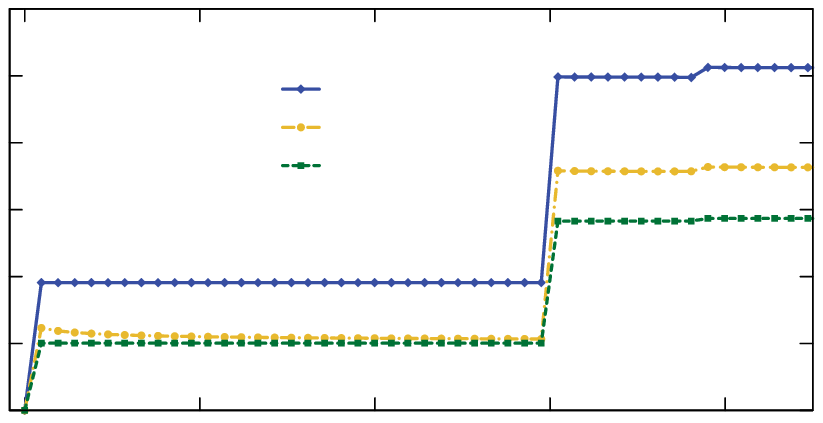}%
\end{picture}%
\begingroup
\setlength{\unitlength}{0.0200bp}%
\begin{picture}(14039,8423)(0,0)%
\put(1375,1650){\makebox(0,0)[r]{\strut{} 0}}%
\put(1375,2614){\makebox(0,0)[r]{\strut{} 2}}%
\put(1375,3577){\makebox(0,0)[r]{\strut{} 4}}%
\put(1375,4541){\makebox(0,0)[r]{\strut{} 6}}%
\put(1375,5505){\makebox(0,0)[r]{\strut{} 8}}%
\put(1375,6468){\makebox(0,0)[r]{\strut{} 10}}%
\put(1375,7432){\makebox(0,0)[r]{\strut{} 12}}%
\put(11954,1100){\makebox(0,0){\strut{}8}}%
\put(9433,1100){\makebox(0,0){\strut{}6}}%
\put(6911,1100){\makebox(0,0){\strut{}4}}%
\put(4390,1100){\makebox(0,0){\strut{}2}}%
\put(1868,1100){\makebox(0,0){\strut{}0}}%
\put(550,4541){\rotatebox{90}{\makebox(0,0){\strut{}Current $I_{01}$ [mA]}}}%
\put(7432,275){\makebox(0,0){\strut{}Bias voltage $eV_\bias$ [meV]}}%
\put(3732,6276){\makebox(0,0)[l]{\strut{}$\theta=$}}%
\put(5307,6276){\makebox(0,0)[r]{\strut{}$0\dgr$}}%
\put(5307,5726){\makebox(0,0)[r]{\strut{}$140\dgr$}}%
\put(5307,5176){\makebox(0,0)[r]{\strut{}$180\dgr$}}%
\end{picture}%
\endgroup
}
\end{footnotesize}
\end{minipage}
\vskip 0.2cm
\imagecaption{\small Nonlinear current $I_{01}(\theta)$ for $\theta\in\{0\dgr,140\dgr,180\dgr\}$ and bias voltages exceeding the first neutral excitation. The numerics respected a maximum transition energy of $3\epsilon_0$. For a non-collinear lead magnetisation, a negative differential conductance appears below the first excitation.}\label{nonlin2_01}\vskip 0.2cm\end{center}
Within the range of the bias voltage fig. \ref{nonlin2_01} shows, the curves for $\theta\in\{0\dgr,180\dgr\}$ reach three different plateaus: below the first excitation, there is a constant current outside the Coulomb blockade diamonds. As soon as the bias voltage is high enough to excite a neutral mode, more states can contribute to transport and the current jumps to a higher value. It is important that from here on, successively by picking up the energy the bias voltage provides, multiple excited states can be generated. The probability to fall back to a less excited state, however, is always larger than the one for creating another additional excitation and therefore it is certainly valid to at an energy of $3\epsilon_0$. Indeed, another slight enlargement takes place at that value of the bias, which provides enough energy to add an additional electron to a state $|N$+$1,1\ket$;  this is another allowed transition above the first neutral excitation.\\The current at non-collinear magnetisations exhibits a further feature: we find a \textit{negative differential conductance} (NDC) \textit{below} the first bosonic excitation. This behaviour becomes more evident for high polarisation, but can still be seen for our passably realistic case of $\mathcal{P}=0.6$. The explanation for the occurrence of the NDC is the decaying influence of the virtual transitions with growing bias voltage (due to the energy arguments appearing under the principal part integral, see Eq. (\ref{eq3:rates})). We learned that the principal part terms narrow the $I_{01}(\theta)$ curve  (fig. \ref{graph2}), which means that the heavier their influence, the close comes the non-collinear current for a certain magnetisation angle $\theta$ to the maximum current $I_{01}(0\dgr)$. Thus it is clear that $I_{01}(140\dgr)$ approaches the minimal current $I_{01}(180\dgr)$ for higher bias voltages where the principal part is more and more suppressed. Notice that here $R$ has an influence, but as it does not depend on any external voltage, it tends to wipe out the effect. 

\subsection{Resonant tunnelling regime $\Delta=1\leftrightarrow\Delta=2$}

\begin{center}
\vskip -0.4cm
\begin{minipage}{0.5\textwidth}
\begin{footnotesize}
\textsf{
\hspace{-1.2cm}
\begin{picture}(0,0)%
\includegraphics{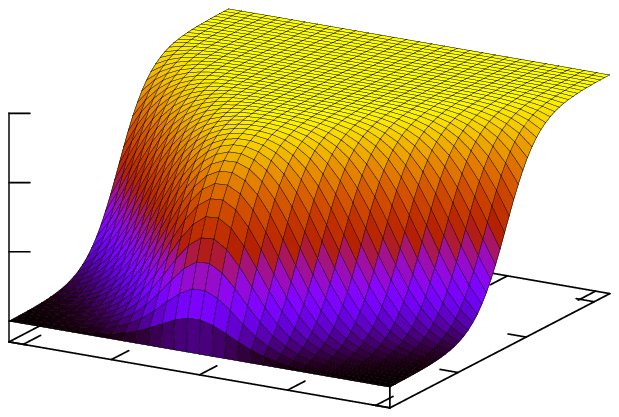}%
\end{picture}%
\begingroup
\setlength{\unitlength}{0.0200bp}%
\begin{picture}(14759,8855)(0,0)%
\put(8104,1556){\makebox(0,0){\strut{}12}}%
\put(6836,1776){\makebox(0,0){\strut{}6}}%
\put(5567,1996){\makebox(0,0){\strut{}0}}%
\put(4298,2215){\makebox(0,0){\strut{}-6}}%
\put(3029,2435){\makebox(0,0){\strut{}-12}}%
\put(11718,3205){\makebox(0,0){\strut{}45}}%
\put(10737,2695){\makebox(0,0){\strut{}30}}%
\put(9756,2186){\makebox(0,0){\strut{}15}}%
\put(8776,1676){\makebox(0,0){\strut{}0}}%
\put(2453,3008){\makebox(0,0)[r]{\strut{} 0}}%
\put(2453,4006){\makebox(0,0)[r]{\strut{} 2}}%
\put(2453,5002){\makebox(0,0)[r]{\strut{} 4}}%
\put(2453,6000){\makebox(0,0)[r]{\strut{} 6}}%
\put(4729,1272){\makebox(0,0){\strut{}Gate voltage $eV_\gate$ [$\mu$eV]}}%
\put(11768,1574){\makebox(0,0){\strut{}Bias voltage $eV_\bias$ [$\mu$eV]}}%
\put(3052,6823){\makebox(0,0){\strut{}Current $I_{12}$ [mA]}}%
\end{picture}%
\endgroup
}
\end{footnotesize}
\end{minipage}
\vskip -0.4cm
\imagecaption{\small Nonlinear current  $I_{12}(\theta=0)$ below the first excitation, transitions with a maximum energy $\epsilon_0$ taken into account.}\label{nonlin1_12}
\vskip 0.4cm
\end{center}
At the $\Delta=1\leftrightarrow\Delta=2$ resonance, most statements from the previous subsection hold true as well. Due to the higher degeneracy of the states involved in transport, the plateau the current reaches is $50\%$ higher than for the $\Delta=0\leftrightarrow\Delta=1$ resonance \cite{LEO}.
\begin{center}
\vskip -0.2cm
\begin{minipage}{0.5\textwidth}
\begin{footnotesize}
\textsf{
\hspace{-2.2cm}
\begin{picture}(0,0)%
\includegraphics{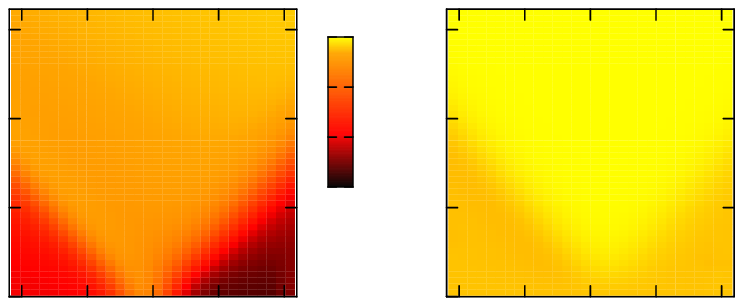}%
\end{picture}%
\begingroup
\setlength{\unitlength}{0.0200bp}%
\begin{picture}(12960,7776)(0,0)%
\put(9635,3672){\makebox(0,0)[l]{\strut{} 0}}%
\put(9635,4391){\makebox(0,0)[l]{\strut{} 0.1}}%
\put(9635,5111){\makebox(0,0)[l]{\strut{} 0.2}}%
\put(9635,5832){\makebox(0,0)[l]{\strut{} 0.3}}%
\put(6480,553){\makebox(0,0){\strut{}Gate voltage $eV_\gate$ [$\mu$eV]}}%
\put(2816,4163){\rotatebox{90}{\makebox(0,0){\strut{}Bias voltage $eV_\bias$ [$\mu$eV]}}}%
\put(4575,6895){\makebox(0,0)[l]{\strut{}\textbf{(a)}\ \,TMR\,$(\theta=140\dgr)$}}%
\put(8370,1378){\makebox(0,0){\strut{}12}}%
\put(7425,1378){\makebox(0,0){\strut{}6}}%
\put(6480,1378){\makebox(0,0){\strut{}0}}%
\put(5535,1378){\makebox(0,0){\strut{}-6}}%
\put(4590,1378){\makebox(0,0){\strut{}-12}}%
\put(4053,5938){\makebox(0,0)[r]{\strut{}45}}%
\put(4053,4656){\makebox(0,0)[r]{\strut{}30}}%
\put(4053,3375){\makebox(0,0)[r]{\strut{}15}}%
\put(4053,2093){\makebox(0,0)[r]{\strut{}0}}%
\put(12779,553){\makebox(0,0){\strut{}Gate voltage $eV_\gate$ [$\mu$eV]}}%
\put(10874,6895){\makebox(0,0)[l]{\strut{}\textbf{(b)}\ \,TMR\,$(\theta=180\dgr)$}}%
\put(14669,1378){\makebox(0,0){\strut{}12}}%
\put(13724,1378){\makebox(0,0){\strut{}6}}%
\put(12779,1378){\makebox(0,0){\strut{}0}}%
\put(11834,1378){\makebox(0,0){\strut{}-6}}%
\put(10889,1378){\makebox(0,0){\strut{}-12}}%
\put(10352,5938){\makebox(0,0)[r]{\strut{}}}%
\put(10352,4656){\makebox(0,0)[r]{\strut{}}}%
\put(10352,3375){\makebox(0,0)[r]{\strut{}}}%
\put(10352,2093){\makebox(0,0)[r]{\strut{}}}%
\end{picture}%
\endgroup
}
\end{footnotesize}
\end{minipage}
\vskip 0.4cm
\imagecaption{\small TMR vs. gate and bias voltage around the $\Delta=1\leftrightarrow\Delta=2$ resonance. Again, the non-collinear TMR for the polarisation angles $\theta=140\dgr$ is similar to the one in the linear bias regime: before dropping to zero after the resonance, it shows a maximum near the conductance maximum. Just as the latter one, the TMR maximum widens with the bias voltage. The collinear TMR is qualitatively equivalent to the $\Delta=0\leftrightarrow\Delta=1$ case of Fig. \ref{nonlin1_01_TMR}, but the increment of the TMR is for $\Delta=1\leftrightarrow\Delta=2$ considerably less, namely about $10\%$, compared to some $40\%$ for Fig. \ref{nonlin1_01_TMR}}\label{nonlin1_12_TMR}
\vskip 0.6cm
\end{center}
For the colourmap TMR plots of Fig. \ref{nonlin1_12_TMR}, the same scale as for Fig. \ref{nonlin1_01_TMR} was applied. Again, at $\theta=140\dgr$ we find an evolution of the TMR we expected from the linear regime, Fig. \ref{8_12}, and for $\theta=180\dgr$ we obtain a qualitatively similar picture as Fig. \ref{nonlin1_01_TMR}b for $\Delta=0\leftrightarrow\Delta=1$. Figs. \ref{nonlin1_01_TMR}b and \ref{nonlin1_12_TMR}b, however, differ in quantity: the maximum value ($\sim0.39$) of the TMR at $\Delta=1\leftrightarrow\Delta=2$ is close to the value $0.36$ for the linear bias, while at $\Delta=0\leftrightarrow\Delta=1$ it still rises with the bias voltage from 0.36 to about 0.5.

For the higher bias voltage regime, Fig. \ref{nonlin2_12}, we can observe the same facts described in section \ref{VIA}. It is worth mentioning that beyond the first bosonic excitation, the differences in the heights of the currents at the two distinct resonances get much smaller, because the number of involved states multiplies in both regimes.
\begin{center}
\begin{minipage}{0.5\textwidth}
\begin{footnotesize}
\textsf{
\hspace{-0.6cm}
\begin{picture}(0,0)%
\includegraphics{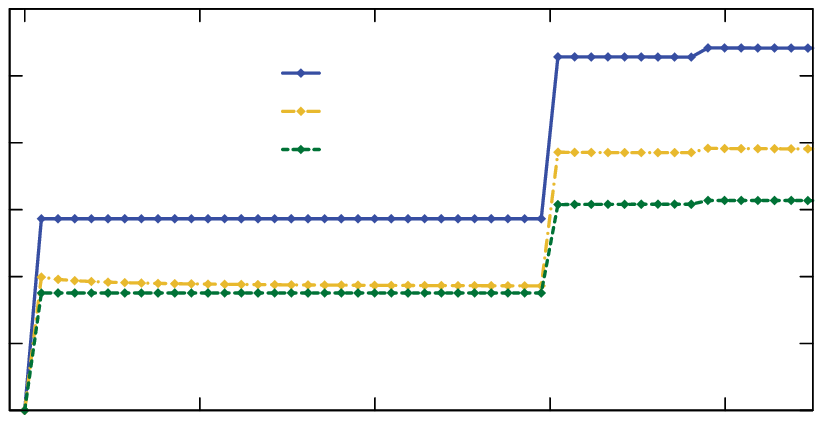}%
\end{picture}%
\begingroup
\setlength{\unitlength}{0.0200bp}%
\begin{picture}(14039,8423)(0,0)%
\put(1375,1650){\makebox(0,0)[r]{\strut{} 0}}%
\put(1375,2614){\makebox(0,0)[r]{\strut{} 2}}%
\put(1375,3577){\makebox(0,0)[r]{\strut{} 4}}%
\put(1375,4541){\makebox(0,0)[r]{\strut{} 6}}%
\put(1375,5505){\makebox(0,0)[r]{\strut{} 8}}%
\put(1375,6468){\makebox(0,0)[r]{\strut{} 10}}%
\put(1375,7432){\makebox(0,0)[r]{\strut{} 12}}%
\put(11954,1100){\makebox(0,0){\strut{}8}}%
\put(9433,1100){\makebox(0,0){\strut{}6}}%
\put(6911,1100){\makebox(0,0){\strut{}4}}%
\put(4390,1100){\makebox(0,0){\strut{}2}}%
\put(1868,1100){\makebox(0,0){\strut{}0}}%
\put(550,4541){\rotatebox{90}{\makebox(0,0){\strut{}Current $I_{12}$ [mA]}}}%
\put(7432,275){\makebox(0,0){\strut{}Bias voltage $eV_\bias$ [meV]}}%
\put(3732,6507){\makebox(0,0)[l]{\strut{}$\theta=$}}%
\put(5307,6507){\makebox(0,0)[r]{\strut{}$0\dgr$}}%
\put(5307,5957){\makebox(0,0)[r]{\strut{}$140\dgr$}}%
\put(5307,5407){\makebox(0,0)[r]{\strut{}$180\dgr$}}%
\end{picture}%
\endgroup
}
\end{footnotesize}
\end{minipage}
\vskip 0.1cm
\imagecaption{\small Nonlinear current $I_{12}(\theta)$ for $\theta\in\{0\dgr,140\dgr,180\dgr\}$ and bias voltages exceeding the first bosonic excitation. The numerics respected a maximum transition energy of $3\epsilon_0$.}\label{nonlin2_12}\vskip 0.1cm
\end{center}

\section{Main results}

Within the framework of our master equation approach, we could deduce equations for spin-dependent transport across carbon nanotube quantum dots. In the tunnelling regimes belonging to tube fillings $4N\leftrightarrow4N$+$1$ and $4N$+$3\leftrightarrow 4(N$+$1)$, the system behaves equivalent to a single level quantum dot. The resonances $4N$+$1\leftrightarrow4N$+$2$ and $4N$+$2\leftrightarrow 4N$+$3$ show similar characteristics; however, due to the larger degeneracy, the currents are increased and the fact that the dot is always populated by at least one electron brings about a more complex evolution of the TMR. In our perturbation approach (low transmission, which involes also only a small spin-splitting produced by boundary reflections), neither for symmetric nor for antisymmetric coupling to the leads, a negative TMR can occur.
\clearpage
\section{Conclusion}\label{conclusion}
We have investigated spin-dependent transport in SWCNT quantum dots. A spin-dependent equation for the reduced density matrix of a SWCNT weakly contacted to ferromagnetic leads of arbitrary magnetisations was presented. We demonstrated that the SWCNT behaves as a spin-valve single electron transistor and showed analytical and numerical results for the current flow.\smallskip\\
Because of the fourfold periodicity for the electron number $N$, it is sufficient to discriminate between tube fillings with different values of $\Delta=N\mod4$ and, due to mirror-symmetries in the SWCNT eigenstates, we could even restrict our examinations to the tunnelling regimes $\Delta=0\leftrightarrow\Delta=1$ and $\Delta=1\leftrightarrow\Delta=2$.\\The analytical analysis in the case of symmetric coupling to the leads resulted for the $\Delta=0\leftrightarrow\Delta=1$ resonance in an equivalent formula for the angular dependence of $I_{01}(\theta)$ as \cite{KOE} and \cite{WOU} obtained, for a single level quantum dot and a metallic island respectively. The total current, due to the existence of the degenerate left and a right mover bands, is twice as large as for a single level quantum dot. The maxima of the current peaks lie slightly off-resonance; the positions for a parallel and an antiparallel magnetisation are identical, but on the way from $\theta=0\dgr$ to $\theta=180\dgr$ the maximum moves a bit, following a curve bent towards the resonance gate voltage (in the $I_{01}(V_\gate)$ diagram). The TMR for $\theta\neq180\dgr$ changes around the resonance: it smoothly drops from a constant value, for gate voltages below the resonance, to zero.\\ We additionally gave the average spin on our quantum dot SWCNT in the $\Delta=1$ state and the two occupation probabilities $P_{\Delta=0}$ and $P_{\Delta=1}$, where we find that the latter solely depend on $V_\gate$. For a nonzero band offset $0<\kBT\ll\delta$ we can apply these results, because it makes the SWCNT at all gate voltages equivalent to a single level quantum dot.\\For $\delta\ll\kBT(\ll\epsilon_0)$, the resonance regimes $\Delta=1\leftrightarrow\Delta=2$ is more complex. The TMR around the resonance not simply decays monotonously from one constant value to another, but shows a peak before decreasing. Nevertheless, all TMR curves are strictly positive and also for an asymmetric coupling to the leads, numerical results in the different regimes at $\theta=140\dgr$ reveal that a non-collinear contact polarisation alone cannot produce a negative TMR.\\Specifically, we could deduce the general law (\ref{PAAP_curr}) for the TMR in the case of collinear (P-AP) magnetisations. It shows that under strict lowest order perturbation treatment of both the tunnelling and the reflection parameters, the linear bias P-AP current is even for an arbitrarily asymmetric coupling independent of the gate voltage. In order to reproduce a negative TMR as observed by \cite{SAH}, a spin-dependent energy shift (which can be obtained from a non-perturbative treatment of $\hat{H}_R$) is necessary. Actually, exchange effects emerging from a distinguishability of inter- and intra-lattice interactions can be source of an intrinsic spin-dependent energy shift, as measurements shown in \cite{MIY} on unpolarised small-diameter tubes and recent theoretical investigations exhibit \cite{LEO3}. The feature, however, is only present at the resonances involving $\Delta=2$ and therefore not responsible for the required gate-voltage independent spin-splitting.\smallskip\\For the nonlinear bias voltage regime of the quantum dot SWCNT, the numerical data revealed that the TMR for P-AP configurations is no longer strictly constant, but rises inside the resonant tunnelling regimes, whereas at $\Delta=0\leftrightarrow\Delta=1$, the effect is much more pronounced than at $\Delta=1\leftrightarrow\Delta=2$. The non-collinear TMR at $\theta=140\dgr$ is similar to the linear bias TMR, but now it becomes obvious that the changes in value take place at the edges of the Coulomb diamonds.\\ Tracing the current at the resonances to bias voltages exceeding the energy of the first possible excitation, we find for both tunnelling regimes a qualitatively equivalent dependence. Besides for the large jump at the first excitation, another small one can be found when the bias voltage reaches a value that provides the energy to enable transitions to states with a second additional electron. A special feature of the non-collinear polarisation is that a negative differential conductance appears below the first resonance due to the decaying influence of virtual processes.\bigskip\\
\centerline{*\quad*\quad*}\\
\centerline{We thank Wouter Wetzels for many useful suggestions and discussions.}
\centerline{Support by the DFG under the program SFB 689 is acknowledged.}

\clearpage

\appendix
\section*{APPENDIX}{\bf Decomposition of the correlation functions}\label{App_A}
\setcounter{section}{1}\\

To rewrite the rates (\ref{eq1:rates}),
\begin{eqnarray*}
\fl\Gamma^{(\alpha)NN+1}_{l(nkk'j)}&:=&\frac{1}{\hbar^2}\sum_{\sigl}\!\int\!\!\rmd{}^3r\!\int\!\!\rmd{}^3r'\!\left(\!\hPsi_{\odot\sigl}(\vr)\!\right)_{nk}\!\left(\!\Psid_{\odot\sigl}(\vr\,')\!\right)_{k'j}\int_0^\infty\!\!\!\!\rmd{}t_2 \ \mathcal{F}_{l\sigl}(\vec{r},\vec{r}\,'\!\!,t_2) \ \rme^{\alpha\frac{\rmi}{\hbar}\left(E_j-E_k\right)t_2}\!,\end{eqnarray*}\begin{eqnarray*}
\fl\Gamma^{(\alpha)NN-1}_{l(nii'j)}&:=&\frac{1}{\hbar^2}\sum_{\sigl}\!\int\!\!\rmd{}^3r\!\int\!\!\rmd{}^3r'\!\left(\!\Psid_{\odot\sigl}(\vr)\!\right)_{ni}\!\left(\!\hPsi_{\odot\sigl}(\vr\,')\!\right)_{i'j}\int_0^\infty\!\!\!\!\rmd{}t_2 \ \mathcal{E}_{l\sigl}(\vec{r},\vec{r}\,'\!\!,t_2) \ \rme^{\alpha\frac{\rmi}{\hbar}\left(E_j-E_i\right)t_2}-\nonumber\\\fl&&-\ \alpha\,\frac{\rmi}{\hbar}\sum_{\sigl}\!\int\!\rmd{}^3r\ \Delta_l(\vr)\ \sgn(\sigl)\left(\!\Psid_{\odot\sigl}(\vr)\!\right)_{ni}\!\left(\!\hPsi_{\odot\sigl}(\vr)\!\right)_{i'j}\,,\end{eqnarray*}
we first have to determine the explicit form of the correlation functions (\ref{correl_func}):
\begin{eqnarray*}\fl\left\langle \hPsi_{l\sigma_l}(\vr)\hPsi_{l\sigma_l}^{\dagger}(\vr\,',-t_2)\right\rangle _{\mathsf{th}}&=&\!\!\int\!\rmd{}\epsilon\, D_{l\sigl}(E^{l\sigl}_{tot}|_{\epsilon})\sum\limits_{\vq|_{\epsilon}}\phi_{l\vq}(\vr)\phi^*_{l\vq}(\vr\,')\left\langle\can_{l\sigl\vq}\,\,\rme^{-\frac{\rmi}{\hbar} \hat{H}_l t_2}\ccr_{l\sigl\vq}\,\,\rme^{+\frac{\rmi}{\hbar}\hat{H}_l t_2}\right\rangle=\\
\fl&=&\!\!\int\!\rmd{}\epsilon\, D_{l\sigl}(E^{l\sigl}_{tot}|_{\epsilon})\sum\limits_{\vq|_{\epsilon}}\phi_{l\vq}(\vr)\phi^{*}_{l\vq}(\vr\,')\left(1-f_l(E^{l\sigl}_{tot}|_{\epsilon})\right)\,\rme^{-\frac{\rmi}{\hbar}E^{l\sigl}_{tot}|_{\epsilon}\,t_2}\,,\end{eqnarray*}and in analogue\begin{equation*}\fl\left\langle \hPsi_{l\sigl}^{\dagger}(\vec{r})\hPsi_{l\sigl}(\vec{r}\,'\!\!,-t_2)\right\rangle _{\rm{th}}=\int\!\rmd{}\epsilon\, D_{l\sigl}(E^{l\sigl}_{tot}|_{\epsilon})\sum\limits_{\vq|_{\epsilon}}\phi^{*}_{l\vq}(\vr)\phi^{}_{l\vq}(\vr\,')\,f_l(E^{l\sigl}_{tot}|_{\epsilon})\,\rme^{+\frac{\rmi}{\hbar}E^{l\sigl}_{tot}|_{\epsilon}\,t_2}\,.
\end{equation*}
Here,\begin{eqnarray*}f_l(E^{l\sigl}_{tot}|_{\epsilon})=\Biggl(1+\exp{\frac{E^{l\sigl}_{tot}|_{\epsilon}+eV_l-\tilde{E}^{(zf)}_{F,l}}{\kBT}}\Biggr)^{-1}\end{eqnarray*} is the Fermi function in lead $l$, where $\tilde{E}^{(zf)}_{F,l}$ is the common Fermi level for the two spin species $\sigl=+_l$ and $\sigl=-_l$ in contact $l$ without any bias voltage applied.\\
Additionally inserting the decomposition of the SWCNT electron operator, Eq. (\ref{hPsi_r2}), and introducing the quantities
\begin{equation*}\fl\Phi_{lrr'}(\epsilon)\!:=\!\!\sum_{FF'}\!\sgn(FF')\!\int\!\!\rmd{}^{3}r\!\!\int\!\!\rmd{}^{3}r'\,\,\mathcal{T}^{*}_l(\vr)\mathcal{T}_l(\vr\,')\sum\limits_{\vq|_{\epsilon}}\phi^{*}_{l\vq}(\vr)\phi_{l\vq}(\vr\,')\,{\varphi}_{[\sgn(F)r]F}^{}(\vr){\varphi}_{[\sgn(F')r']F'}^{*}(\vec{r}\,')
\end{equation*}\begin{equation}\fl \quad {\rm{and}} \qquad R_{lrr'}(\epsilon):=\sum_{FF'}\sgn(FF')\int\!\!\rmd{}^{3}r\,\,\Delta_l(\vr)\,{\varphi}_{[\sgn(F)r]F}^{*}(\vr){\varphi}_{[\sgn(F')r']F'}(\vec{r})\,,\label{eq:transparencyreflection}\end{equation}
the rates change to
\begin{eqnarray*}\fl\Gamma^{(\alpha)NN+1}_{l(nkk'j)}&:=&\frac{L_t}{\hbar^2}\sum_{rr'}\sum_{\sigl}\int\rmd{}\epsilon\ \Phi_{lrr'}(\epsilon)\,\left(\hpsi_{\tilde{r}\sigl}\right)_{nk}\left(\psid_{\tilde{r}'\sigl}\right)_{k'j}\times\nonumber\\\fl&&\hspace{5.8cm}\times D_{l\sigl}(E^{l\sigl}_{tot}|_{\epsilon})\,f_l(E^{l\sigl}_{tot}|_{\epsilon})\ e^{+\alpha\frac{i}{\hbar}\left(E^{l\sigl}_{tot}|_{\epsilon}-E_{kj}\right)t_2},\nonumber\\\fl\nonumber\\\fl
\Gamma^{(\alpha)NN-1}_{l(nii'j)}&:=&\frac{L_t}{\hbar^2}\sum_{rr'}\sum_{\sigl}\int\rmd{}\epsilon\ \Phi^*_{lrr'}(\epsilon)\,\left(\psid_{\tilde{r}\sigl}\right)_{ni}\left(\hpsi_{\tilde{r}'\sigl}\right)_{i'j}\times\nonumber\\\fl&&\hspace{2.3cm}\times D_{l\sigl}(E^{l\sigl}_{tot}|_{\epsilon})\,\left(1-f_l(E^{l\sigl}_{tot}|_{\epsilon})\right)\ e^{-\alpha\frac{i}{\hbar}\left(E^{l\sigl}_{tot}|_{\epsilon}-E_{ji}\right)t_2}-\nonumber\\\fl&&\hspace{5cm}-\alpha\frac{\rmi L_t}{\hbar}\sum_{rr'}R_{lrr'}\sum_{\sigl}\sgn(\sigl)\left(\psid_{\tilde{r}\sigl}\right)_{ni}\left(\hpsi_{\tilde{r}'\sigl}\right)_{i'j},\end{eqnarray*}where $E_a-E_b:=E_{ab}$\,, and it was assumed that the tunnelling and the reflection processes take mainly place close to the leads, which justifies to drop the position dependence of the electron operators.\medskip\\
More details on this can be found in the appendix of \cite{LEO}, where it is also nicely explained how some considerations about the main contributions under the integrals in (\ref{eq:transparencyreflection}) and the  fact that the SWCNT is unpolarised with respect to the $\tilde{L}$- and $\tilde{R}$-bands allow to set $\Phi_{lrr'}(\epsilon)=\delta_{rr'}\Phi_l$, and actually in analogue $R_{lrr'}(\epsilon)=\delta_{rr'}R_l$. Then it is rather easy to carry out the integration $\int\rmd{}\epsilon\int_{0}^{\infty}\rmd{}t_2$.\medskip\\
Any formulary tells us for some real function $G(\epsilon)$:
\begin{eqnarray*}
\mathsf{Re}\left(\int\rmd{}\epsilon\,G(\epsilon)\int_{0}^{\infty}\rmd{}t_2\,\,\rme^{\pm \frac{\rmi}{\hbar}(\epsilon-E)t_2}\right)&=&\pi\hbar G(E)\,,\\
\mathsf{Im}\left(\int\rmd{}\epsilon\,G(\epsilon)\int_{0}^{\infty}\rmd{}t_2\,\,\rme^{\pm\frac{\rmi}{\hbar}\left(\epsilon-E\right)t_2}\right)&=&\pm \hbar\pint_{-\infty}^{\infty}\!\!\rmd{}\epsilon\ \frac{G(\epsilon)}{\epsilon-E}\,,\end{eqnarray*}where {\small{$\mathcal{P}$}}\hspace{-0.8em}{\Large{$\int$}} denotes a principal part integration.\\

So finally:\\
\begin{eqnarray}\fl\Gamma^{(\alpha)NN+1}_{l(nkk'j)}&:=&\frac{\pi L_t}{\hbar}\sum_{\tilde{r}}\sum_{\sigl}\,\Phi_{l}\left(\hpsi_{\tilde{r}\sigl}\right)_{nk}\left(\psid_{\tilde{r}\sigl}\right)_{k'j}\times\nonumber\\\fl&&\hspace{3cm}\times\left[D_{l\sigl}(E_{kj})\,f_l(E_{kj})+\alpha\frac{\rmi}{\pi}\left(\pint_{-\infty}^{\infty}\!\!\rmd{}\epsilon\ \frac{D_{l\sigl}(\epsilon)\,f_l(\epsilon)}{\epsilon-E_{kj}}\right)\right]\,,\nonumber\\\fl&&\quad\nonumber\\\fl
\Gamma^{(\alpha)NN-1}_{l(nii'j)}&:=&\frac{\pi L_t}{\hbar}\sum_{\tilde{r}}\sum_{\sigl}\,\Phi_l\left(\psid_{\tilde{r}\sigl}\right)_{ni}\left(\hpsi_{\tilde{r}\sigl}\right)_{i'j}\times\nonumber\\\fl&&\hspace{0.2cm}\times\Biggl[D_{l\sigl}(E_{ji})\,\left(1-f_l(E_{ji})\right)-\alpha\frac{\rmi}{\pi}\left(\pint_{-\infty}^{\infty}\!\!\rmd{}\epsilon\ \frac{D_{l\sigl}(\epsilon)\,\left(1-f_l(\epsilon)\right)}{\epsilon-E_{ji}}+\frac{1}{\Phi_l}R_l\right)\Biggr]\,.\fl\nonumber\\\fl\nonumber\\\fl\label{app_res}\end{eqnarray}


\newpage 

\end{document}